\documentclass[a4paper,11pt]{article}
\pdfoutput=0
\usepackage{jheppub}
\usepackage{tikz}

\usepackage{enumitem}
\usepackage[T1]{fontenc} 
\usepackage{floatrow}
\usepackage{appendix}
\usepackage{tabularx}
\usepackage[normalem]{ulem}

\allowdisplaybreaks

\usepackage{epsf}
\usepackage{amsmath}
\usepackage{amsfonts}
\usepackage{amssymb}
\usepackage{psfrag,epsfig,graphicx,graphics}

\newcommand\numberthis[1][]{%
    \refstepcounter{equation}%
    \ifx#1\empty\else\label{eq:#1}\fi%
    \tag{\theequation}%
}

\usepackage{xargs} 
\usepackage[colorinlistoftodos,prependcaption,textsize=tiny]{todonotes}
\newcommandx{\SW}[2][1=]{\todo[linecolor=OliveGreen,backgroundcolor=OliveGreen!25,bordercolor=OliveGreen,#1]{#2}}

\newcommandx{\LS}[2][1=]{\todo[linecolor=Plum,backgroundcolor=Plum!25,bordercolor=Plum,#1]{#2}}

\newcommandx{\EL}[2][1=]{\todo[linecolor=orange,backgroundcolor=orange!25,bordercolor=orange,#1]{#2}}

\newcommandx{\MF}[2][1=]{\todo[linecolor=blue,backgroundcolor=blue!25,bordercolor=blue,#1]{#2}}

\providecommand{\U}[1]{\protect\rule{.1in}{.1in}}

\def\slashchar#1{\setbox0=\hbox{$#1$}
   \dimen0=\wd0
   \setbox1=\hbox{/} \dimen1=\wd1
   \ifdim\dimen0>\dimen1
      \rlap{\hbox to \dimen0{\hfil/\hfil}}
      #1
   \else
      \rlap{\hbox to \dimen1{\hfil$#1$\hfil}}
      /
   \fi}

\def\bei{\begin{itemize}}
\def\ei{\end{itemize}}

\def\beeq{\begin{eqnarray}} 
\def\beqa{\begin{eqnarray}}
\def\bea{\begin{eqnarray}}

\def\eea{\end{eqnarray}}
\def\eqa{\end{eqnarray}}
\def\eeeq{\end{eqnarray}}

\def\eqar{\end{array}}
\def\beqar{\begin{array}}

\def\beas{\begin{eqnarray*}}
\def\beqas{\begin{eqnarray*}}

\def\eqas{\end{eqnarray*}}
\def\eeas{\end{eqnarray*}}

\def\beq{\begin{equation}} 
\def\be{\begin{equation}}

\def\ee{\end{equation}}
\def\eq{\end{equation}}
\def\eeq{\end{equation}}

\def\beqd{\begin{displaymath}}
\def\eeqd{\end{displaymath}}
\def\eqd{\end{displaymath}}

\def\beeq{\begin{eqnarray}} \def\eeeq{\end{eqnarray}}

\newcommand{\fin}{\end{document}}

\newcommand{\Arrow}[1]{%
\parbox{#1}{\tikz{\draw[->](0,0)--(#1,0);}}
}

\title{\boldmath Diffractive single hadron production in a saturation framework at the NLO}

\author[a,1]{Michael Fucilla, \note{Corresponding author.}}
\author[b,c]{Andrey Grabovsky,}
\author[a]{Emilie Li,}
\author[d]{Lech Szymanowski,}
\author[a]{Samuel Wallon}

\affiliation[a]{Université Paris-Saclay, CNRS/IN2P3, IJCLab, 91405, Orsay, France}
\affiliation[b]{Budker Institute of Nuclear Physics, 11, Lavrenteva avenue, 630090, Novosibirsk, Russia}
\affiliation[c]{Novosibirsk State University, 630090, 2, Pirogova street, Novosibirsk, Russia}
\affiliation[d]{National Centre for Nuclear Research (NCBJ),Pasteura 7, 02-093 Warsaw,  Poland}

\emailAdd{Michael.Fucilla@unical.it}
\emailAdd{A.V.Grabovsky@inp.nsk.su}
\emailAdd{Emilie.Li@ijclab.in2p3.fr}
\emailAdd{Lech.Szymanowski@ncbj.gov.pl}
\emailAdd{Samuel.Wallon@ijclab.in2p3.fr}

\abstract{We calculate the cross-sections of diffractive single hadron photo- or electroproduction with large $p_T$, on a nucleon or a nucleus in the shockwave formalism. We use the hybrid formalism mixing collinear factorization with high energy small-$x$ factorization with the impact factors computed at next-to-leading order accuracy. We prove the cancellation of divergence and we determine the finite parts of the differential cross-sections. We work in general kinematics such that both photoproduction and leptoproduction are considered. The results can be used to detect saturation effects, at both the future EIC or already at LHC, using Ultra-Peripheral Collisions.}

\begin{document} 
\maketitle
\flushbottom

\section{Introduction}
\label{sec:intro}
Gluonic saturation effects in scattering on nucleons and nuclei at small-$x$ represent one of the most intriguing phenomena of strong interactions. In the small-$x$ kinematics, the BFKL dynamics\footnote{For a recent review on tests of BFKL through semi-hard processes involving jets and hadrons, see Ref.~\cite{Celiberto:2020wpk}.}~\cite{Fadin:1975cb,Kuraev:1976ge,
Kuraev:1977fs,
Balitsky:1978ic,
Fadin:1998py,Ciafaloni:1998gs,Fadin:2004zq,Fadin:2005zj} predicts a power-like increase of total cross sections at low values of the Bjorken variable $x=Q^2/s$, where $s$ is the center of mass and $Q^2$ an hard scale. This rise of cross-section is physically interpretable as a constant growth of the gluon density inside the proton. Although this growth has been experimentally observed, confirming the robustness of the BFKL approach, it is equally clear that it must necessarily be interpreted as a pre-asymptotic regime. In fact, at very low values of the $x$ variable, the parton density, per unit of transversal area, in the hadronic wave functions becomes very large leading to the so-called \textit{recombination effects} (not included in the BFKL dynamics). When gluon recombination balances gluon splitting, the density of the latter reaches a \textit{saturation} point, producing new and universal properties of hadronic matter. The state of gluonic matter that is formed is known as \textit{color-glass condensate}\footnote{This state is characterized not only by a high density of particles possessing a color charge (color condensate), but also by a slow evolution compared to the natural time of the interaction and by a disordered field distribution (properties similar to those of a glass).}~\cite{McLerran:1993ni}. The evolution of parton densities must then be described by nonlinear generalizations of the BFKL equation, i.e. the Balitsky --- Jalilian Marian-Iancu-McLerran-Weigert-Leonidov-Kovner (B---JIMWLK) equations~\cite{Balitsky:1995ub, Balitsky:1998kc, Balitsky:1998ya, Balitsky:2001re,JalilianMarian:1997jx,JalilianMarian:1997gr,JalilianMarian:1997dw,JalilianMarian:1998cb,Kovner:2000pt,Weigert:2000gi,Iancu:2000hn,Iancu:2001ad,Ferreiro:2001qy}. In practice, we will rely on the Balitsky shockwave formulation. \\

In the present article, we extend a series of works by us devoted to a complete Next-to-Leading Order (NLO) description of the direct coupling of the Pomeron to several kinds of  diffractive states, namely exclusive diffractive dijet production~\cite{Boussarie:2014lxa, Boussarie:2016ogo, Boussarie:2019ero}, exclusive $\rho$-meson production~\cite{Boussarie:2016bkq}, double hadron production at large $p_T$~\cite{Fucilla:2022wcg}.

In the same spirit as in Ref.~\cite{Fucilla:2022wcg}, this study is  motivated by present and future possibilities of accessing gluonic saturation through large-$p_T$ single hadron production. 
The novelty of the present study, as we will show in detail in this article, is that passing from dihadron to single hadron production, i.e. increasing the level of inclusivity, changes rather significantly the structure of the cancellation of IR divergencies. At the parton level, one indeed faces contributions with 
 one (at LO and NLO) or two (at NLO) spectator partons, contrarily to the case of dihadron production.

Similarly to the case of dihadron production, this process could be studied both in photoproduction and leptoproduction. One should focus on the window which is both perturbative, the hard scale being provided either by the large virtuality $Q^2$ of the virtual photon  (in the leptoproduction case) and/or the large $p_T$ of the produced hadron, and subject to saturation effects, characterized by the scale $Q_s^2 \simeq (A/x)^{1/3}$ where $A$ is the mass number of the nucleus. This could thus be achieved at the LHC in $pA$ and $AA$ scattering, using Ultra-Peripheral Collisions (UPC), as well as at the EIC, where both  photoproduction and leptoproduction could be considered.

\section{Theoretical framework}
\label{sec: framework}
\subsection{Hybrid collinear/high-energy factorization}
\begin{figure}
\begin{picture}(430,120)
\put(0,0){\includegraphics[scale=0.36]{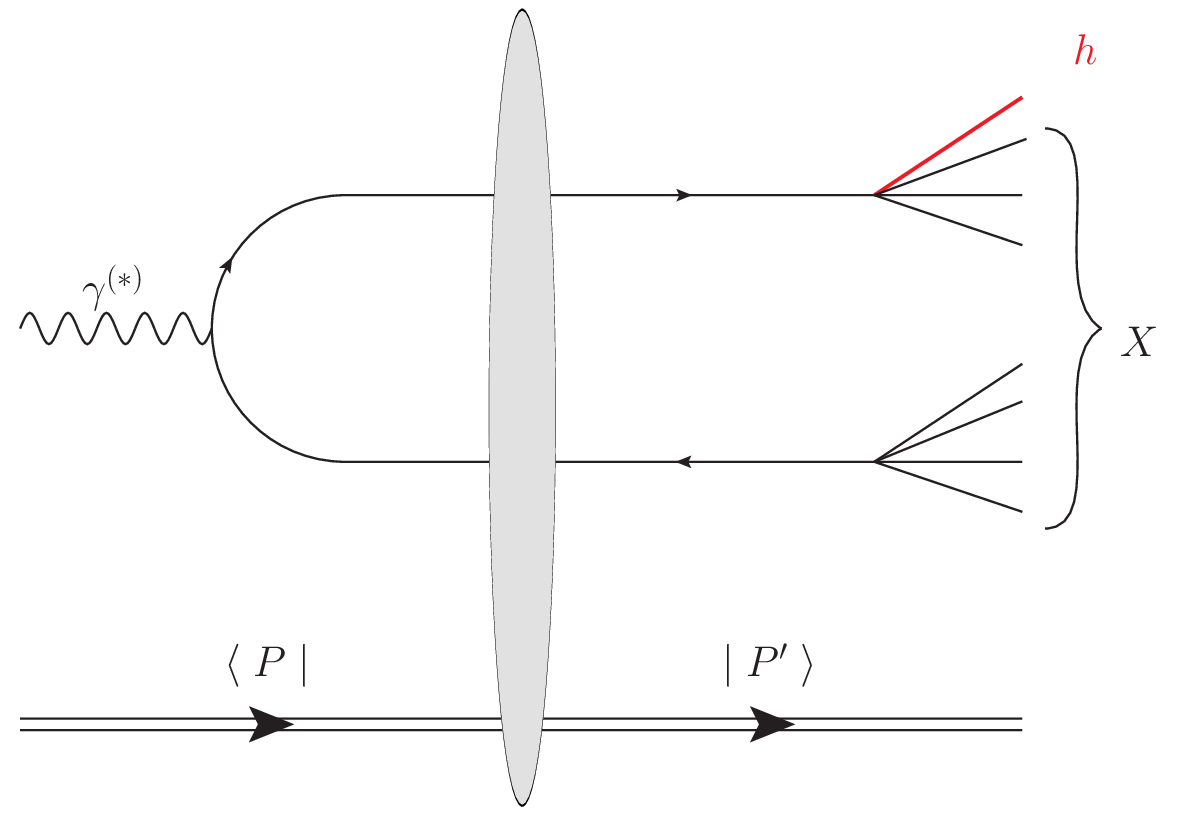}}
\put(230,0){\includegraphics[scale=0.36]{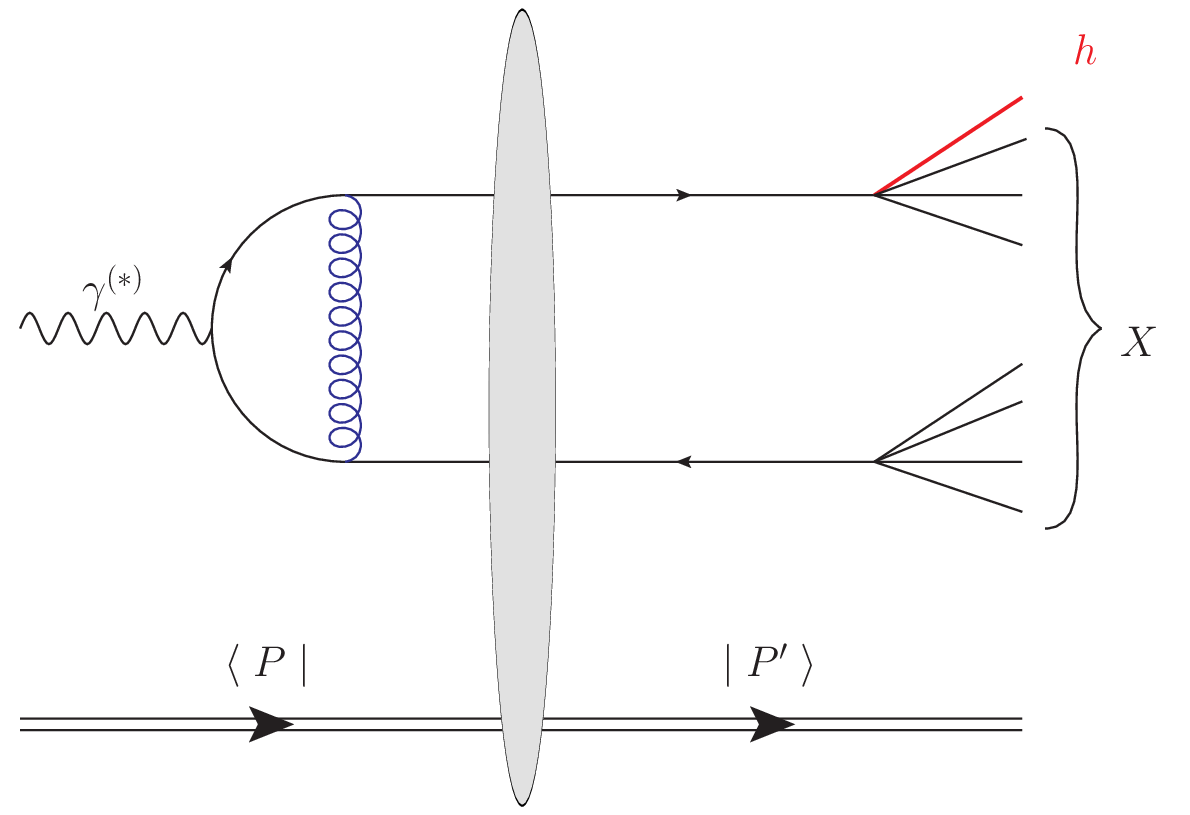}}
\put(145,30){rapidity gap}
\put(375,30){rapidity gap}
\end{picture} 
 \caption{Left: An example of amplitude of the process (\ref{Eq:process}) at LO. Right: An example of amplitude contributing to the process (\ref{Eq:process}) at NLO. The grey blob symbolizes the QCD shockwave. The double line symbolizes the target, which remains intact in the figure, but could just as well break. The quark and the antiquark fragment into the systems $(h X)$ (in the specific diagram $h$ is produced by the quark, but can be as well produced by the anti-quark). The tagged hadron $h$ is drawn in red.}
\label{fig:process}
\end{figure}

In the present paper, we focus on the computation  at full NLO of the semi-inclusive diffractive hadron production in the high energy limit, namely
\begin{equation}
\label{Eq:process}
    \gamma^{(*)}(p_\gamma) + P(p_0) \rightarrow h(p_{h}) + X + P'(p_{0'})
\end{equation}
where $P$ is a nucleon or a nucleus target, generically called proton in the following. The initial photon plays the role of a probe (also named projectile). Our computation applies both to the photoproduction case (including ultra-peripheral collisions) and to the electroproduction case (e.g. at EIC). A gap in rapidity is assumed between the outgoing nucleon/nucleus $(P')$ and the diffractive system $(X h)$. This is illustrated by Fig.~\ref{fig:process}.

We will be working in a combination of collinear factorization and small-$x$ factorization, more precisely in the shockwave formalism for the latter. 

\subsubsection*{Kinematics}

We introduce a light-cone basis composed of $n_1 $ and $n_2$, with $n_1 \cdot n_2 = 1$ defining the $+/-$ direction.
We write the Sudakov decomposition for any vector as 
\begin{equation}
p^\mu = p^+ n_1^\mu + p^- n_2^\mu + p_\perp^\mu
\end{equation}
and the scalar product of two vectors as\footnote{Any transverse momentum in Euclidean space will be denoted with an arrow, while a $\perp$ index will be used in Minkowski space.} 

\begin{equation}
    \begin{aligned}
    p \cdot q &= p^+ q^- + p^- q^+ + p_\perp \cdot q_\perp \\ 
    &= p^+ q^- + p^- q^+ -\vec{p} \cdot \vec{q}\,.
    \end{aligned} 
\end{equation}
We work in a reference frame, called probe frame\footnote{Although the probe may itself move relativistically in this frame.} such that the target moves ultra-relativistically and such that 
$s = (p_\gamma + p_0)^2 \sim 2 p_\gamma^+ p_0^- \gg \Lambda_{\text{QCD}}^2$, $s$ also being larger than any other scale and $p_\gamma^+ \sim p_0^- \sim \sqrt{s}$. Particles on the projectile side are moving in the $n_1$ (i.e. $+$) direction while particles on the target side have a large component along $n_2$ (i.e. $-$ direction).

We will use kinematics such that the photon with virtuality $Q$ is forward, and thus it does not carry any transverse momentum:
\begin{equation}
\vec{p}_{\gamma}=0,\quad p_{\gamma}^{\mu}=p_{\gamma}^{+}n_{1}^{\mu}+\frac{p_{\gamma}^{2}}{2p_{\gamma}^{+}}n_{2}^{\mu},\quad-p_{\gamma}^{2}\equiv Q^{2}\geq 0. \label{photonk}
\end{equation}
We will denote its transverse polarization $\varepsilon_T$. Its longitudinal polarization vector reads
\begin{equation}
\varepsilon_{L}^{\alpha}=\frac{1}{\sqrt{-p_{\gamma}^{2}}}\left(  p_{\gamma
}^{+}n_{1}^{\alpha}-\frac{p_{\gamma}^{2}}{2p_{\gamma}^{+}}n_{2}^{\alpha
}\right)  ,\quad\varepsilon_{L}^{+}=\frac{p_{\gamma}^{+}}{Q},\quad
\varepsilon_{L}^{-}=\frac{Q}{2p_{\gamma}^{+}}.
\end{equation}
We write the momentum of the produced hadron as
\beqa
\label{ph}
p^\mu_{h}=p^+_{h} n_1^\mu + \frac{m_{h}^2 + \vec{p}_{h}^{\,2}}{2 p^+_{h}} n_2^\mu + p^\mu_{h \perp} \,.
\eqa
The momenta of the fragmenting quark of virtuality $p_q^2$ reads
\beqa
\label{pq}
p^\mu_q=p^+_q n_1^\mu + \frac{p_q^2+\vec{p}_{q}^{\,2}}{2 p^+_{q}} n_2^\mu + p^\mu_{q\perp}\, ,
\eqa
similarly for an antiquark of virtuality $p_{\bar{q}}^2$ we have
\beqa
\label{pqbar}
p^\mu_{\bar{q}}=p^+_{\bar{q}} n_1^\mu + \frac{p^2_{\bar{q}}+\vec{p}_{\bar{q}}^{\,2}}{2 p^+_{\bar{q}}} n_2^\mu + p^\mu_{\bar{q}\perp}\, ,
\eqa
and, finally, for a gluon appearing at NLO level, we can write
\beqa
\label{g}
p^\mu_{g}=p^+_{g} n_1^\mu + \frac{p^2_{g}+\vec{p}_{g}^{\,2}}{2 p^+_{g}} n_2^\mu + p^\mu_{g \perp}\, .
\eqa
From now, we will use the notation $p_{ij}=p_i-p_j$ and $z_{ij} = z_i - z_j$.

\subsubsection*{Collinear factorization}

We consider the 
 kinematical region in which $\vec{p}_{h}^{\,2} \gg \Lambda_{\text{QCD}}^2$. This transverse hadron momentum provides  the hard scale, justifying the use of perturbative QCD and collinear factorization. In the hard part, after collinear factorization, the quark and antiquark  can be treated as on-shell particles. We later on use  the longitudinal momentum fraction $x_q$ and $x_{\bar{q}}$, defined as
\beqa
\label{xq-xqbar}
p_q^+ = x_q p^+_{\gamma} \quad \hbox{ and } \quad p_{\bar{q}}^+ = x_{\bar{q}} p^+_{\gamma}\,.
\eqa
We also denote
\beqa
\label{xh}
p_{h}^+ = x_{h} p^+_{\gamma} \,.
\eqa

\subsubsection*{Shockwave approach}

We now shortly present the shockwave formalism, an effective approach to deal with gluonic saturation.\\

In this effective field theory, the gluonic field $A$ is separated into external background fields $b$ (resp. internal fields $\mathcal{A}$) depending on whether their $+$-momentum is below (resp. above) the arbitrary rapidity cut-off $e^\eta p_\gamma^+$, with $\eta < 0$. This effective field theory dramatically simplifies when using the light-cone gauge $n_2 \cdot A = 0$.
The external field, after being highly boosted from the target rest frame to the probe  frame, take the form 
\begin{equation}
    b^\mu (x) = b^-(x_\perp) \delta (x^+) n_2^\mu \,.
\end{equation}

The resummation of all order interactions with those fields leads to a high-energy Wilson line, that represents the shockwave and is located exactly at $x^+ =0$:
\begin{equation}
    U_{\vec{z}} = \mathcal{P} \exp \left(i g \int d z^+ b^-(z)\right)\,,
\end{equation}
where $\mathcal{P}$ is the usual path ordering operator for the $+$ direction.

Relying on the small-$x$ factorization, the scattering amplitude can be written as the convolution of the projectile impact factor with the non-perturbative matrix element of operators from the Wilson line operators on the target states. 

For the present process, we will deal with two kinds of operators. 
The first one is the dipole operator, which in the fundamental representation of $SU(N_c)$ takes the form:
\begin{equation}
\left[\operatorname{Tr} \left(U_1 U_2^\dag\right)-N_c\right]\left(\vec{p_1},\vec{p}_2\right) = \int d^d \vec{z}_{1} d^d \vec{z}_{2\perp} e^{- i \vec{p}_1 \cdot \vec{z}_1} e^{- i \vec{p}_2 \cdot \vec{z}_2} \left[\operatorname{Tr} \left(U_{\vec{z}_1} U_{\vec{z}_2}^\dag\right)-N_c\right]\,,
\end{equation}
where
$\vec{z}_{1,2}$ are the transverse positions of the $q,\bar{q}$ coming from the photon and $\vec{p}_{1,2}$ their respective transverse momentums kicks from the shockwave.

The proton matrix element can be parameterized through a generic function $F$, following the definition of
Ref.~\cite{Boussarie:2016ogo} 
\begin{eqnarray}
\left\langle P^{\prime}\left(p_{0^{\prime}}\right)\left|T\left(\operatorname{Tr}\left(U_{\frac{z_{\perp}}{2}} U_{-\frac{z_{\perp}}{2}}^{\dagger}\right)-N_{c}\right)\right| P\left(p_{0}\right)\right\rangle
& \equiv & 2 \pi \delta\left(p_{00^{\prime}}^{-}\right) F_{p_{0 \perp} p_{0^{\prime} \perp}}\left(z_{\perp}\right) \nonumber \\
   & \equiv & 2 \pi \delta\left(p_{00^{\prime}}^{-}\right) F\left(z_{\perp}\right) 
\end{eqnarray}
and its Fourier Transform (FT) is
\begin{equation}
\label{eq:FTF}
\int d^{d} z_{\perp} e^{i\left(z_{\perp} \cdot p_{\perp}\right)} F\left(z_{\perp}\right) \equiv \mathbf{F}\left(p_{\perp}\right).
\end{equation}

The second operator we will deal with is
 the double dipole operator. Its action on proton states, as can be seen with eqs.~(5.3) and (5.6) in \cite{Boussarie:2016ogo}, can be written as
\begin{eqnarray}
    && \left\langle P^{\prime}\left(p_{0^{\prime}} \right)\left|\left(\operatorname{Tr}\left(U_{\frac{z_\perp}{2}} U_{x_\perp}^{\dagger}\right) \operatorname{Tr}\left(U_{x_\perp} U_{-\frac{z_\perp}{2}}^{\dagger}\right)-N_c \operatorname{Tr}\left(U_{\frac{z_\perp}{2}} U_{-\frac{z_\perp}{2}}^{\dagger}\right)\right)\right| P\left(p_0\right)\right\rangle  \nonumber \\
     &&\equiv 2 \pi \delta\left(p_{00^{\prime}}^{-}\right) \tilde{F}_{p_{0 \perp} p_{0^{\prime} \perp}}\left(z_{\perp}, x_{\perp}\right) \equiv 2 \pi \delta\left(p_{00^{\prime}}^{-}\right) \tilde{F}\left(z_{\perp}, x_{\perp}\right)
\end{eqnarray}
and its FT is
\begin{equation}
\int d^d z_{\perp} d^d x_{\perp} e^{i\left(p_{\perp} \cdot x_{\perp}\right)+i\left(z_{\perp} \cdot q_{\perp}\right)} \tilde{F}\left(z_{\perp}, x_{\perp}\right) \equiv \tilde{\mathbf{F}}\left(q_{\perp}, p_{\perp}\right).
\end{equation}

In this paper, dimensional regularization will be used with $D=2 + d$, where $d= 2+2\epsilon$ is the transverse dimension.

\subsection{LO computation}
\label{Sec:LO_computation}
\begin{figure}
\begin{picture}(430,70)
\put(130,1){\includegraphics[scale=0.35]{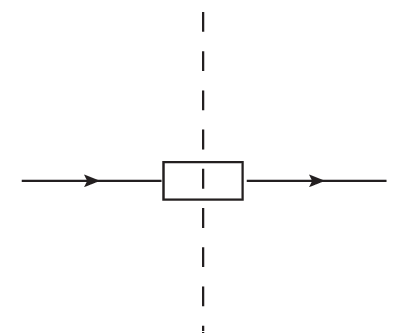}}
\put(212,24){$\equiv$}
\put(230,0){\includegraphics[scale=0.35]{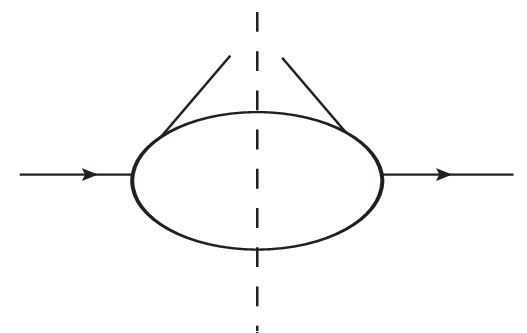}}
\put(255,45){$h$}
\end{picture} 
 \caption{Graphical convention for the fragmentation function of a parton (here a quark for illustration) to a hadron $h$ plus spectators. In the rest of this article, we will use the left-hand side of this drawing.}
    \label{fig:fragmentation}
\end{figure}

We start from the usual collinear factorization of the hadronic cross section for the production of a single hadron which, at LO and leading twist, reads~\cite{Altarelli:1979kv} 
\begin{equation}
\label{eq: coll facto}
\frac{d \sigma_{0JI}^{q \rightarrow h}}{d x_{h}} = \sum_{q} \int_{x_{h}}^1 \frac{d x_q}{x_q} D_q^{h}\left(\frac{x_{h_1}}{x_q},\mu_F\right) \frac{d\hat{\sigma}_{0JI}}{d x_q} \; ,
\end{equation}
where $q$ specifies the quark or anti-quark flavor types ($q=u,\bar{u},d,\bar{d},s,\bar{s},c,\bar{c},b,\bar{b}$), and $J,I=L,T$ specify the photon polarization since we deal here with a modulus square amplitude ($J$ labels the photon polarization in the complex conjugated amplitude and $I$ in the amplitude). Here $\mu_F$ is the factorization scale, $D_{q}^{h}$ denotes the quark (or antiquark) Fragmentation Function (FF) and $d\hat{\sigma}$ is the cross-section for the production of partons\footnote{To be precise, it contains proton matrix elements and hence it is not exactly the partonic cross section, but it is the cross section for the production of the parton pair from which the fragmentation in the identified hadron subsequently occurs.}, i.e. the cross-section for the subprocess 
\begin{equation}
\label{partonic_LO}
     \gamma^{(*)}(p_\gamma) + P(p_0) \rightarrow q (p_q) + \bar{q}(p_{\bar{q}})  + P'(p_{0'}) \,.
\end{equation}
Following the convention of our previous work~\cite{Fucilla:2022wcg} we denote the fragmentation process by a small rectangle as in Fig.~\ref{fig:fragmentation}. \\

For illustrative purposes, and simplicity of notation, let us consider the case in which the hadron fragments starting from a quark. The case of anti-quark is completely identical. Collinear factorization means that the produced hadron should fly collinearly to the fragmenting parton, we then have the following constraints

\beqa
\label{constraint-collinear-q}
p^+_q &=& \frac{x_q}{x_{h}} p^+_{h},    \quad \vec{p}_q = \frac{x_q}{x_{h}} \vec{p}_{h} \, .
\eqa
To keep things quite general with regard to photon polarization, and therefore to be able to describe photo- and electroproduction, we build the polarization matrix
\begin{equation}
\label{eq:density_matrix}
d\sigma_{JI}=
\begin{pmatrix}
d\sigma_{LL} & d\sigma_{LT}\\
d\sigma_{TL} & d\sigma_{TT}
\end{pmatrix}
,\qquad d\sigma_{TL}=d\sigma_{LT}^{\ast}\,.
\end{equation}
Each element of this matrix has a LO contribution $d\sigma_{0JI}$.
This Born order result, see Eq.~(5.14) of Ref.~\cite{Boussarie:2016ogo}, has the following structure:
\begin{align}  
d \hat{\sigma}_{0JI}  & =  \frac{\alpha_{\mathrm{em}}Q_{q}^{2}}{2 \left(2\pi\right)^{4d}N_{c}}\frac{\left(p_{0}^{-}\right)^{2}}{2x_q x_{\bar{q}} s^{2}}d x_q d x_{\bar{q}} d^{d}p_{q\perp}d^{d}p_{\bar{q}\perp}\delta\left(1-x_q-x_{\bar{q}} \right)\left(\varepsilon_{I\beta}\varepsilon_{J\gamma}^\ast\right)\nonumber \\
& \quad  \times  \int d^{d}p_{1\perp}d^{d}p_{2\perp}d^{d}p_{1^{\prime}\perp}d^{d}p_{2^{\prime}\perp}\delta\left(p_{q1\perp}+p_{\bar{q}2\perp}\right)\delta\left(p_{11^{\prime}\perp}+p_{22^{\prime}\perp}\right)\nonumber \\
& \quad  \times  \sum_{\lambda_q,\lambda_{\bar{q}}}\Phi_{0}^{\beta}\left(p_{1\perp},\, p_{2\perp}\right)\Phi_{0}^{\gamma*}\left(p_{1^{\prime}\perp},\, p_{2^{\prime}\perp}\right)\mathbf{F}\left(\frac{p_{12\perp}}{2}\right)\mathbf{F^{*}}\left(\frac{p_{1^{\prime}2^{\prime}\perp}}{2}\right) .
\label{dsigma0}
\end{align}
\begin{figure}
\centering
 \includegraphics[scale=0.35]{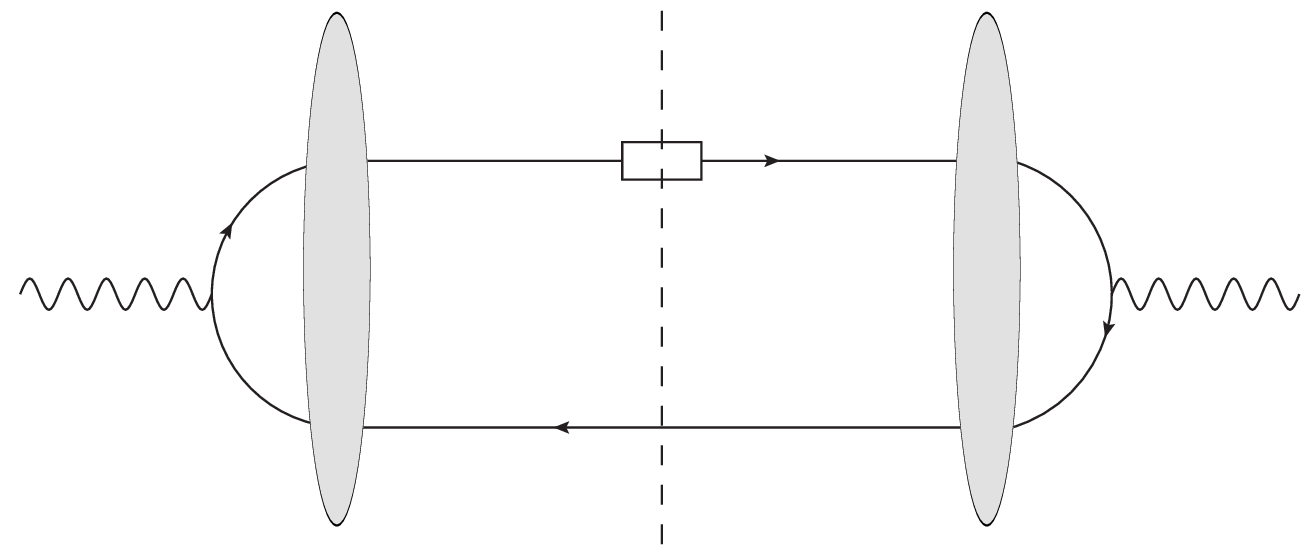}
 \caption{Diagram of the LO process at cross-section level. The blob is the shockwave (we do not draw the coupling with the target for clarity) and the square the FF, see Fig.~\ref{fig:fragmentation}. The dashed line  represents the integration over phase space.}
    \label{fig:LO}
\end{figure}
It is important to note that the formula (\ref{dsigma0}) is divided by a factor of $1/(2 (2 \pi)^4)$ with respect to Eq.~(5.14) of Ref.~\cite{Boussarie:2016ogo}. This is necessary to get proper normalization which is missing in Ref.~\cite{Boussarie:2016ogo} due to a misprint. This same division must be applied to the cross section expressions in Ref.~\cite{Fucilla:2022wcg}, where this misprint propagated. Using the explicit expressions of the product $\Phi_{0}^{\beta}\Phi_{0}^{\gamma*}$, see Eqs.~(\ref{eq:LeadingCS_LL}, \ref{eq:LeadingCS_TL}, \ref{eq:LeadingCS_TT}) in Appendix~\ref{LO impact factor squared}, the LO cross-sections are obtained and read \vspace{0.1 cm} 
\begin{equation}
   \label{eq:LL-LO}
\frac{d \sigma_{0 J I}^{q \rightarrow h}}{d x_{h} d^d p_{h \perp} }  = \frac{2 \alpha_{\mathrm{em}} Q^2 }{(2\pi)^{4d}N_c \; x_h^d } \sum_{q}  Q_q^2 \int_{x_{h}}^1 \hspace{- 0.25 cm} d x_q \;  x_q^{1+d} (1-x_{q})^2 D_q^{h} \left(\frac{x_{h}}{x_q} \right) f_{JI} \,, 
\end{equation}
where
\begin{equation}
\label{f-LL}
f_{LL} = \int d^{d} p_{ \bar{q} \perp} \int d^{d} p_{2 \perp} \frac{\mathbf{F}\left(\frac{x_q}{2x_{h}}  p_{h \perp} + \frac{1}{2} p_{ \bar{q} \perp} - p_{2 \perp} \right)}{ \vec{p}_{\bar{q} 2}^{\; 2} + x_q (1-x_{q}) Q^{2}} \int d^{d} p_{2' \perp} \frac{\mathbf{F}^{*} \left(\frac{x_q}{2x_{h}}  p_{h \perp} + \frac{1}{2} p_{ \bar{q} \perp} - p_{2' \perp} \right)}{ \vec{p}_{\bar{q} 2'}^{\; 2} + x_q (1-x_{q}) Q^{2}} \, ,
\end{equation}
\begin{align}
f_{TL} & = \int d^{d} p_{ \bar{q} \perp} \int d^{d} p_{2 \perp} \frac{\mathbf{F}\left(\frac{x_q}{2x_{h}}  p_{h \perp} + \frac{1}{2} p_{ \bar{q} \perp} - p_{2 \perp} \right)}{ \vec{p}_{\bar{q} 2}^{\; 2} + x_q (1-x_{q}) Q^{2}} \int d^{d} p_{2' \perp} \frac{\mathbf{F}^{*} \left(\frac{x_q}{2x_{h}}  p_{h \perp} + \frac{1}{2} p_{ \bar{q} \perp} - p_{2' \perp} \right)}{ \vec{p}_{\bar{q} 2'}^{\; 2} + x_q (1-x_{q}) Q^{2}} \nonumber \\
& \times \frac{(1 - 2 x_q)}{2 x_q (1-x_q) Q} \left( \vec{p}_{\bar{q} 2'} \cdot \vec{\varepsilon}_T \right)^{*} \; ,
\label{f-TL}
\end{align}
\begin{align}
f_{TT} & = \int d^{d} p_{ \bar{q} \perp} \int d^{d} p_{2 \perp} \frac{\mathbf{F} \left(\frac{x_q}{2x_{h}}  p_{h \perp} + \frac{1}{2} p_{ \bar{q} \perp} - p_{2 \perp} \right)}{ \vec{p}_{\bar{q} 2}^{\; 2} + x_q (1-x_{q}) Q^{2}} \int d^{d} p_{2' \perp} \frac{\mathbf{F}^{*} \left(\frac{x_q}{2x_{h}}  p_{h \perp} + \frac{1}{2} p_{ \bar{q} \perp} - p_{2' \perp} \right)}{ \vec{p}_{\bar{q} 2'}^{\; 2} + x_q (1-x_{q}) Q^{2}} \nonumber \\ 
& \times \left[ (1 -2x_q )^2 g_{\perp}^{ri} g_\perp^{lk} - g_\perp^{rk} g_\perp^{li} + g_{\perp}^{rl} g_\perp^{ik} \right] \frac{ \varepsilon_{T i} \; p_{\bar{q} 2 \perp r} \left( \varepsilon_{T k} \; p_{\bar{q} 2' \perp l} \right)^{*} }{4 x_q^2 (1-x_q)^2 Q^2} \; ,
\label{f-TT}
\end{align}
are the three different functions for the $LL$, $TL$ and $TT$ cross-sections, respectively and the sum over $q$ is extended to the five quark flavor species ($q=u,d,s,c,b$). For compactness, we use the short notation 
\begin{equation}
    f_{JI} (x_q, x_h, \vec{p}_h, Q^2) \equiv f_{JI} \; .
\end{equation}
The correct cross section, in the case of anti-quark fragmentation, is obtained by including a minus sign in the argument of the function \textbf{F}, extending the sum over $q$ to the five anti-quark flavor species ($q=\bar{u},\bar{d},\bar{s},\bar{c},\bar{b}$) and performing the relabelling $(x_q, \vec{p}_q, p_2, p_{2'}) \leftrightarrow (x_{\bar{q}}, \vec{p}_{\bar{q}}, p_1, p_{1'})$\footnote{Since the variables involved are all integration variables, this last operation is not necessary at the LO level and in some NLO contributions, but we will always do it for clarity of notation.}. We will call this last operation $(q \leftrightarrow \bar{q})$ relabelling.

\subsection{NLO computations in a nutshell}
\subsubsection{Different mechanisms of fragmentation}
At the next-to-leading order there are six kinds of contributions to the cross-section
\begin{itemize}
    \item[(a)] $\gamma^{*} + P \rightarrow h + \bar{q} + X + P$ cross-section at one-loop (i.e. virtual contribution and fragmentation from a quark) ,
    \item[(b)] $\gamma^{*} + P \rightarrow h + q + X + P$ cross-section at one-loop (i.e. virtual contribution and fragmentation from an anti-quark) , 
    \item[(c)] $\gamma^{*} + P \rightarrow h + \bar{q} + g + X + P$ cross-section at Born level (i.e. real contribution and fragmentation from a quark) ,
    \item[(d)] $\gamma^{*} + P \rightarrow h + q + g + X + P$ cross-section at Born level (i.e. real contribution and fragmentation from an anti-quark) ,
    \item[(e)] $\gamma^{*} + P \rightarrow h + q + \bar{q} + X + P$ cross-section at Born level (i.e. real contribution and fragmentation from a gluon) ,
    \item[(f)] FFs counterterms .
\end{itemize}
\subsubsection{Hard cross section}
\begin{figure}
\begin{picture}(430,470)
\put(180,470){\fbox{virtual contributions}}
\put(10,400){\includegraphics[scale=0.25]{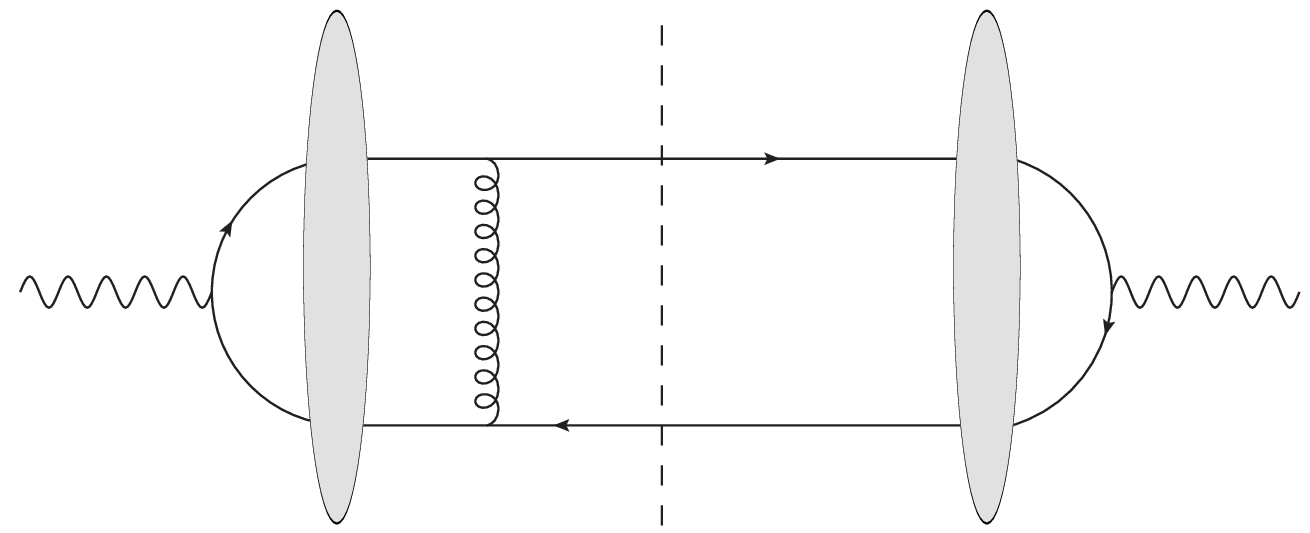}}
\put(180,425){$\Arrow{1cm}$}
\put(230,425){$d\sigma_{1IJ}$}
\put(280,425){dipole $\times$ dipole}

\put(10,300){\includegraphics[scale=0.25]{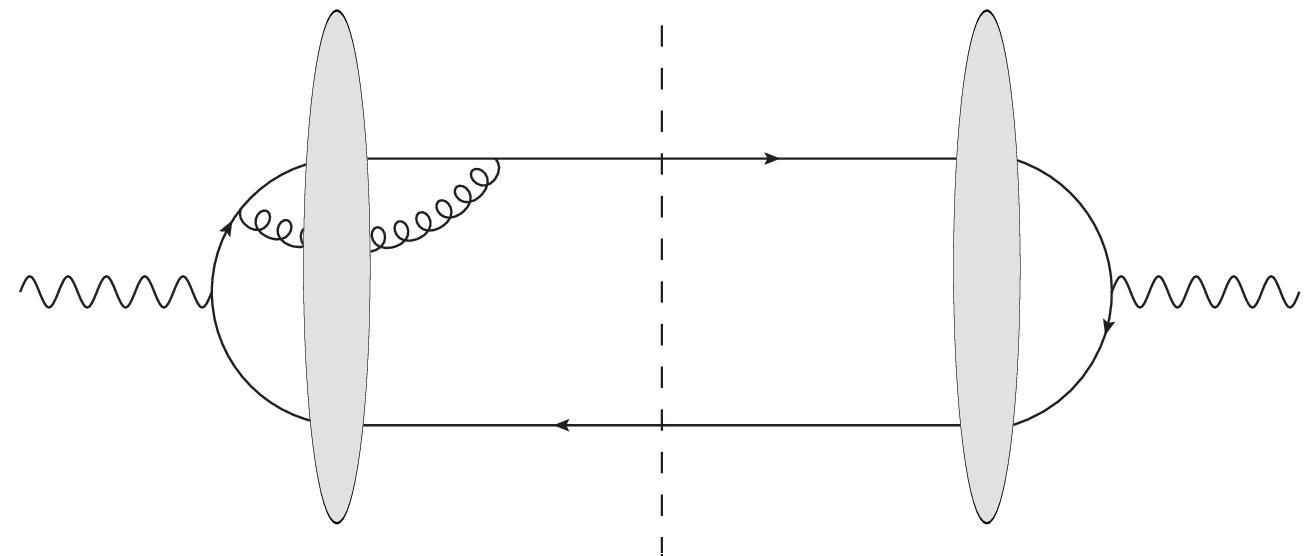}}
\put(180,350){\rotatebox{58}{$\Arrow{1.7cm}$}}
\put(180,325){$\Arrow{1cm}$}
\put(230,325){$d\sigma_{2IJ}$}
\put(280,325){double dipole $\times$ dipole}

\put(180,270){\fbox{real contributions}}

\put(10,200){\includegraphics[scale=0.25]{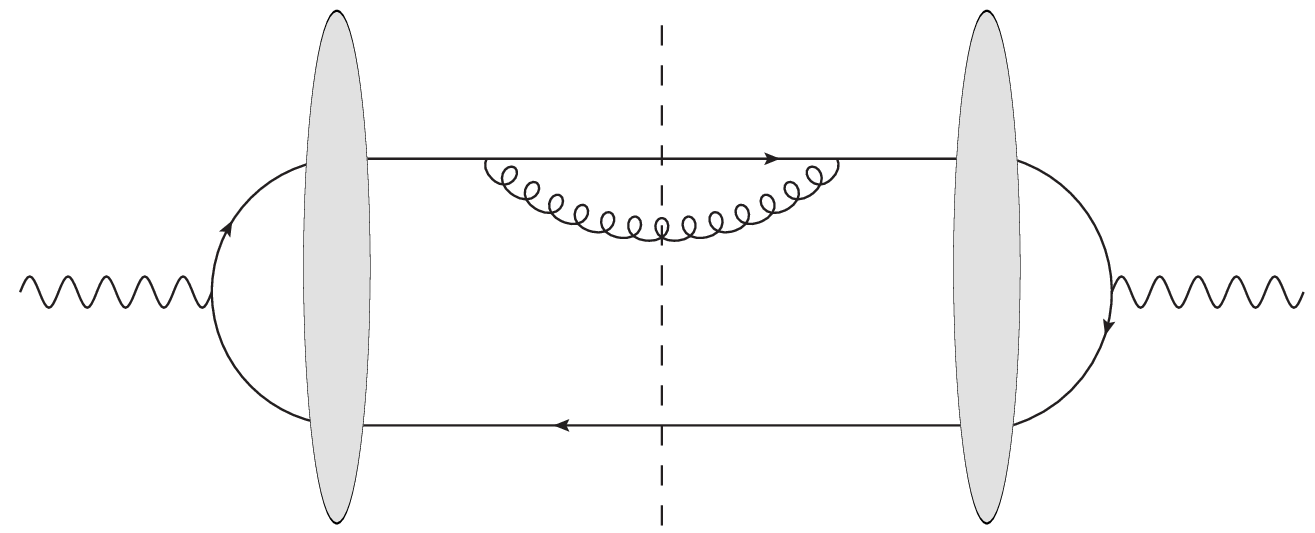}}
\put(180,225){$\Arrow{1cm}$}
\put(230,225){$d\sigma_{3IJ}$}
\put(280,225){dipole $\times$ dipole}

\put(10,100){\includegraphics[scale=0.25]{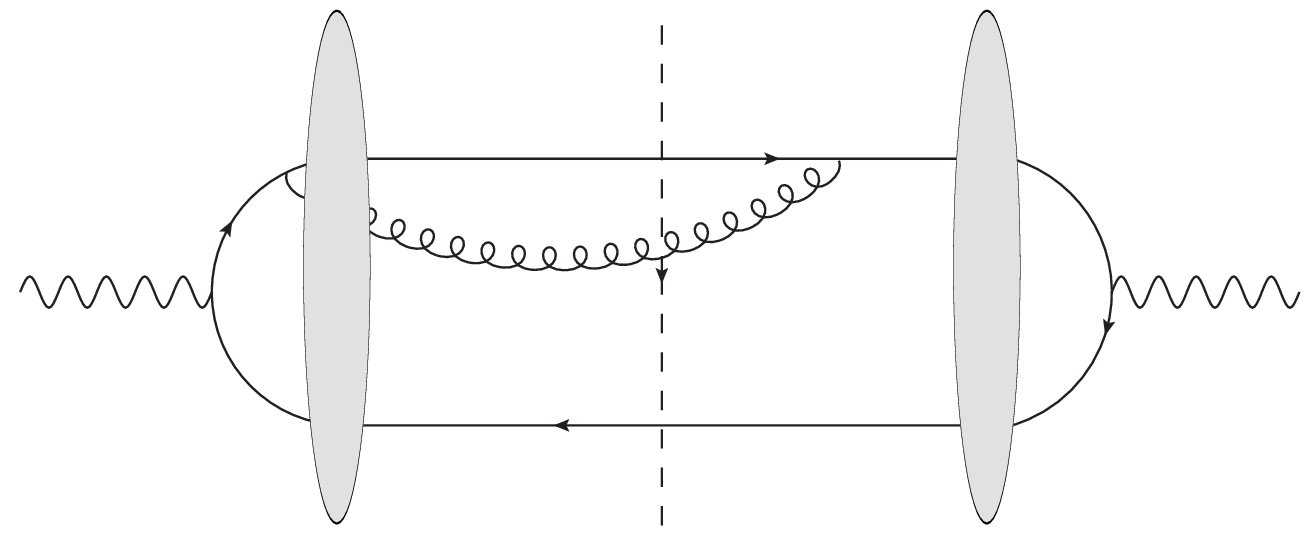}}
\put(180,125){$\Arrow{1cm}$}
\put(180,150){\rotatebox{58}{$\Arrow{1.7cm}$}}
\put(230,125){$d\sigma_{4IJ}$}
\put(280,125){double dipole $\times$ dipole}

\put(10,0){\includegraphics[scale=0.25]{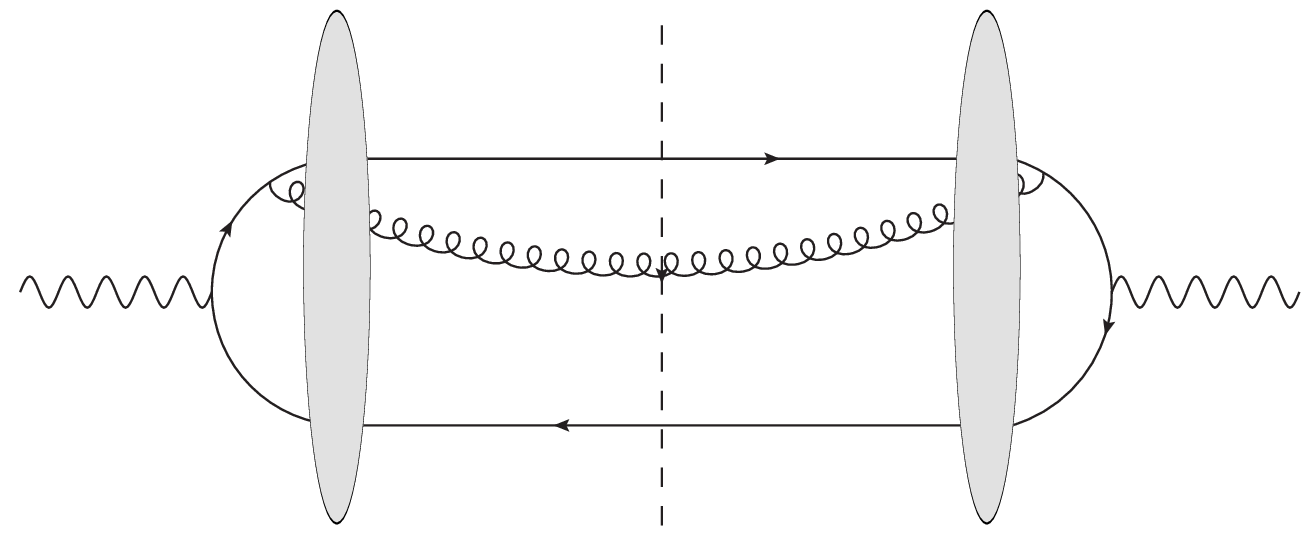}}
\put(180,25){$\Arrow{1cm}$}
\put(180,50){\rotatebox{58}{$\Arrow{1.7cm}$}}
\put(180,70){\rotatebox{75}{$\Arrow{4cm}$}}
\put(230,25){$d\sigma_{5IJ}$}
\put(280,25){double dipole $\times$ double dipole}

\end{picture}
\vspace{.1cm}
\caption{Illustration of the 5 kinds of  contributions to the NLO hard cross-section from the dipole point of view. Arrows show to which combination of dipole structures each type of diagrams contributes.}
  \label{fig:sigma-NLO-dipole}
\end{figure}
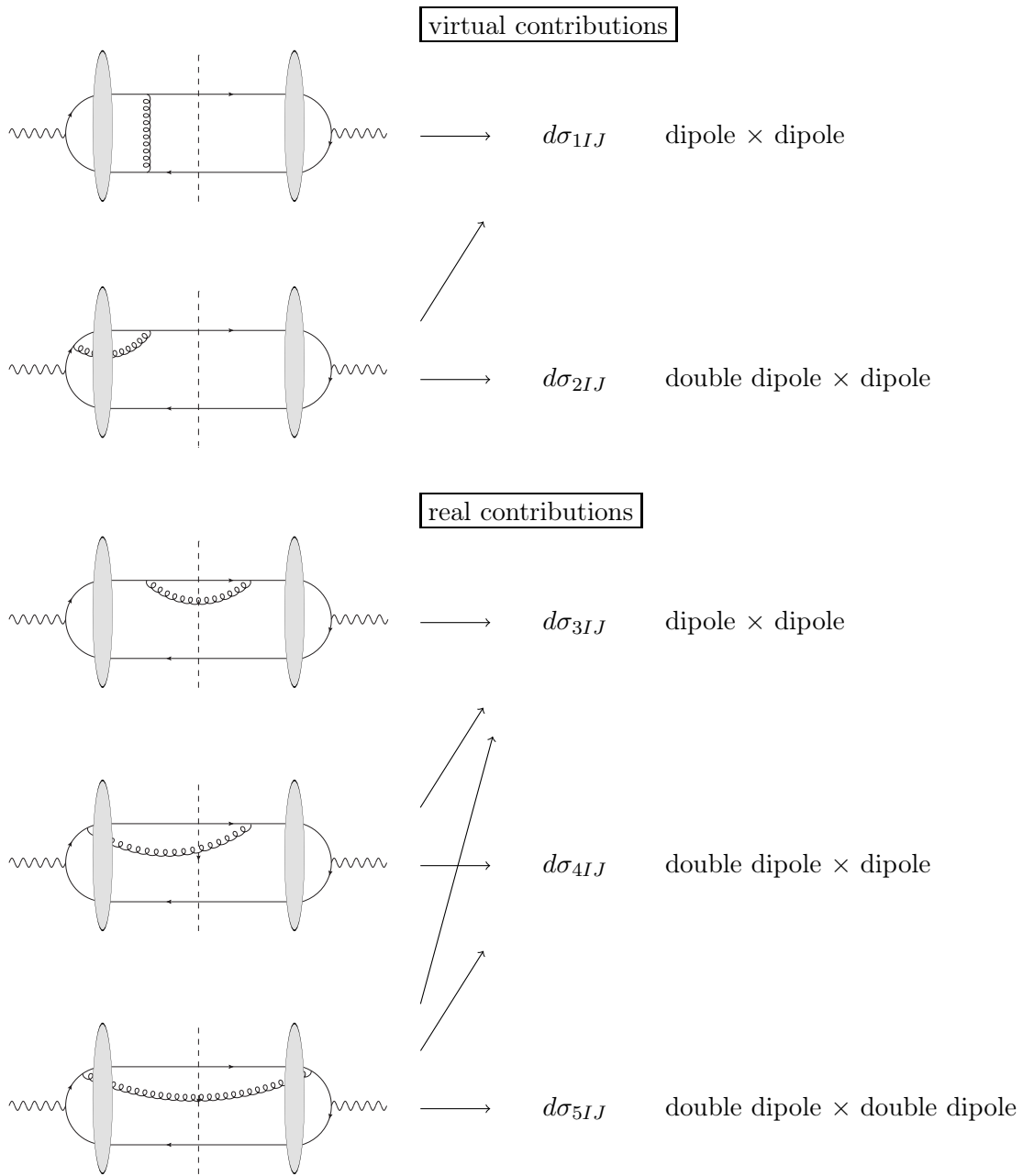

At NLO, since we rely on the shockwave approach, it is convenient to separate the various contributions from the dipole point of view, as illustrated in Fig.~\ref{fig:sigma-NLO-dipole}. In this figure, we exhibit a few examples of diagrams, either virtual or real, as a representative of each 5 classes of diagrams. There are indeed 5 classes of contributions from the dipole point of view, namely $d \sigma_{iJI}\, (i=1,\cdots 5)$, so that the NLO polarization matrix can be written as
\begin{equation}
d\sigma_{JI}=d\sigma_{0JI}+d\sigma_{1JI}+d\sigma_{2JI}+d\sigma_{3JI}+d\sigma_{4JI}+d\sigma_{5JI}. 
\label{sigmaNLO}
\end{equation}
Now, we will shortly discuss each of these 5 NLO corrections. \\

For the virtual diagrams, there are two classes of diagrams: the diagrams in which the virtual gluon does not cross the shockwave, thus contributing to $d \sigma_{1IJ}$, purely made of dipole $\times$ dipole terms; the diagrams in which the virtual gluon does cross the shockwave, contributing both to $d \sigma_{1IJ}$, made of dipole $\times$ dipole terms as well as to $d \sigma_{2IJ}$, made of double dipole $\times$ dipole  (and dipole $\times$ double dipole) terms. \\

For the real diagrams, there are three classes of diagrams: the diagrams in which the real gluon does not cross the shockwave, thus contributing to $d \sigma_{3IJ}$, purely made of dipole $\times$ dipole terms; the diagrams in which the real gluon crosses  exactly once the shockwave, contributing both to $d \sigma_{3IJ}$, made of dipole $\times$ dipole terms as well as to $d \sigma_{4IJ}$, made of double dipole $\times$ dipole  (and dipole $\times$ double dipole) terms; the diagrams in which the real gluon crosses  exactly twice the shockwave, contributing to $d \sigma_{3IJ}$, made of dipole $\times$ dipole terms, to $d \sigma_{4IJ}$, made of double dipole $\times$ dipole  (and dipole $\times$ double dipole) terms, and to $d \sigma_{5IJ}$, made of double dipole $\times$ double dipole terms. \\

We stress that in Fig.~\ref{fig:sigma-NLO-dipole} we show the hard cross-section\footnote{With respect to the fragmentation mechanism.}. In order to construct the quark (anti-quark) part of the physical cross-section, the five contributions must be convoluted with the quark (anti-quark) $\rightarrow$ hadron FF. To include the gluon contribution to the physical cross-section, only the three kinds of real corrections must be convoluted with the gluon $\rightarrow$ hadron FF. 

\subsubsection{Rapidity divergences and UV-sector}
The dipole $\times$ double dipole part of the virtual amplitude contains a rapidity divergences of the form $\ln \alpha$. The presence of the divergence in rapidity is a natural consequence of the separation between the impact factor and the target. Intuitively, a gluon crossing the shockwave cannot have arbitrarily small fraction of longitudinal momentum (and hence arbitrary small rapidity), because only gluon with positive $+$-momentum above the cut-off $\alpha p_{\gamma}^+$ can contribute to the quantum corrections to the impact factor. The rapidity divergent-terms have to be absorbed into the renormalized Wilson operators with the help of the B-JIMWLK equation. We thus have to use the B-JIMWLK evolution for these operators from the cutoff $\alpha$ to the rapidity divide $e^{\eta}$, by writing
\begin{equation}
    \widetilde{\mathcal{U}}_{12}^\alpha  = \widetilde{\mathcal{U}}_{12}^{e^\eta} - \int_\alpha^{e^\eta} \hspace{-0.3 cm} d \rho \frac{\partial \widetilde{\mathcal{U}}_{12}}{\partial \rho} \; .
    \label{Eq:BKnonExp}
\end{equation}
This operation, applied to the leading term, produces an additional next-to-leading contribution which cancels the rapidity divergences. In next-to-leading term, the effect is simply to replace the scale $\alpha$ with the scale $e^{\eta}$. We refer the reader to Appendix~\ref{App_Virtual_Corrections} or to Ref.~\cite{Boussarie:2016ogo} for more details. \\

In principle, we should deal with ultraviolet renormalization, which is very challenging in non-covariant gauges, however, in the shockwave approach, the only UV-divergences at NLO\footnote{Those that are included in the impact factor.} are associated with the dressing of external states (e.g. quark self-energy). Since we treat both the ultraviolet (UV) and the infrared (IR) divergences using dimensional regularization, these singularities are of the type
\begin{equation}
    \frac{1}{\epsilon_{UV}} - \frac{1}{\epsilon_{IR}} 
\end{equation}
and can be set to zero by choosing $\epsilon_{UV}=\epsilon_{IR}$. Then, in practice, some UV divergences will cancel out some infrared divergences in the calculation.

\subsubsection{Treatment of the IR-sector}
When generically decomposing any on-shell parton momentum in the Sudakov basis as\footnote{Here $p^+$ is a large fixed momentum, e.g. $p_\gamma^+$ in our present case.}
\begin{equation}
\label{p-sudakov}
p^\mu = z p^+ n_1^\mu + \frac{\vec{p}^{\,2}}{2 z p^+} n_2^\mu + p_\perp^\mu\,,
\end{equation} 
in the IR sector, we face three kinds of divergences: 
\begin{itemize}
    \item \textbf{Rapidity}: $x_g$ goes to zero while the value of $p_{g, \perp}$ is arbitrary but strongly suppressed with respect to $p_{\gamma}^{+} \sim \sqrt{s}$.
    \item \textbf{Collinear}: $p_{g, \perp} \rightarrow (x_g/x_q) p_{q, \perp}$ (collinear to the quark line) or $p_{g, \perp} \rightarrow (x_g/x_{\bar{q}}) p_{\bar{q}, \perp}$ (collinear to the anti-quark line) while $x_g$ is arbitrary.
    \item \textbf{Soft}: all components linearly vanishing (both $x_g$ and $p_{g, \perp}$ go linearly to zero). Parameterizing the transverse momenta of the gluon as $p_{g, \perp} = x_g u_{\perp}$, with $|u_{\perp}|$ fixed in the limit $x_g$ goes to zero, we can then define the soft limit as $x_g$ goes to zero with $u_{\perp}$ generic\footnote{Please note that, when computing the soft contribution, we often relabel $\vec{u}$ as $\vec{p}_g$ after the rescaling.}. 
\end{itemize}
The superposition of the last two types generates a 
\begin{itemize}
    \item \textbf{Soft and collinear divergence}: soft as defined above and additionaly with $u_{\perp} \rightarrow (1/x_q) p_{q, \perp}$ (to ensure that the gluons becomes soft and collinear to the quark line) or $u_{\perp} \rightarrow (1/x_{\bar{q}}) p_{\bar{q}, \perp}$ (to ensure that the gluons becomes soft and collinear to the anti-quark line).
\end{itemize} 
Technically, as the integration over  $z$  is regulated through a lower cut-off ($\alpha$), care must be taken that the appearance of $\ln \alpha$ can arise from both rapidity and soft divergences.  \\

The calculation is organized as follows. First, the rapidity divergences, which appear only in the virtual corrections in the present computation, are regularized at the amplitude level by absorbing them in the shockwave through one step of B-JIMWLK evolution. Part of terms with $\ln \alpha$, the one related to pure rapidity divergences, are then removed. Soft divergences must cancel in the combination between real and virtual contributions as guaranteed by the Kinoshita-Lee-Naurenberg theorem. To observe easily the cancellation we separate the real cross-section into soft-divergent and soft-free part. Then, when the cancellation takes place, any dependence on $\alpha$ disappears. Finally, the remaining type of divergences, which are of purely collinear nature, will be cancelled performing the renormalization of FFs~\cite{Gribov:1972ri,Lipatov:1974qm,Altarelli:1977zs,Dokshitzer:1977sg}. \\

Before calculating all contributions, we explicitly show how the final cross section is organized. We strongly rely on the separation of the hard cross-section in Eq.~(\ref{sigmaNLO}).\\

\noindent \textbf{Quark fragmentation} \\

\begin{figure}[h]
\begin{picture}(430,220)
\put(20,140){\includegraphics[scale=0.28]{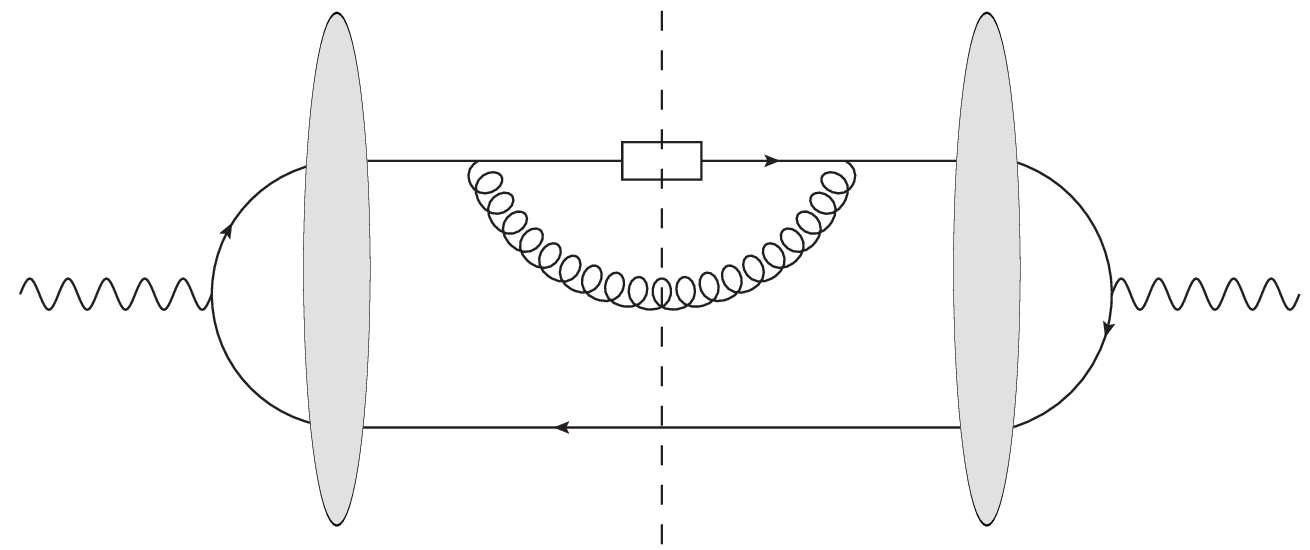}}
\put(102,120){(1)}
\put(240,140){\includegraphics[scale=0.28]{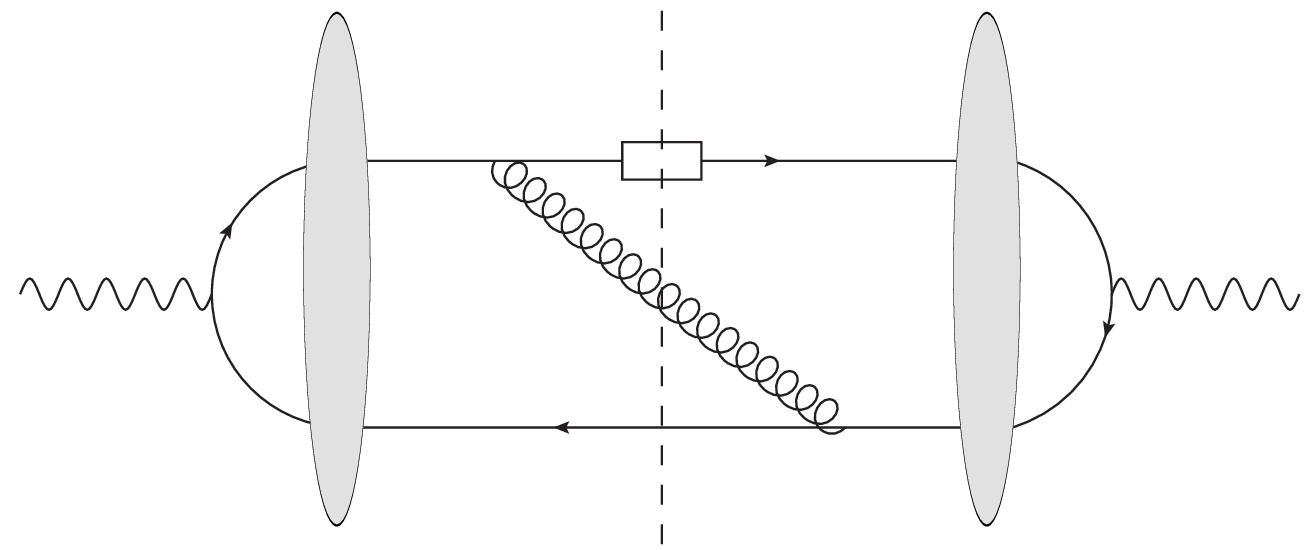}}
\put(322,120){(2)}
\put(20,30){\includegraphics[scale=0.28]{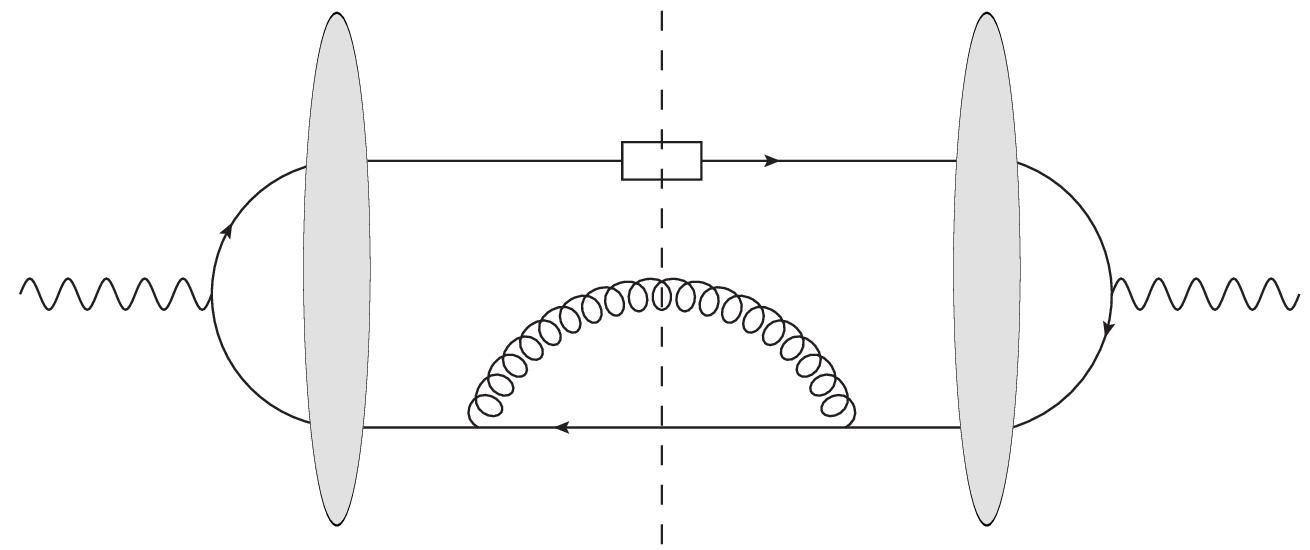}}
\put(102,10){(3)}
\put(240,30){\includegraphics[scale=0.28]{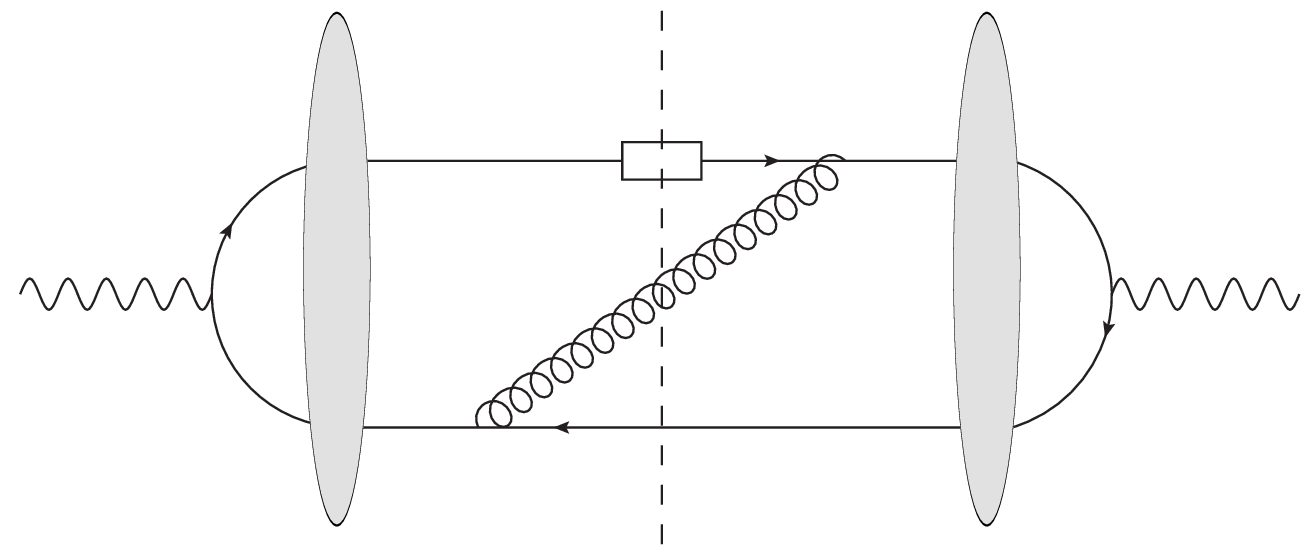}}
\put(322,10){(4)}
\end{picture} 
 \caption{Real diagrams with the gluon emitted after the shockwave, in the quark fragmentation case.}
\label{fig:QuarkInANutshell}
\end{figure}
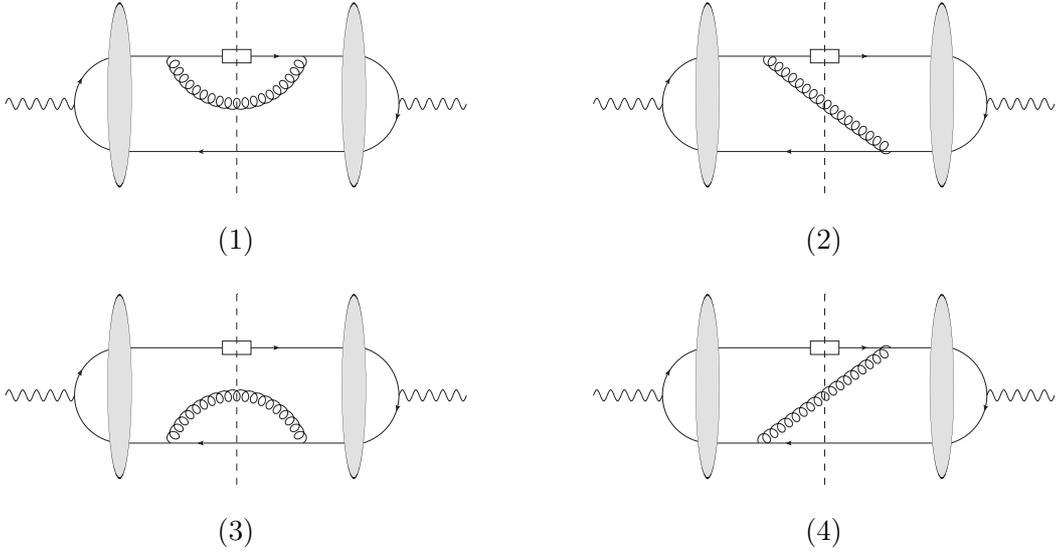

The virtual part of the cross-section, in the quark fragmentation case, can be split as
\begin{equation}
    \frac{d \sigma_{ J I}^{q \rightarrow h}}{d x_{h} d^d p_{h \perp} } \bigg |_{\text{virt., NLO}} \hspace{-0.2 cm} = \frac{d \sigma_{ J I}^{q \rightarrow h}}{d x_{h} d^d p_{h \perp} } \bigg |_{S_V} \hspace{-0.2 cm} + \frac{d \sigma_{ J I}^{q \rightarrow h}}{d x_{h} d^d p_{h \perp} } \bigg |_{\text{virt., dip. $\times$ dip.}} \hspace{-0.2 cm} + \frac{d \sigma_{ J I}^{q \rightarrow h}}{d x_{h} d^d p_{h \perp} } \bigg |_{\text{virt., dip. $\times$ d. dip.}} ,
\end{equation}
where the first term contains the singular virtual dipole $\times$ dipole contribution, the second one contains the finite virtual dipole $\times$ dipole contribution and, finally, the last term contains the finite dipole $\times$ double dipole contribution (see the two top diagrams in Fig.~\ref{fig:sigma-NLO-dipole}). We observe that the latter contribution becomes completely finite once the rapidity divergences have been removed. \\

The real part of the cross-section in the quark fragmentation case can be split as
\begin{gather}
    \frac{d \sigma_{ J I}^{q \rightarrow h}}{d x_{h} d^d p_{h \perp} } \bigg |_{\text{real, NLO}} = \frac{d \sigma_{ J I}^{q \rightarrow h}}{d x_{h} d^d p_{h \perp} } \bigg |_{\text{real, dip. $\times$ dip.}} \nonumber 
    \\ + \frac{d \sigma_{ J I}^{q \rightarrow h}}{d x_{h} d^d p_{h \perp} } \bigg |_{\text{real, dip. $\times$ d. dip.}} + \frac{d \sigma_{ J I}^{q \rightarrow h}}{d x_{h} d^d p_{h \perp} } \bigg |_{\substack{\text{real,} \\ \text{d. dip. $\times$ d. dip.}}} \; ,
\end{gather}
where the splitting follows the separation illustrated in the bottom diagrams in Fig.~\ref{fig:sigma-NLO-dipole}. The last two contributions are finite, while the first one can be further divided into a singular and finite contribution, 
\begin{equation}
\frac{d \sigma_{ J I}^{q \rightarrow h}}{d x_{h} d^d p_{h \perp} } \bigg |_{\text{real, dip. $\times$ dip.}} = \frac{d \sigma_{ J I}^{q \rightarrow h}}{d x_{h} d^d p_{h \perp} } \bigg |_{\substack{\text{real, singular} \\ \text{dip. $\times$ dip.}}} + \frac{d \sigma_{ J I}^{q \rightarrow h}}{d x_{h} d^d p_{h \perp} } \bigg |_{\substack{\text{real, finite} \\ \text{dip. $\times$ dip.}}} \; .
\end{equation}
The singular contribution is generated by the diagrams shown in Fig.~\ref{fig:QuarkInANutshell}. This contribution contains both soft and collinear singularity that can be promptly separated by casting the contribution into the following form (the labels $(i)$ refer to Fig.~\ref{fig:QuarkInANutshell})
\begin{align}
    & \frac{d \sigma_{ J I}^{q \rightarrow h}}{d x_{h} d^d p_{h \perp} } \bigg |_{\substack{\text{real, singular} \\ \text{dip. $\times$ dip.}}} = \frac{d \sigma_{ J I}^{q \rightarrow h}}{d x_{h} d^d p_{h \perp} } \bigg |_{(1)} + \frac{d \sigma_{ J I}^{q \rightarrow h}}{d x_{h} d^d p_{h \perp} } \bigg |_{(2)} + \frac{d \sigma_{ J I}^{q \rightarrow h}}{d x_{h} d^d p_{h \perp} } \bigg |_{(3)} + \frac{d \sigma_{ J I}^{q \rightarrow h}}{d x_{h} d^d p_{h \perp} } \bigg |_{(4)} \nonumber \\
        & = \frac{d \sigma_{ J I}^{q \rightarrow h}}{d x_{h} d^d p_{h \perp} } \bigg |_{\text{Soft}} + \frac{d \sigma_{ J I}^{q \rightarrow h}}{d x_{h} d^d p_{h \perp} } \bigg |_{\text{coll}(qg)} + \frac{d \sigma_{ J I}^{q \rightarrow h}}{d x_{h} d^d p_{h \perp} } \bigg |_{\text{coll}(\bar{q}g)} + \frac{d \sigma_{ J I}^{q \rightarrow h}}{d x_{h} d^d p_{h \perp} } \bigg |_{\text{real, fin. sub.}} \, .
    \end{align}
    The first contribution contains the sum of the four diagrams in the soft limit, i.e.
    \begin{equation}
        \frac{d \sigma_{ J I}^{q \rightarrow h}}{d x_{h} d^d p_{h \perp} } \bigg |_{\text{Soft}} \hspace{-0.2 cm} \equiv \frac{d \sigma_{ J I}^{q \rightarrow h}}{d x_{h} d^d p_{h \perp} } \bigg |_{(1), \text{Soft}} \hspace{-0.2 cm} + \frac{d \sigma_{ J I}^{q \rightarrow h}}{d x_{h} d^d p_{h \perp} } \bigg |_{(2), \text{Soft}} \hspace{-0.2 cm} + \frac{d \sigma_{ J I}^{q \rightarrow h}}{d x_{h} d^d p_{h \perp} } \bigg |_{(3), \text{Soft}} \hspace{-0.2 cm} + \frac{d \sigma_{ J I}^{q \rightarrow h}}{d x_{h} d^d p_{h \perp} } \bigg |_{(4), \text{Soft}} \; ,
    \end{equation}
and hence the complete soft singular part. The second (third) contribution contains the difference between the first (third) diagram and its soft limit, i.e. 
\begin{equation}
    \frac{d \sigma_{ J I}^{q \rightarrow h}}{d x_{h} d^d p_{h \perp} } \bigg |_{\text{coll}(qg)} \equiv \frac{d \sigma_{ J I}^{q \rightarrow h}}{d x_{h} d^d p_{h \perp} } \bigg |_{(1)} - \frac{d \sigma_{ J I}^{q \rightarrow h}}{d x_{h} d^d p_{h \perp} } \bigg |_{(1), \text{Soft}} 
    \label{eq:QFragcollqg}
\end{equation}
and
\begin{equation}
    \frac{d \sigma_{ J I}^{q \rightarrow h}}{d x_{h} d^d p_{h \perp} } \bigg |_{\text{coll}(\bar{q}g)} \equiv \frac{d \sigma_{ J I}^{q \rightarrow h}}{d x_{h} d^d p_{h \perp} } \bigg |_{(3)} - \frac{d \sigma_{ J I}^{q \rightarrow h}}{d x_{h} d^d p_{h \perp} } \bigg |_{(3), \text{Soft}} \; .
\end{equation}
These contributions are collinearly divergent. Finally, the sum of the remaining contributions constitutes the last term, i.e.
\begin{equation}
    \frac{d \sigma_{ J I}^{q \rightarrow h}}{d x_{h} d^d p_{h \perp} } \bigg |_{\text{real, fin. sub.}} \hspace{-0.2 cm} \equiv \frac{d \sigma_{ J I}^{q \rightarrow h}}{d x_{h} d^d p_{h \perp} } \bigg |_{(2)} \hspace{-0.2 cm} - \frac{d \sigma_{ J I}^{q \rightarrow h}}{d x_{h} d^d p_{h \perp} } \bigg |_{(2), \text{Soft}} \hspace{-0.2 cm} + \frac{d \sigma_{ J I}^{q \rightarrow h}}{d x_{h} d^d p_{h \perp} } \bigg |_{(4)} \hspace{-0.2 cm} - \frac{d \sigma_{ J I}^{q \rightarrow h}}{d x_{h} d^d p_{h \perp} } \bigg |_{(4), \text{Soft}} \; .
\end{equation}
This term is finite since in diagrams $(2)$ and $(4)$, because of topology, there is no space for pure collinear divergences.\\
The case of anti-quark fragmentation is treated in a completely identical way. \\

\noindent
\textbf{Gluon fragmentation} \\

\begin{figure}[h]
\begin{picture}(430,220)
\put(20,140){\includegraphics[scale=0.28]{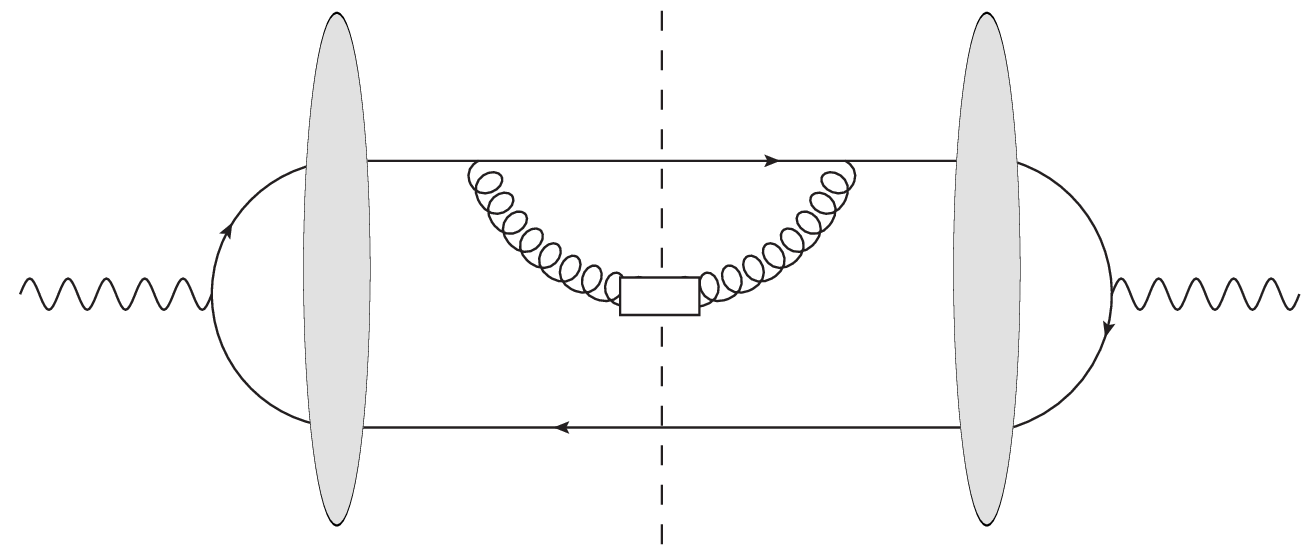}}
\put(102,120){(1)}
\put(240,140){\includegraphics[scale=0.28]{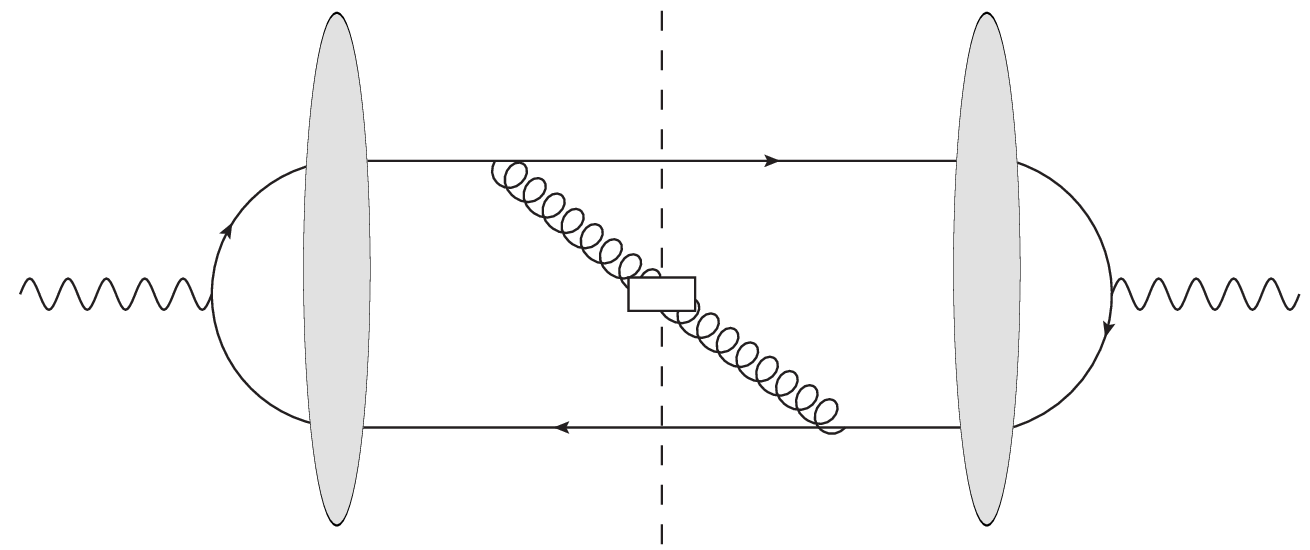}}
\put(322,120){(2)}
\put(20,30){\includegraphics[scale=0.28]{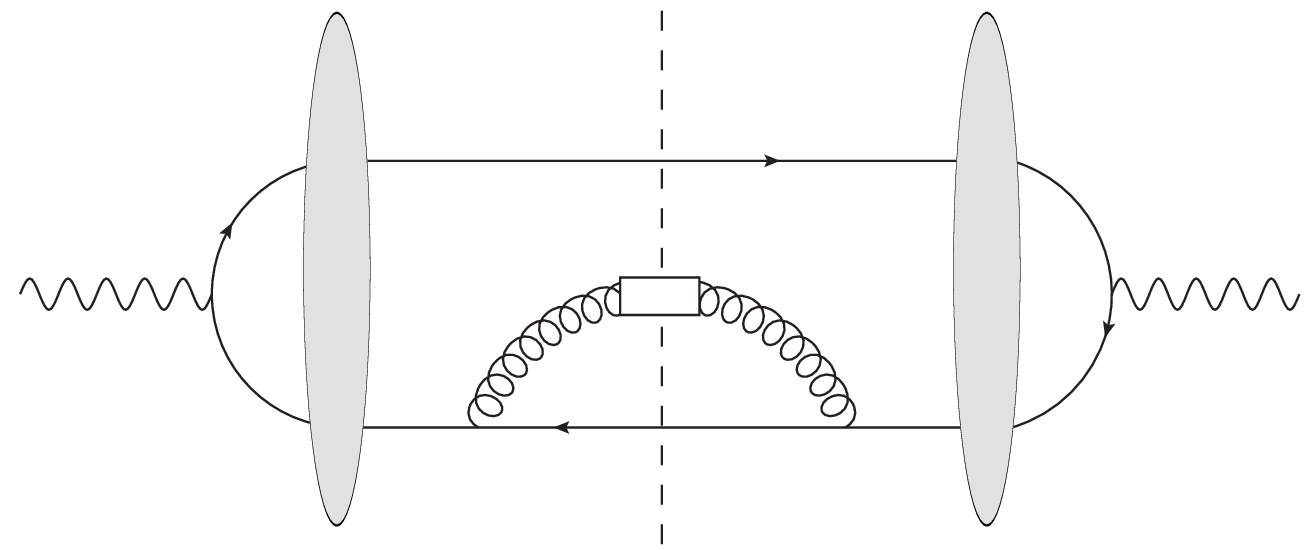}}
\put(102,10){(3)}
\put(240,30){\includegraphics[scale=0.28]{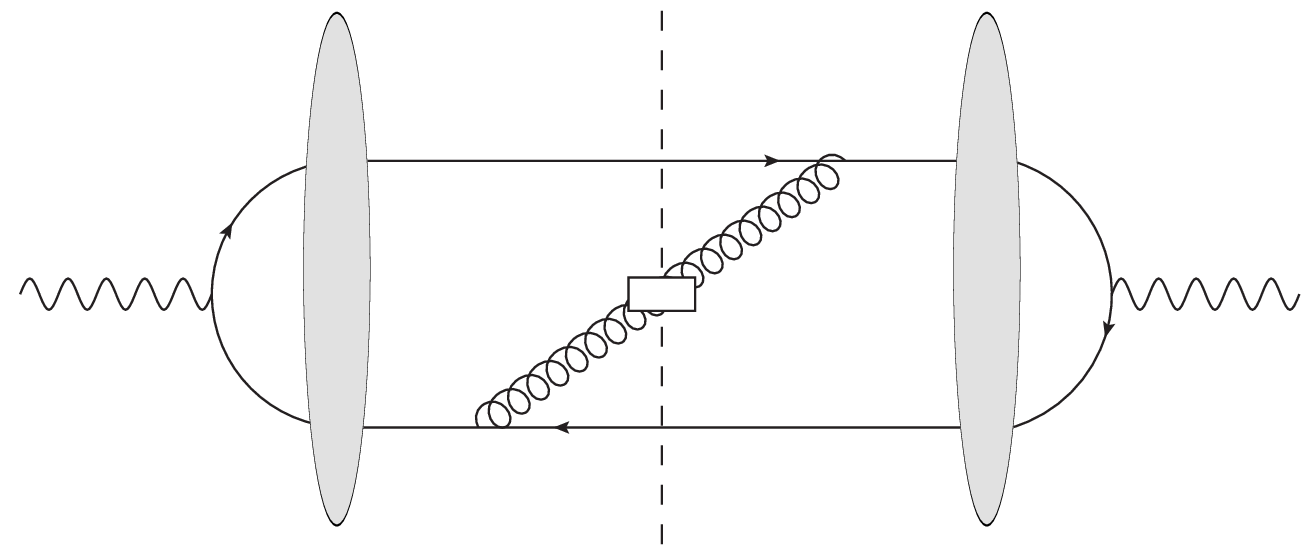}}
\put(322,10){(4)}
\end{picture} 
 \caption{Real diagrams with the gluon emitted after the shockwave, in the gluon fragmentation case.}
\label{fig:GluonInANutshell}
\end{figure}
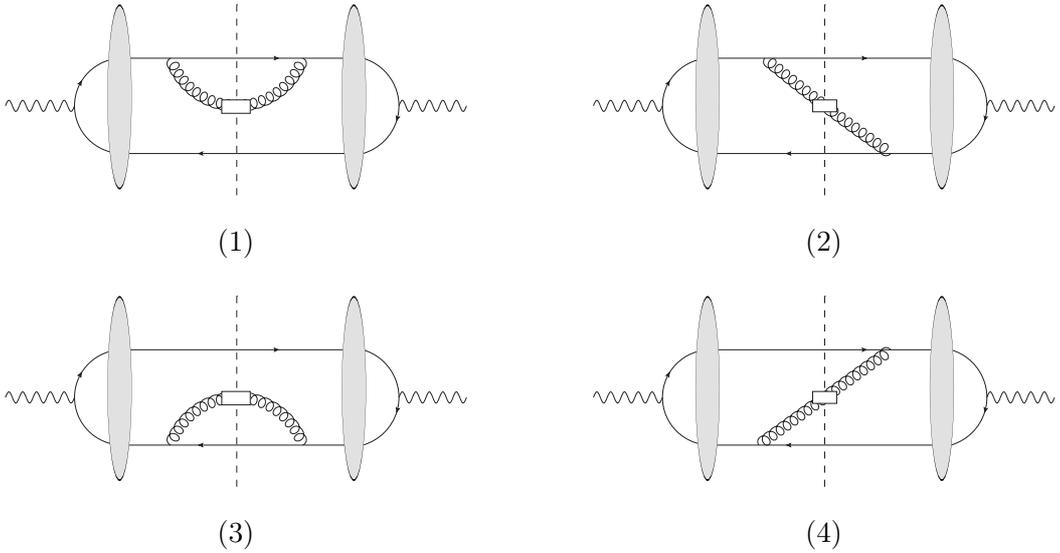

This fragmentation mechanism is possible only when a real gluon is produced, therefore we only deal with real corrections which can be arranged as
\begin{gather}
    \frac{d \sigma_{ J I}^{g \rightarrow h}}{d x_{h} d^d p_{h \perp} } \bigg |_{\text{real, NLO}} = \frac{d \sigma_{ J I}^{g \rightarrow h}}{d x_{h} d^d p_{h \perp} } \bigg |_{\text{dip. $\times$ dip.}} \nonumber \\ + \frac{d \sigma_{ J I}^{g \rightarrow h}}{d x_{h} d^d p_{h \perp} } \bigg |_{\text{dip. $\times$ d. dip.}} + \frac{d \sigma_{ J I}^{g \rightarrow h}}{d x_{h} d^d p_{h \perp} } \bigg |_{\substack{ \text{d. dip. $\times$ d. dip.}}} \; .
\end{gather}
The last two contributions are finite, while the first one can be split as
\begin{gather}
   \frac{d \sigma_{ J I}^{g \rightarrow h}}{d x_{h} d^d p_{h \perp} } \bigg |_{\text{dip. $\times$ dip.}} = \frac{d \sigma_{ J I}^{g \rightarrow h}}{d x_{h} d^d p_{h \perp} } \bigg |_{\substack{\text{singular} \\ \text{dip. $\times$ dip.}}} + \frac{d \sigma_{ J I}^{g \rightarrow h}}{d x_{h} d^d p_{h \perp} } \bigg |_{\text{dip. $\times$ dip.},1} + \frac{d \sigma_{ J I}^{g \rightarrow h}}{d x_{h} d^d p_{h \perp} } \bigg |_{\text{dip. $\times$ dip.,2}} \; .
\label{eq:GluonRealCrossSplit}
\end{gather}
The singular part contains contributions coming from diagrams (1) and (3) in Fig.~\ref{fig:GluonInANutshell}, 
\begin{gather}
    \frac{d \sigma_{ J I}^{g \rightarrow h}}{d x_{h} d^d p_{h \perp} } \bigg |_{\substack{\text{singular} \\ \text{dip. $\times$ dip.}}} = \frac{d \sigma_{ J I}^{g \rightarrow h}}{d x_{h} d^d p_{h \perp} } \bigg |_{(1)} + \frac{d \sigma_{ J I}^{g \rightarrow h}}{d x_{h} d^d p_{h \perp} } \bigg |_{(3)} \; .
\end{gather}
Since these contributions are collinearly divergent, we relabel them as
\begin{gather}
    \frac{d \sigma_{ J I}^{g \rightarrow h}}{d x_{h} d^d p_{h \perp} } \bigg |_{\text{coll}(qg)} \equiv \frac{d \sigma_{ J I}^{g \rightarrow h}}{d x_{h} d^d p_{h \perp} } \bigg |_{(1)} \; , \hspace{2 cm} \frac{d \sigma_{ J I}^{g \rightarrow h}}{d x_{h} d^d p_{h \perp} } \bigg |_{\text{coll}(\bar{q}g)} \equiv \frac{d \sigma_{ J I}^{g \rightarrow h}}{d x_{h} d^d p_{h \perp} } \bigg |_{(3)} \; .
\end{gather}
The second contribution in Eq.~(\ref{eq:GluonRealCrossSplit}) contains the finite diagrams (2) and (4) in Fig.~\ref{fig:GluonInANutshell} ,
\begin{gather}
   \frac{d \sigma_{ J I}^{g \rightarrow h}}{d x_{h} d^d p_{h \perp} } \bigg |_{\text{dip. $\times$ dip.},1} = \frac{d \sigma_{ J I}^{g \rightarrow h}}{d x_{h} d^d p_{h \perp} } \bigg |_{(2)} + \frac{d \sigma_{ J I}^{g \rightarrow h}}{d x_{h} d^d p_{h \perp} } \bigg |_{(4)} \; ,
\end{gather}
while the last contains the rest of the dipole $\times$ dipole contribution. The nature of this last term arises from the fact that, technically, a dipole $\times$ dipole contribution can be produced by a gluon emitted before the shockwave, but passes through it without receiving any transverse kick ($\vec{p}_3=0$)\,. We discuss this situation with more details in the subsection \ref{sec:RealDipxDip2}.

\section{NLO cross-section: FF counterterms}

\begin{figure}[h]
\centering
 \includegraphics[scale=0.35]{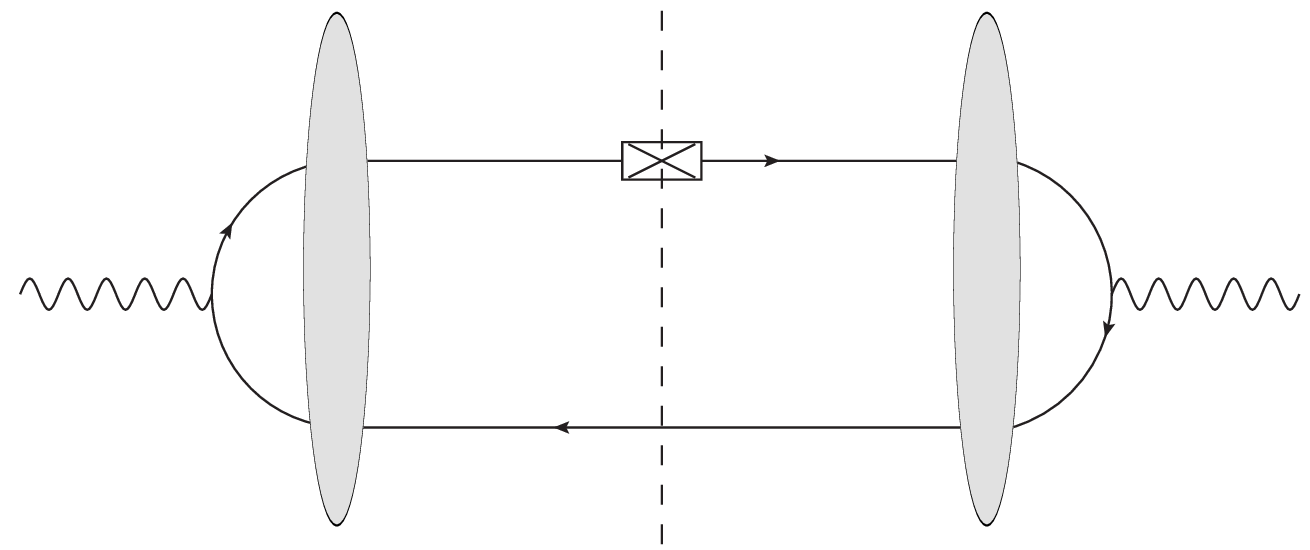}
 \caption{Diagrammatic representation of the quark counterterm.}
    \label{fig:NLO_CounterTerm}
\end{figure}
At the next-to-leading order, the quark/anti-quark FFs should be renormalized, i.e.
\begin{equation}
\label{eq: FF evolution} 
\begin{aligned}
    D_{q}^{h}(x)& =D_{q}^{h}\left(x, \mu_{F}\right)-\frac{\alpha_{s}}{2 \pi}\left(\frac{1}{\hat{\epsilon}}+\ln \frac{\mu_{F}^{2}}{\mu^{2}}\right) \int_{x}^{1} \frac{d z}{z}\left[D_{q}^{h}\left(\frac{x}{z}, \mu_{F}\right) P_{q q}(z)+D_{g}^{h}\left(\frac{x}{z}, \mu_{F}\right) P_{gq}(z)\right], 
\end{aligned}
\end{equation}
where $\frac{1}{\hat{\epsilon}} = \frac{\Gamma (1- \epsilon)}{\epsilon (4 \pi )^\epsilon} \sim \frac{1}{\epsilon} + \gamma_E - \ln (4 \pi)$, $\mu_F$ is the factorization scale and $\mu$ is an arbitrary parameter introduced by dimensional regularization. 
The LO splitting functions are given by  
\begin{eqnarray}
    P_{qq}(z) &=& C_F \left[ \frac{1 + z^2}{(1-z)_+} + \frac{3}{2} \delta (1-z) \right], \\
    P_{gq}(z) &=& C_F \frac{1 + (1-z)^2}{z}\,, 
\end{eqnarray}
where the + prescription is defined as 
\begin{equation}
\label{eq: plus prescription}
    \int_a^1 d \beta \frac{f(\beta)}{(1-\beta)_+} = \int_a^1 d \beta \frac{f(\beta)- f(1)}{(1-\beta)} - \int_0^{a} d\beta \frac{f(1)}{1-\beta} \,.\\ 
\end{equation}
The effect of the renormalization means that the leading cross section (\ref{eq:LL-LO}) is now calculated at the factorization scale $\mu_F$ and a divergent NLO contribution is produced. In the case of fragmentation from a quark, the renormalization of the FF, $D_q^h$, produces the following NLO contribution: 
\begin{gather}
\frac{d \sigma_{ J I}^{q \rightarrow h}}{d x_{h} d^d p_{h \perp} } \bigg |_{\text{ct}} \hspace{- 0.25 cm} = - \frac{2 \alpha_{\mathrm{em}} Q^2 }{(2\pi)^{4d}N_c \; x_h^d } \sum_{q}  Q_q^2 \int_{x_{h}}^1 \hspace{- 0.25 cm} d x_q \;  x_q^{1+d} (1-x_{q})^2 f_{JI} \frac{\alpha_s}{2 \pi} \left( \frac{1}{\hat{\epsilon}} + \ln \left( \frac{\mu_F^2}{\mu^2} \right) \right) \int_{\frac{x_h}{x_q}}^{1} \hspace{- 0.1 cm} \frac{d \beta}{\beta} \nonumber \\ \times \left[D_{q}^{h} \left(\frac{x_h}{ \beta x_q}, \mu_{F} \right) P_{q q}( \beta ) + D_{g}^{h}\left(\frac{x_h}{ \beta x_q}, \mu_{F}\right) P_{gq}( \beta ) \right] \equiv \frac{d \sigma_{J I}^{q \rightarrow h}}{d x_{h} d^d p_{h \perp} } \bigg |_{\text{ct, div}} + \frac{d \sigma_{ J I}^{q \rightarrow h}}{d x_{h} d^d p_{h \perp} } \bigg |_{\text{ct, fin}} \; ,
\label{eq:FFNLOCounterTerm}
\end{gather}
which we will call counterterm (ct). In Eq.~(\ref{eq:FFNLOCounterTerm}) the term labelled as ct, div contains the $1/ \hat{\epsilon}$ contribution, while, the one labelled as ct, fin contains the $\ln ( \mu_F^2 / \mu^2 )$ part. It is also useful to separate the divergent part accordingly to the two different FF splitting functions involved, i.e.  
\begin{gather}
    \frac{d \sigma_{J I}^{q \rightarrow h}}{d x_{h} d^d p_{h \perp} } \bigg |_{\text{ct, div}} \equiv \frac{d \sigma_{J I}^{q \rightarrow h}}{d x_{h} d^d p_{h \perp} } \bigg |_{\text{ct, div, }P_{qq}} + \frac{d \sigma_{J I}^{q \rightarrow h}}{d x_{h} d^d p_{h \perp} } \bigg |_{\text{ct, div, }P_{gq}} \, .
    \label{eq:FFNLOCounterTerm2}
\end{gather}

Since this contribution is completely proportional to the LO cross-section, the counterterm for the anti-quark fragmentation case is obtained as before, by including a minus sign in the argument of the function \textbf{F}, extending the sum over $q$ to the five anti-quark flavor species and performing the $(q \leftrightarrow \bar{q})$ relabelling.   

\section{NLO cross-section: Virtual corrections}
We discuss in this section the virtual corrections to the process in Eq.~(\ref{Eq:process}). In deriving these results, we rely on Ref.~\cite{Boussarie:2016ogo}, where the one loop corrections to the $\gamma^{*} \rightarrow q \bar{q}$ impact factor were computed. However, for completeness, we provide a brief summary of this calculation in the Appendix~\ref{App_Virtual_Corrections}. As mentioned earlier, it is necessary to separate the dipole $\times$ dipole contribution into a finite and a divergent part. The dipole $\times$ double dipole contribution is instead completely finite once the divergences in rapidity have been removed~\cite{Boussarie:2016ogo}. 

\subsection{Divergent part of the dipole $\times$ dipole contribution}
\label{SubSec:DivVirt}
\begin{figure}[h]
\centering
 \includegraphics[scale=0.35]{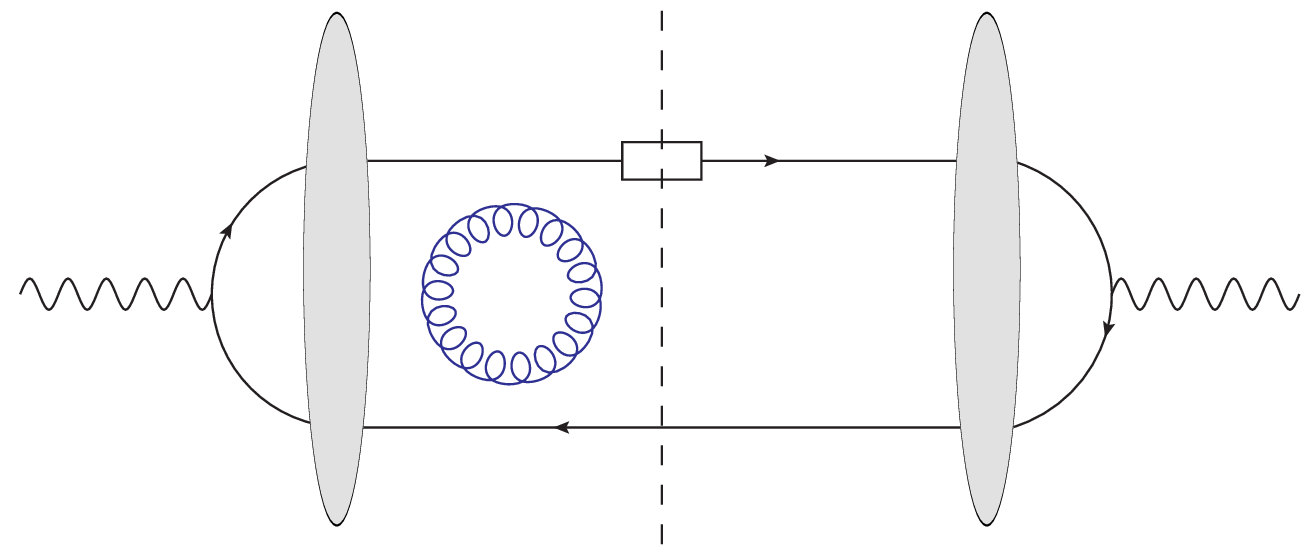}
 \caption{Diagrammatic representation of the divergent part of virtual contributions. The blue gluon loop represents a generic divergence (UV or IR).}
    \label{fig:NLO_Virt_Div}
\end{figure}
Starting from Eq.~(5.16) of Ref.~\cite{Boussarie:2016ogo}, the divergent part of the one-loop cross-section, symbolically illustrated in Fig.~\ref{fig:NLO_Virt_Div} can be written as 
\begin{equation}
    \frac{d \hat{\sigma}_{1JI}}{d x_q d^d \vec{p}_q}  = \frac{\alpha_{s}}{2\pi} \frac{\Gamma(1-\epsilon)}{(4 \pi)^{\epsilon}} C_F \left( \frac{S_{V}+S_{V}^{*}}{2} \right) \frac{d \hat{\sigma}_{0 J I}}{d x_q d^d \vec{p}_q} \; ,
\end{equation}
where
\begin{gather}
\frac{S_V + S_V^*}{2}  =  \frac{1}{\epsilon} \Bigg [ - 4 \epsilon \ln (\alpha) \ln \left(\frac{x_q ^2 x_{\bar{q}}^2 \mu^{2}}{\left(x_q \vec{p}_{\bar{q}}-x_{\bar{q}} \vec{p}_{q}\right)^{2}}\right) + 4 \ln (\alpha) + 4 \epsilon \ln^2(\alpha)  - 2 \ln (x_q x_{\bar{q}})+ 3 \nonumber \\+  2 \epsilon \ln \left(\frac{x_q x_{\bar{q}} \mu^{2}}{\left(x_q \vec{p}_{\bar{q}}-x_{\bar{q}}\vec{p}_{q}\right)^{2}}\right) \ln (x_q x_{\bar{q}}) + \epsilon \ln^2 (x_q x_{\bar{q}})  - 3 \epsilon \ln \left(\frac{x_q x_{\bar{q}} \mu^{2}}{\left(x_q \vec{p}_{\bar{q}}-x_{\bar{q}} \vec{p}_{q}\right)^{2}}\right)  - \frac{\pi^2}{3}\epsilon + 6 \epsilon \Bigg ]  \,
.\label{SV}
\end{gather}
Performing the convolution with FFs as in~(\ref{eq: coll facto}) and using (\ref{constraint-collinear-q}), we can get the final contribution separated into a divergent and a finite part, i.e.
\begin{gather}
\frac{d \sigma_{ J I}^{q \rightarrow h}}{d x_{h} d^d p_{h \perp} } \bigg |_{S_V} \hspace{- 0.25 cm} = \frac{2 \alpha_{\mathrm{em}} Q^2 }{(2\pi)^{4d}N_c \; x_h^d } \sum_{q}  Q_q^2 \int_{x_{h}}^1 \hspace{- 0.25 cm} d x_q \;  x_q^{1+d} (1-x_{q})^2 \int d^{d} p_{ \bar{q} \perp} \int d^{d} p_{2 \perp} \int d^{d} p_{2' \perp} \nonumber \\ \times \frac{\mathbf{F}\left(\frac{x_q}{2x_{h}}  p_{h \perp} + \frac{1}{2} p_{ \bar{q} \perp} - p_{2 \perp} \right)}{ \vec{p}_{\bar{q} 2}^{\; 2} + x_q (1-x_{q}) Q^{2}} \; \frac{\mathbf{F}^{*} \left(\frac{x_q}{2x_{h}}  p_{h \perp} + \frac{1}{2} p_{ \bar{q} \perp} - p_{2' \perp} \right)}{ \vec{p}_{\bar{q} 2'}^{\; 2} + x_q (1-x_{q}) Q^{2}} D_q^h \left( \frac{x_h}{x_q} , \mu_F \right) \delta_{JI} \nonumber \\ \times \frac{\alpha_s}{2 \pi} C_F \left( \frac{1}{\hat{\epsilon}} A_{V, \text{div}} + A_{V, \text{fin}} \right) \equiv \frac{d \sigma_{ J I}^{q \rightarrow h}}{d x_{h} d^d p_{h \perp} } \bigg |_{S_V\text{,div}} + \frac{d \sigma_{ J I}^{q \rightarrow h}}{d x_{h} d^d p_{h \perp} } \bigg |_{S_V\text{,fin}} \; ,
\label{eq:SVpart}
\end{gather}
where 
\begin{equation}
 A_{V, \text{div}} = 4 \ln \alpha - 2 \ln x_q (1-x_q) + 3 - 4 \epsilon \ln \alpha \ln \left( \frac{x_h^2 (1-x_q)^2 \mu^2}{ \left( x_h \vec{p}_{\bar{q}} - (1-x_q) \vec{p}_h \right)^2} \right) + \epsilon \ln^2 \alpha^2 \; ,    
\end{equation}
\begin{gather}
 A_{V, \text{fin}} = \left( 2 \ln x_q (1-x_q) -3 \right) \ln \left( \frac{x_h^2 (1-x_q) \mu^2}{ x_q \left( x_h \vec{p}_{\bar{q}} - (1-x_q) \vec{p}_h \right)^2} \right) \nonumber \\ + 2 \left( 3 - \zeta (2) + \frac{1}{2} \ln^2 (x_q (1-x_q)) \right) 
\end{gather}
and
\begin{equation*}
 \delta_{LL} = 1 \; , \hspace{1 cm} \delta_{TL} = \frac{(1 - 2 x_q)}{2 x_q (1-x_q) Q} \left( \vec{p}_{\bar{q} 2'} \cdot \vec{\varepsilon}_T \right)^{*} \; ,
\end{equation*}
\begin{equation}
\label{deltaTT}
   \delta_{TT} = \left[ (1-2x_q)^2 g_{\perp}^{ri} g_\perp^{lk} - g_\perp^{rk} g_\perp^{li} + g_{\perp}^{rl} g_\perp^{ik} \right] \frac{ \varepsilon_{T i} \; p_{\bar{q} 2 \perp  r} \left( \varepsilon_{T k} \; p_{\bar{q} 2'  \perp l} \right)^{*} }{4 x_q^2 (1-x_q)^2 Q^2} \; .
\end{equation}
Again, the corresponding contribution in the case of fragmentation from anti-quark is obtained by including a minus sign in the argument of the function \textbf{F}, extending the sum over $q$ to the five anti-quark flavor species and performing the $(q \leftrightarrow \bar{q})$ relabelling (also inside the functions $A_{V,{\rm{div}}}$, $A_{V,{\rm{fin}}}$, $\delta_{JI}$). 
\subsection{Finite part of the dipole $\times$ dipole contribution}
\begin{figure}[h]
\centering
 \includegraphics[scale=0.35]{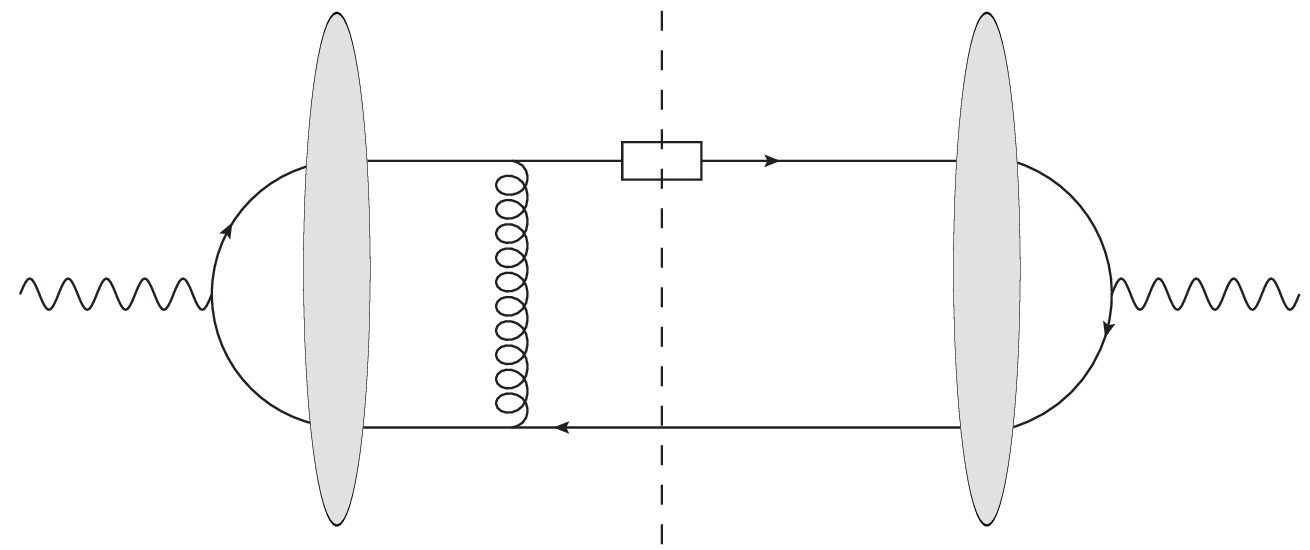}
 \caption{An example of diagram which only contributes to the dipole $\times$ dipole part of virtual contributions.}
    \label{fig:NLO_Virt_DipxDip}
\end{figure}
An example of diagram contributing to the dipole $\times$ dipole part is shown in Fig.~\ref{fig:NLO_Virt_DipxDip}. Let us now consider the finite part associated with these diagrams. We can build this contribution starting from Eqs.~(5.24), (5.28) and (5.35)\footnote{These equations refer respectively to LL, TL and TT case.} of Ref.~\cite{Boussarie:2016ogo}, it reads
\begin{gather}
\frac{d \sigma_{LL}^{q \rightarrow h}}{d x_{h} d^2 p_{h \perp} } \bigg |_{\text{virt., dip. $\times$ dip.}}   =  \frac{2 \alpha_{\mathrm{em}} Q^{2} }{\left(  2\pi\right)^{8}N_c \; x_h^2} \sum_q Q_{q}^{2} \int_{x_h}^1 \hspace{-0.15 cm} d x_q  x_q D_q^h \left( \frac{x_h}{x_q}, \mu_F \right)   \frac{\alpha_{s}}{2 \pi} \frac{C_F}{4} \nonumber \\ 
 \times  \int d^{2} \vec{p}_{\bar{q}} d^{2}\vec{p}_{1} d^2 \vec{p}_{2} d^{2} \vec{p}_{1^{\prime}} d^2 \vec{p}_{2^\prime}
\delta \left( \frac{x_q}{x_h} \vec{p}_h - \vec{p}_{1}+ \vec{p}_{\bar{q}2} \right) \delta(\vec{p}_{11^\prime}+ \vec{p}_{22^\prime})\mathbf{F}\left( \frac{\vec{p}_{12}}{2}\right)\mathbf{F}^\ast\left( \frac{\vec{p}_{1^\prime 2^\prime }}{2}\right) \nonumber \\ 
 \times  \left\{ \frac{1}{{\vec{p}_{q1^{\prime}}^{\; 2} +x_q x_{\bar{q}}  Q^{2}} }  \left[  \frac{6 x_q^{2} x_{\bar{q}}^2}{\vec{p}_{q1}^{\; 2}+x_q x_{\bar{q}} Q^{2}}\ln\left(  \frac
{x_q^{2}x_{\bar{q}}^2 \mu^{4} Q^{2}}{(x_q \vec{p}_{\bar{q}}-x_{\bar{q}} \vec{p}_{q})^{2}(\vec{p}_{q1}^{\; 2}+x_q x_{\bar{q}} Q^{2})^{2}}\right) \right. \right. \nonumber \\
 \left. \left. + \left( \int_0^{x_q} dz \left( \left[(\phi_4)_{LL}\right]_+ + \sum_{n= 5,6} \left[(\phi_n)_{LL}\right]_+ |_{\vec{p}_{3} = \vec{0}} \right) + (q \leftrightarrow \bar{q}) \right ) \right] + h.c.|_{p_1, p_2 \leftrightarrow p_{1'}, p_{2'}} \right\}_{ \substack{x_{\bar{q}} = 1-x_q \\ \vec{p}_q = \frac{x_q}{x_h} \vec{p}_h }} \hspace{-0.3 cm} ,
 \label{eq:VirtDipXDip}
\end{gather} 
in the LL case,
\begin{gather}
 \frac{d \sigma_{TL}^{q \rightarrow h}}{d x_{h} d^2 p_{h \perp} } \bigg |_{\text{virt., dip. $\times$ dip.}} =  \frac{ \alpha_{\mathrm{em}} Q}{(2\pi)^{8}N_c \; x_h^2 } \sum_q Q_{q}^{2} \int_{x_h}^1 d x_q x_q D_q^h \left( \frac{x_h}{x_q}, \mu_F \right) \frac{\alpha_{s}}{ 4 \pi } C_F  \varepsilon_{Ti}^{\ast} \nonumber \\ 
 \times  \int d^2 \vec{p}_{\bar{q}} d^{2} \vec{p}_{1} d^2 \vec{p}_{2} d^{2} \vec{p}_{1^{\prime}} d^2 \vec{p}_{2^\prime } \delta \left( \frac{x_q}{x_h} \vec{p}_h -\vec{p}_{1}+\vec{p}_{\bar{q}2} \right) \delta(\vec{p}_{11^\prime}+\vec{p}_{22^\prime}) \mathbf{F}\left(\frac{\vec{p}_{12}}{2}\right)\mathbf{F}^\ast\left(\frac{\vec{p}_{1^\prime 2^\prime }}{2}\right) \nonumber \\ 
 \times  \left \{  \frac{ \displaystyle \left[ \left( \int_0^{x_q} dz \left(\left[(\phi_4)^i_{TL}\right]_+ + \sum\limits_{n=5,6} \left[(\phi_n)^i_{TL}\right]_+|_{\vec{p}_3 = \vec{0}} \right) \right) + (q \leftrightarrow \bar{q})\right]^\ast }{\vec{p}_{q1}^{\; 2}+x_q x_{\bar{q}} Q^{2}}  + \frac{3x_q x_{\bar{q}}(1-2x_q ) p_{q1^{\prime}\bot}^{i}}{(\vec{p}_{q1}^{\; 2}+x_q x_{\bar{q}} Q^{2})} \right. \nonumber\\
 \times \frac{1}{(\vec{p}_{q1^{\prime}}^{\; 2}+x_q x_{\bar{q}} Q^{2})} \left(  \ln\left(  \frac{x_q^{3} x_{\bar{q}}^3 \mu^{8}Q^{2}(x_q \vec{p}_{\bar{q}}-x_{\bar{q}} \vec{p}_{q})^{-4}}{(\vec{p}_{q1}^{\; 2}+x_q x_{\bar{q}} Q^{2})^{2}(\vec{p}_{q1^{\prime}}^{\; 2}+x_q x_{\bar{q}} Q^{2})}\right)  \right.  -\left.\frac{x_q x_{\bar{q}} Q^{2}}{\vec{p}_{q1^{\prime}}^{\; 2}} \ln\left(  \frac{x_q x_{\bar{q}} Q^{2}}{\vec{p}_{q1^{\prime}}^{\; 2}+x_q x_{\bar{q}} Q^{2}}\right)  \right) \nonumber \\
 +  \left.  \frac{ \displaystyle \left[ \left( \int_0^{x_q} dz  \left(\left[(\phi_4)_{LT}^i \right]_+ + \sum\limits_{n= 5,6} \left[(\phi_n)_{LT}^i \right]_+ |_{\vec{p}_3 = \vec{0}} \right) \right) + \left(q \leftrightarrow \bar{q}\right) \right] }{2x_q x_{\bar{q}} \left( \vec{p}_{q1^{\prime}}^{\; 2} +x_q x_{\bar{q}} Q^{2}\right)} \right \}_{ \substack{x_{\bar{q}} = 1-x_q \\ \vec{p}_q = \frac{x_q}{x_h} \vec{p}_h }}  \; , 
 \label{eq:VirtDipXDipTL}
\end{gather}
in the TL case, and
\begin{gather} 
\frac{d \sigma_{TT}^{q \rightarrow h}}{d x_{h} d^2 p_{h \perp} } \bigg |_{\text{virt., dip. $\times$ dip.}}  = \frac{\alpha_{\mathrm{em}} 
}{2 \left(  2\pi\right)^{8}N_c \; x_h^2} \sum_q Q_{q}^{2} \int_{x_h}^1 d x_q x_q D_q^h \left( \frac{x_h}{x_q}, \mu_F \right) (\varepsilon_{Ti}%
\varepsilon_{Tk}^{\ast}) \nonumber \\
\times \int d^{2} \vec{p}_{\bar{q}} d^{2} \vec{p}_{1} d^d \vec{p}_{2} d^{2} \vec{p}_{1^{\prime}} d^2 \vec{p}_{2^\prime} \delta \left( \frac{x_q}{x_h} \vec{p}_h - \vec{p}_{1} + \vec{p}_{\bar{q}2} \right) \delta(\vec{p}_{11^\prime}+ \vec{p}_{22^\prime}) \mathbf{F}\left(\frac{\vec{p}_{12}}{2}\right)\mathbf{F}^\ast\left(\frac{\vec{p}_{1^\prime 2^\prime }}{2}\right) \nonumber \\ 
\times  \frac{\alpha_{s}}{2 \pi} C_F \Bigg \{  \frac{3}{2}\frac{ p_{q1\bot r}p_{q1^{\prime}\bot l}}{(\vec{p}_{q1}^{\; 2}+x_q x_{\bar{q}} Q^{2})(\vec{p}_{q1^{\prime}}^{\; 2}+x_q x_{\bar{q}} Q^{2})} \left[(1-2x_q)^{2} g_{\bot}^{ri} g_{\bot}^{lk}- g_{\bot}^{rk} g_{\bot}^{li}+ g_{\bot}^{rl} g_{\bot}^{ik} \right] \nonumber \\
\times \left[  \ln\left(  \frac{x_q x_{\bar{q}} \mu^{4}}{(x_q \vec{p}_{\bar{q}}-x_{\bar{q}} \vec{p}_{q})^{2}(\vec{p}_{q1}^{\; 2}+x_q x_{\bar{q}} Q^{2})}\right)-\frac{x_q x_{\bar{q}} Q^{2}}{\vec{p}_{q1}^{\; 2}}\ln\left(  \frac{x_q x_{\bar{q}} Q^{2}}{\vec{p}_{q1}^{\; 2} + x_q x_{\bar{q}} Q^{2}}\right)  \right] \nonumber \\
+\left.   \frac{ \displaystyle \left[ \left( \int_0^{x_q} dz \left(\left[(\phi_4)^{ik}_{TT}\right]_+ + \sum\limits_{n=5,6} \left[(\phi_n)^{ik}_{TT}\right]_+|_{\vec{p}_3 = \vec{0}}  \right) \right) + \left(q \leftrightarrow \bar{q}\right) \right]}{ x_q x_{\bar{q}} \left( \vec{p}_{q1^{\prime}}^{\,\,2}+x\bar{x}Q^{2} \right) } + h.c.|_{\substack{p_{1}, p_2\leftrightarrow p_{1^{\prime}}, p_{2'} \\ i \leftrightarrow k}}\right\}_{ \substack{x_{\bar{q}} = 1-x_q \\ \vec{p}_q = \frac{x_q}{x_h} \vec{p}_h }}  \; ,  
\label{eq:VirtDipXDipTT}
\end{gather}
in the TT case. The explicit expression for the functions $(\phi_n)_{JI}$ can be found in appendix~\ref{AppendixB}.

\subsection{Dipole $\times$ double dipole contribution}

\begin{figure}[h]
\centering
 \includegraphics[scale=0.35]{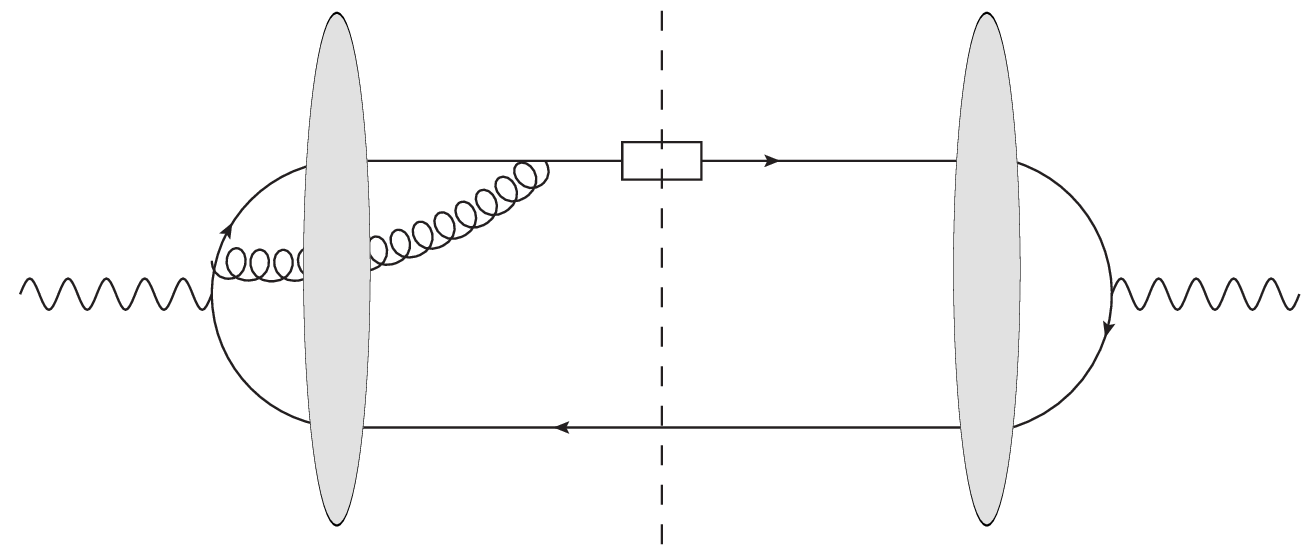}
 \caption{An example of diagram which contributes to the dipole $\times$ double dipole part of virtual contributions.}
\label{fig:NLO_Virt_DipxDouble_Dip}
\end{figure}
In the dipole $\times$ double dipole contribution, the virtual gluon crosses the shockwave (see Fig.~\ref{fig:NLO_Virt_DipxDouble_Dip}). This contribution, in the various cases, can be obtained from Eqs. (5.39), (5.41) and (5.44)\footnote{These equations refer respectively to LL, TL and TT case.}  of Ref.~\cite{Boussarie:2016ogo} and it reads
\begin{align}
& \frac{d \sigma_{LL}^{q \rightarrow h}}{d x_{h} d^2 p_{h \perp} } \bigg |_{\text{virt., dip. $\times$ d. dip.}} \hspace{-0.4 cm} = \frac{2 \alpha_{\mathrm{em}} Q^{2}}{\left(  2\pi\right)^{8} N_{c} \, x_h^2} \sum_q Q_q^2 \int_{x_h}^1  \hspace{-0.25 cm} d x_q x_q D_q^h \left( \frac{x_h}{x_q}, \mu_F \right)  \hspace{-0.1 cm} \int  \hspace{-0.1 cm} d^{2} \vec{p}_{\bar{q}} d^2 \vec{p}_{1}  d^2 \vec{p}_{2}  d^2 \vec{p}_{1^\prime } d^2 \vec{p}_{2^\prime} \nonumber \\
& \times \int\frac{d^{2} \vec{p}_{3}}{\left(  2\pi\right)^{2}} \delta \left( \frac{x_q}{x_h} \vec{p}_h - \vec{p}_{1}+\vec{p}_{\bar{q}2}-\vec{p}_{3} \right) \delta(\vec{p}_{11^\prime}+\vec{p}_{22^\prime}+\vec{p}_{3}) \frac{\alpha_{s}}{2\pi} \frac{1}{8}  \left\{ \frac{1}{\vec{p}_{q1^{\prime}}^{\; 2}+ x_q x_{\bar{q}} Q^{2}} \widetilde{\mathbf{F}}\left( \frac{\vec{p}_{12}}{2},\vec{p}_{3}\right)  \right. \nonumber \\
& \times \mathbf{F}^{\ast}\left( \frac{\vec{p}_{1^\prime 2^\prime}}{2} \right) \left[  4x_q x_{\bar{q}} \left[  \frac{x_q x_{\bar{q}} (\vec{p}_{3}^{\; 2}-\vec{p}_{\bar{q}2}^{\; 2}-\vec{p}_{q1}^{\; 2}-2x_q x_{\bar{q}} Q^{2})}{(\vec{p}_{\bar{q}2}^{\; 2}+x_q x_{\bar{q}} Q^{2})( \vec{p}_{q1}^{\; 2} + x_q x_{\bar{q}} Q^{2}) - x_q x_{\bar{q}} Q^{2}\vec{p}_{3}^{\;2}} \ln\left(  \frac{x_q x_{\bar{q}}}{e^{2\eta}}\right) \right.  \right. \nonumber\\
& \times  \ln\left(  \frac{(\vec{p}_{\bar{q}2}^{\; 2}+x_q x_{\bar{q}} Q^{2})\left(  \vec{p}_{q1}^{\;2}+x_q x_{\bar{q}} Q^{2}\right)  }{x_q x_{\bar{q}} Q^{2}\vec{p}_{3}^{\; 2}}\right) 
-   \left.  \left(  \frac{2x_q x_{\bar{q}} }{\vec{p}_{q1}^{\; 2} + x_q x_{\bar{q}} Q^{2}} \ln\left(  \frac{x_{\bar{q}}}{e^{\eta}}\right) 
\ln\left(\frac{\vec{p}_{3}^{\; 2}}{\mu^{2}}\right)  +\left(  q \leftrightarrow \bar{q} \right)  \right)  \right] \nonumber\\
& + \left. \left.  Q^{2} \left[ \left( \int_{0}^{x_q} dz \sum_{n= 5,6 }\left[(\phi_n)_{LL}\right]_+ \right) + \left(q \leftrightarrow \bar{q}\right) \right] \right]  +h.c.|_{p_1, p_1 \leftrightarrow p_{1'}, p_{2'}}  \right\}_{ \substack{x_{\bar{q}} = 1-x_q \\ \vec{p}_q = \frac{x_q}{x_h} \vec{p}_h }} \; ,  
\label{eq:VirtDoubleDipXDipLL}
\end{align}
in the LL case,
\begin{align} 
& \frac{d \sigma_{TL}^{q \rightarrow h}}{d x_{h} d^2 p_{h \perp} } \bigg |_{\text{virt., dip. $\times$ d. dip.}} \hspace{-0.4 cm} = \frac{ \alpha_{\mathrm{em}} Q}{\left(  2\pi\right)^{8}N_{c} \, x_h^2} \sum_q Q_{q}^{2} \int_{x_h}^1 \hspace{-0.15 cm} dx_q x_q D_q^h \left( \frac{x_h}{x_q}, \mu_F \right) \hspace{-0.10 cm} \int \hspace{-0.15 cm} d^{2} \vec{p}_{\bar{q}} d^2 \vec{p}_{1} d^2 \vec{p}_{2} d^2 \vec{p}_{1^\prime} d^2 \vec{p}_{2^\prime} \nonumber \\
& \times \hspace{-0.15 cm} \int \hspace{-0.1 cm} \frac{d^2 \vec{p}_{3} d^2 \vec{p}_{3^\prime}}{(2\pi)^2} \delta \left( \frac{x_q}{x_h} \vec{p}_h - \vec{p}_{1}+ \vec{p}_{\bar{q}2 } - \vec{p}_{3} \right) \delta(\vec{p}_{11^\prime}+ \vec{p}_{22^\prime}+ \vec{p}_{33^\prime})  \frac{\alpha_{s}}{8 \pi}  \varepsilon_{Ti}^{\ast} \hspace{-0.05 cm} \left \{ \frac{\delta(\vec{p}_{3^\prime})}{\vec{p}_{q1^{\prime}}^{\; 2}+x_q x_{\bar{q}} Q^{2}} \widetilde{\mathbf{F}} \hspace{-0.05 cm} \left( \frac{\vec{p}_{12}}{2}, \vec{p}_{3}\right) \hspace{-0.05 cm} \right. \nonumber \\
& \times \mathbf{F}^{\ast}\left( \frac{\vec{p}_{1^\prime 2^\prime}}{2}\right) \left[  2(1-2x_q) p_{q1^{\prime}\bot}^{i}\left[  \frac{x_q x_{\bar{q}} (\vec
{p}_{3}^{\; 2}-\vec{p}_{\bar{q}2}^{\;  2}-\vec{p}_{q1}^{\;  2}-2x_q x_{\bar{q}} Q^{2})}{(\vec{p}_{\bar{q}2}^{\;  2} + x_q x_{\bar{q}} Q^{2})(\vec{p}_{q1}^{\;  2}+ x_q x_{\bar{q}}  Q^{2}) - x_q x_{\bar{q}} Q^{2}\vec{p}_{3}^{\; 2}}\right.
\right. \ln\left(  \frac{x_q x_{\bar{q}} }{e^{2\eta}}\right)  \nonumber \\
& \times  \ln\left(  \frac{(\vec{p}_{\bar{q}2}^{\; 2}+x_q x_{\bar{q}} Q^{2})\left(  \vec{p}_{q1}^{\; 2} + x_q x_{\bar{q}} Q^{2}\right)  }{x_q x_{\bar{q}} Q^{2}\vec{p}_{3}^{\; 2}}\right) - \left.  \left(  \frac{2x_q x_{\bar{q}} }{\vec{p}_{q1}^{\; 2} + x_q x_{\bar{q}} Q^{2} }
\ln\left(  \frac{x_{\bar{q}}}{e^{\eta}}\right)  \ln\left( \frac{\vec{p}_{3}^{\; 2}}{\mu^{2}}\right) +\left(  q\leftrightarrow\bar{q}\right)  \right)  \right] \nonumber \\
& +  \left.  \frac{1}{2x_q x_{\bar{q}} } \left[ \left( \int_0^{x_q} dz \sum_{n=5,6} \left[(\phi_n)^i_{TL}\right]_+ \right) + \left(q \leftrightarrow \bar{q} \right) \right] \right] +  \frac{\delta(\vec{p}_{3})}{\vec{p}_{q1}^{\; 2}+x_q x_{\bar{q}} Q^{2}} \widetilde{\mathbf{F}}^\ast\left( \frac{\vec{p}_{1^\prime 2^\prime}}{2}, \vec{p}_{3^{\prime}}\right) \nonumber \\
& \times \mathbf{F}\left( \frac{\vec{p}_{12}}{2}\right) \left[  \left[ 2x_q x_{\bar{q}} (1-2x_q ) p_{q1^{\prime}\bot}^{ i} \left( \frac{-2}{\vec{p}_{q1^{\prime}}^{\;  2} + x_q x_{\bar{q}} Q^{2}} \ln\left(  \frac{x_{\bar{q}}}{e^{\eta}}\right)\ln\left(  \frac{\vec{p}_{3'}^{\; 2}}{\mu^{2}}\right) +  \ln\left(  \frac{x_q x_{\bar{q}} }{e^{2\eta}} \right)   \right. \right.  \right. \nonumber \\
& \times \left(  \frac{ -( \vec{p}_{\bar{q}2^\prime}^{\; 2}+x_q x_{\bar{q}} Q^{2})}{( \vec{p}_{q1^{\prime}}^{\; 2} +x_q x_{\bar{q}} Q^{2})  ( \vec{p}_{\bar{q}2^\prime}^{\; 2} +x_q x_{\bar{q}}  Q^{2})-x_q x_{\bar{q}} Q^{2}\vec{p}_{3'}^{\; 2}}\right.  \ln\left(  \frac{( \vec{p}_{q1^{\prime}}^{\;  2} +x_q x_{\bar{q}} Q^{2})  ( \vec{p}_{\bar{q}2^\prime}^{\; 2} + x_q x_{\bar{q}} Q^{2} )  }{x_q x_{\bar{q}} Q^{2}\vec{p}_{3'}^{\; 2}}\right) \nonumber \\
& +  \left.  \left.  \left.  \frac{1}{\vec{p}_{q1^{\prime}}^{\;2}}\ln\left( \frac{\vec{p}_{q1^{\prime}}^{\;  2}+x_q x_{\bar{q}} Q^{2}}{x_q x_{\bar{q}} Q^{2}}\right)  \right)  \right) +(q\leftrightarrow\bar{q})\right]  + \left.  \left.  \left[ \left( \int_0^{x_q}dz \sum_{n=5,6} \left[ (\phi_n)_{LT}^i\right]_+ \right) + \left(q \leftrightarrow \bar{q}\right)\right]^{\ast} \right]  \right \}_{ \substack{x_{\bar{q}} = 1-x_q \\ \vec{p}_q = \frac{x_q}{x_h} \vec{p}_h }} \;
\label{eq:VirtDoubleDipXDipTL}
\end{align}
in the TL case, and
\begin{align} 
&  \frac{d \sigma_{TT}^{q \rightarrow h}}{d x_{h} d^2 p_{h \perp} } \bigg |_{\text{virt., dip. $\times$ d. dip.}} \hspace{-0.4 cm}  = \frac{   \alpha_{\mathrm{em}}}{2 \left(  2\pi\right)^{8}N_{c} \, x_h^2 } \sum_q Q_q^2 \int_{x_h}^1 \hspace{-0.2 cm} dx_q x_q D_q^h \left( \frac{x_h}{x_q}, \mu_F \right) \int \, d^2 \vec{p}_{1} \,  d^2\vec{p}_{2} \,  d^2 \vec{p}_{1^\prime} \,  d^2 \vec{p}_{2^\prime} \nonumber \\
& \times \int d^{2} \vec{p}_{\bar{q}} \int\frac{d^{2}\vec{p}_{3}}{\left(2\pi\right)^{2}} \delta \left( \frac{x_q}{x_h} \vec{p}_h-\vec{p}_{1}+ \vec{p}_{\bar{q}2}-\vec{p}_{3} \right) \delta(\vec{p}_{11^\prime}+ \vec{p}_{22^\prime}+\vec{p}_{3})  \frac{\alpha_{s}}{4\pi} (\varepsilon_{Ti}\varepsilon_{Tj}^{\ast}) \Bigg \{ \widetilde{\mathbf{F}}\left( \frac{\vec{p}_{12}}{2}, \vec{p}_{3}\right) \mathbf{F}^{\ast}\left( \frac{\vec{p}_{1^\prime 2^\prime}}{2}\right)  \nonumber \\
& \times \frac{1}{\vec{p}_{q1^{\prime}}^{\, \,2}+x_q x_{\bar{q}} Q^{2}} \left[  \left[  p_{q1^{\prime}\bot l}p_{q1\bot k} \left[(1-2x)^{2}g_{\bot}^{ki}g_{\bot}^{lj}-g_{\bot}^{kj}g_{\bot}^{li}+g_{\bot}^{kl}g_{\bot}^{ij}\right]\left[ \frac{ \displaystyle -2 \ln\left(  \frac{x_{\bar{q}}}{e^{\eta}}\right) \ln\left(  \frac{\vec{p}_{3}^{\; 2}}{\mu^{2}}\right) }{\vec{p}_{q1}^{\; 2} + x_q x_{\bar{q}} Q^{2}} \right.  \right. \right. \nonumber \\
& + \ln\left(  \frac{x_q x_{\bar{q}}}{e^{2\eta}}\right) \left(  \frac{1}{\vec{p}_{q1}^{\; 2}}\ln\left(  \frac{\vec{p}_{q1}^{\; 2} + x_q x_{\bar{q}} Q^2}{x_q x_{\bar{q}} Q^{2}} \right)  \right. -  \frac{\vec{p}_{\bar{q}2}^{\; 2}+x_q x_{\bar{q}} Q^{2}}{( \vec{p}_{q1}^{\; 2} + x_q x_{\bar{q}} Q^{2})  (\vec{p}_{\bar{q}2}^{\; 2}+x_q x_{\bar{q}} Q^{2})-x_q x_{\bar{q}} Q^{2}\vec{p}_{3}^{\; 2}} \nonumber \\
& \times  \left.  \left.  \left.  \ln\left(  \frac{( \vec{p}_{q1}^{\;  2} +x_q x_{\bar{q}} Q^{2})  ( \vec{p}_{\bar{q}2}^{\; 2} +x_q x_{\bar{q}} Q^{2} )  }{x_q x_{\bar{q}} Q^{2} \vec{p}_{3}^{\; 2}}\right)  \right)  \right] + (q\leftrightarrow\bar{q}) \right] \nonumber \\
&  + \left. \frac{1}{x_q x_{\bar{q}} } \left[ \left( \int_0^{x_q} dz \sum_{n = 5,6} \left[(\phi_n)^{ij}_{TT}\right]_+ \right) + \left(q \leftrightarrow \bar{q}\right)\right] \right]  
+  h.c.|_{\substack{p_{1},p_{2}\leftrightarrow p_{1'},p_{2'} \\ i\leftrightarrow j}} \Bigg \}_{ \substack{x_{\bar{q}} = 1-x_q \\ \vec{p}_q = \frac{x_q}{x_h} \vec{p}_h }}  \; , 
\label{eq:VirtDoubleDipXDipTT}
\end{align}
in the TT case. \\

The corresponding finite virtual contributions in the case of anti-quark fragmentation are obtained as follows: 1) by changing the integration variables $x_q, \vec{p}_{\bar{q}}$ to $x_{\bar{q}}, \vec{p}_{q}$, 2) extending the sum over $q$ to anti-quark flavor types (in order to have the FFs of anti-quarks), 3) computing the objects in the curly brackets fixing $x_q = 1-x_{\bar{q}}$ and $\vec{p}_{\bar{q}} = \frac{x_{\bar{q}}}{x_h} \vec{p}_h$ and 4) making the changes $x_q \rightarrow x_{\bar{q}}$ and $\vec{p}_{\bar{q}} \rightarrow \vec{p}_q$ in the argument of the first delta function. 

\section{NLO cross-section: Divergent part of real corrections}

\subsection{Divergent part of the dipole $\times$ dipole cross-section}

The dipole-dipole partonic cross-section is given by Eq.~(6.6) of Ref.~\cite{Boussarie:2016ogo}:\footnote{Again the expression is dived by a factor $1/(2 (2 \pi)^4)$ with respect to Ref.~\cite{Boussarie:2016ogo} in order to get the correct normalization.}
\begin{align}
   &  d \hat{\sigma}_{3JI}   = \frac{\alpha_s}{\mu^{2\epsilon}} \left( \frac{N_c^2 -1}{N_c}\right) \frac{\alpha_{\mathrm{em}}Q_q^2}{2 (2\pi)^{4d}N_c} \frac{(p_0^-)^2}{s^2 x_q'x_{\bar{q}}'} \varepsilon_{I\alpha} \varepsilon_{J\beta}^*  d x_q' d x_{\bar{q}}'   \delta (1-x_q'-x_{\bar{q}}'-x_g) d^d p_{q\perp}  d^d p_{\bar{q}\perp} \nonumber \\ 
   & \times  \frac{d x_g  d^d p_{g\perp}}{x_g (2\pi)^d} \int d^d p_{1\perp} d^d p_{2\perp} \mathbf{F} \left(\frac{p_{12\perp}}{2}\right)\delta (p_{q1\perp} + p_{\bar{q}2\perp} + p_{g\perp}) \int d^d p_{1'\perp} d^d p_{2'\perp} \mathbf{F}^*\left(\frac{p_{1'2'\perp}}{2}\right) \nonumber \\ 
   & \times \delta (p_{q1'\perp} + p_{\bar{q}2'\perp} + p_{g\perp}) \Phi_3^\alpha (p_{1\perp}, p_{2\perp}) \Phi_3^{\beta*} (p_{1'\perp}, p_{2'\perp}) \; . 
     \label{eq:PartonicDipxDip}
\end{align}
where we introduce shorthand notation by suppressing summation over helicities of partons
\begin{equation}
    \Phi_3^\alpha (p_{1\perp}, p_{2\perp}) \Phi_3^{\beta*} (p_{1'\perp}, p_{2'\perp}) \equiv \sum_{\lambda_q, \lambda_g, \lambda_{\bar{q}} } \Phi_3^\alpha (p_{1\perp}, p_{2\perp}) \Phi_3^{\beta*} (p_{1'\perp}, p_{2'\perp}) \; .
    \label{eq:ShortHand}
\end{equation}
For later convenience, we have also relabelled $x_q$ and $x_{\bar{q}}$ of Ref.~\cite{Boussarie:2016ogo} as $x_q'$ and $x_{\bar{q}}'$. This is because, in dealing with real contributions containing a $qg$ splitting (or even a $\bar{q}g$ splitting), we need to distinguish the longitudinal momentum fraction of the quark transported before and after the splitting.  \\

The dipole part of the $\gamma^{(*)} \rightarrow q \bar{q} g$ impact factor has the form $\Phi_3^\alpha = \Phi_4^\alpha |_{\vec{p}_3 = 0} + \Tilde{\Phi}_3^\alpha$, where $\Tilde{\Phi}_3^\alpha$ is the contribution in which the gluon is emitted after the shockwave and $\Phi_4^{\alpha}$ is the contribution in which the gluon cross the shockwave with $\vec{p}_3$, the transverse momentum exchanged between the gluon and the shockwave, vanishes. Only the square of $\Tilde{\Phi}_3^\alpha$ provides divergences in the cross-section and it is given by (B.3) in Ref.~\cite{Boussarie:2016ogo}. \\

The $LL$ contribution reads
\begin{equation}
\label{eq: div real impact factor}
\begin{aligned}
   &    \Tilde{\Phi}_3^+(\vec{p}_1, \vec{p}_2) \Tilde{\Phi}_3^{+*}(\vec{p}_{1'}, \vec{p}_{2'}) \\
   &= \frac{8 x_q' x_{\bar{q}}' (p_\gamma^+)^4 \left( d x_g^2 + 4 x_q' (x_q' + x_g) \right)}{\left(Q^2 + \frac{\vec{p}_{\bar{q}2}^{\,2}}{x_{\bar{q}}'(1-x_{\bar{q}}')} \right) \left(Q^2 + \frac{\vec{p}_{\bar{q}2'}^{\,2}}{x_{\bar{q}}'(1-x_{\bar{q}}')} \right) (x_q' \vec{p}_g -x_g \vec{p}_q)^2 } \\ 
   & - \frac{8 x_q' x_{\bar{q}}'(p_\gamma^+)^4 \left(2 x_g -d x_g^2 + 4 x_q' x_{\bar{q}}' \right) \left(x_q' \vec{p}_g - x_g \vec{p}_q \right) \cdot \left( x_{\bar{q}}' \vec{p}_g - x_g \vec{p}_{\bar{q}} \right)}{\left(Q^2 + \frac{\vec{p}_{\bar{q}2'}^{\,2}}{x_{\bar{q}}'(1-x_{\bar{q}}')} \right) \left(Q^2 + \frac{\vec{p}_{q1}^{\,2}}{x_q' (1-x_q')} \right)  \left(x_q' \vec{p}_g - x_g \vec{p}_q \right)^2  \left( x_{\bar{q}}' \vec{p}_g - x_g \vec{p}_{\bar{q}} \right)^2 }
   \\
   & +  \frac{8 x_q' x_{\bar{q}}' (p_\gamma^+)^4 \left( d x_g^2 + 4 x_{\bar{q}}' (x_{\bar{q}}' + x_g) \right)}{\left(Q^2 + \frac{\vec{p}_{q1}^{\,2}}{x_q' (1-x_q')} \right) \left(Q^2 + \frac{\vec{p}_{q1'}^{\,2}}{x_q' (1-x_q')} \right) (x_{\bar{q}}' \vec{p}_g -x_g \vec{p}_{\bar{q}})^2 } \\ 
   & - \frac{8 x_q' x_{\bar{q}}' (p_\gamma^+)^4 \left(2 x_g -d x_g^2 + 4 x_q' x_{\bar{q}}' \right) \left(x_q'  \vec{p}_g - x_g \vec{p}_q \right) \cdot \left( x_{\bar{q}}' \vec{p}_g - x_g \vec{p}_{\bar{q}} \right)}{\left(Q^2 + \frac{\vec{p}_{q1'}^{\,2}}{x_q' (1-x_q')} \right) \left(Q^2 + \frac{\vec{p}_{\bar{q}2}^{\,2}}{x_{\bar{q}}' (1-x_{\bar{q}}')} \right)  \left(x_q' \vec{p}_g - x_g \vec{p}_q \right)^2  \left( x_{\bar{q}}' \vec{p}_g - x_g \vec{p}_{\bar{q}} \right)^2 } \, ,
\end{aligned}
\end{equation}
while, the $TL$ contribution is
{\allowdisplaybreaks
\begin{align*}
&  \tilde{\Phi}_3^{+}(\vec{p}_{1},\vec{p}_{2}) \tilde{\Phi}_3^{i*}(\vec{p}_{1'},\vec{p}_{2'}) \\*
& = \frac{4 x_q'\left(p_\gamma^{+}\right)^3}{\left(x_q' +x_g\right)\left(Q^2+\frac{\vec{p}_{\bar{q} 2'}^{\,2}}{x_{\bar{q}}'\left(1-x_{\bar{q}}'\right)}\right)\left(Q^2+\frac{\vec{p}_{q1}^{\,2}}{x_q'\left(1-x_q'\right)}\right)}\left(\frac{ \left(x_q' p_{g\perp} - x_g p_{q\perp}\right)_\mu \left(x_{\bar{q}}' p_{g\perp} - x_g p_{\bar{q}\perp}\right)_\nu}{\left(x_q' \vec{p}_g - x_g \vec{p}_q\right)^2 \left(x_{\bar{q}}' \vec{p}_g - x_g \vec{p}_{\bar{q}}\right)^2}\right) \\ 
& \times \left[x_g\left(4 x_{\bar{q}}' +x_g d-2\right)\left(p_{\bar{q} 2' \perp}^\mu g_{\perp}^{i \nu}-p_{\bar{q} 2' \perp}^\nu g_{\perp}^{\mu i}\right)-\left(2 x_{\bar{q}}'-1\right)\left(4 x_q' x_{\bar{q}}'+x_g\left(2-x_g d\right)\right) g_{\perp}^{\mu \nu} p_{\bar{q} 2' \perp}^i\right] \\
& -\frac{4 x_q ' \left(p_\gamma^{+}\right)^3\left(2 x_{\bar{q}}' -1\right)\left(x_g^2 d+4 x_q' \left(x_q' +x_g\right)\right) p_{\bar{q} 2' \perp}^i}{\left(x_q'+x_g\right)\left(Q^2+\frac{\vec{p}_{\bar{q} 2'}^{\,2}}{x_{\bar{q}}'\left(1-x_{\bar{q}}'\right)}\right)\left(Q^2+\frac{\vec{p}_{\bar{q} 2}^{\,2}}{x_{\bar{q}}'\left(1-x_{\bar{q}}'\right)}\right) \left(x_q' \vec{p}_g- x_g \vec{p}_q\right)^2}
+(q \leftrightarrow \bar{q}) \,, \numberthis
\label{Divergent_Part_impact_factor_TL}
\end{align*}}
and, finally, the $TT$ contribution reads
{\allowdisplaybreaks
\begin{align*}
&  \tilde{\Phi}_3^i(\vec{p}_1,\vec{p}_2) \tilde{\Phi}_3^{k*}(\vec{p}_{1'},\vec{p}_{2'}) \\
& =\frac{-2\left(p_\gamma^{+}\right)^2}{\left(x_q'+x_g\right)\left(x_{\bar{q}}'+x_g\right)\left(Q^2+\frac{\vec{p}_{\bar{q} 2}^{\,2}}{x_{\bar{q}}'\left(1-x_{\bar{q}}'\right)}\right)\left(Q^2+\frac{\vec{p}_{q 1^{\prime}}^{\,2}}{x_q'\left(1-x_q'\right)}\right)} \\
& \times  \left(\frac{ \left(x_q' p_{g\perp} - x_g p_{q\perp}\right)_\mu \left(x_{\bar{q}}' p_{g\perp} - x_g p_{\bar{q}\perp}\right)_\nu}{\left(x_q' \vec{p}_g - x_g \vec{p}_q\right)^2  \left(x_{\bar{q}}' \vec{p}_g - x_g \vec{p}_{\bar{q}}\right)^2}\right) \left\{ x_g ((d-4)) x_g -2) \left[p_{q 1^{\prime} \perp}^\nu\left(p_{\bar{q} 2 \perp}^\mu g_{\perp}^{i k}+p_{\bar{q} 2 \perp}^k g_{\perp}^{\mu i}\right) \right. \right. \\
& \left. +g_{\perp}^{\mu \nu}\left(\left(\vec{p}_{q 1^{\prime}} \cdot \vec{p}_{\bar{q} 2}\right) g_{\perp}^{i k}+p_{q 1^{\prime} \perp}^i p_{\bar{q} 2 \perp}^k\right) -g_{\perp}^{\nu k} p_{q 1^{\prime} \perp}^i p_{\bar{q} 2 \perp}^\mu -g_{\perp}^{\mu i} g_{\perp}^{\nu k}\left(\vec{p}_{q 1^{\prime}} \cdot \vec{p}_{\bar{q} 2}\right) \right] -g_{\perp}^{\mu \nu} \\
& \times  \left[ \left(2x_q' -1 \right) \left(2 x_{\bar{q}}' - 1\right) p_{q1'\perp}^k p_{\bar{q}2\perp}^i \left( 4 x_q' x_{\bar{q}}' + x_g (2 - x_g d)\right)  + 4 x_q' x_{\bar{q}}' ((\vec{p}_{q1'} \cdot \vec{p}_{\bar{q}2})g_\perp^{ik} + p_{q1'\perp}^i p_{\bar{q}2\perp}^k  )\right] \\
& + \left( p_{q1'\perp}^\mu p_{\bar{q}2\perp}^\nu g_\perp^{ik} - p_{q1'\perp}^\mu p_{\bar{q}2\perp}^k g_\perp^{\nu i } - p_{q1'\perp}^i p_{\bar{q}2\perp}^\nu g_\perp^{\mu k } - g_\perp^{\mu k } g_\perp^{\nu i } (\vec{p}_{q1'} \cdot \vec{p}_{\bar{q}2} ) \right) \\ 
& \times x_g ((d-4)x_g + 2) + x_g (2x_{\bar{q}}' - 1 ) (x_g d + 4 x_q' -2 ) \left( g_\perp^{\mu k } p_{q1'\perp}^\nu - g_\perp^{\nu k} p_{q1'\perp}^\mu \right) p_{\bar{q}2\perp}^i \\
& \left.  + x_g (2 x_q' -1 ) p_{q1'\perp}^k (4 x_{\bar{q}}' + x_g d -2) \left( g_\perp^{\nu i } p_{\bar{q}2\perp}^\mu -g_\perp^{\mu i } p_{\bar{q}2\perp}^\nu \right) \right\} \\
& - \frac{2 x_q' (p_\gamma^+)^2 (x_g^2 d + 4x_q'(x_q'+ x_g)) \left( (\vec{p}_{\bar{q}2} \cdot \vec{p}_{\bar{q}2'}) g_\perp^{ik}-(1-2x_{\bar{q}}')^2 p_{\bar{q}2\perp}^i p_{\bar{q}2'\perp}^k + p_{\bar{q}2'\perp}^i p_{\bar{q}2\perp}^k\right) }{x_{\bar{q}}' (x_q' + x_g)^2 \left(Q^2 + \frac{\vec{p}_{\bar{q}2}^{\,2} }{x_{\bar{q}}' (1 - x_{\bar{q}}')} \right) \left(Q^2 + \frac{\vec{p}_{\bar{q}2'}^{\,2}}{x_{\bar{q}}' (1 - x_{\bar{q}}')} \right) \left(x_q' \vec{p}_{g} - x_g \vec{p}_q \right)^2 }    \\
& + (q \leftrightarrow \bar{q}) \numberthis  \,.
\label{Divergent_Part_impact_factor_TT}
\end{align*}} 
When two partons labeled $i$ and $j$ become collinear, the variable
\beqa
\vec{A}_{ij} = x_i \vec{p}_j - x_j \vec{p}_i
\eqa
vanishes. In the $LL$ case, for instance, the first term on the right-hand side of Eq.~\eqref{eq: div real impact factor} gives the collinear divergence associated to the quark-gluon splitting ($\vec{A}_{qg}^{\; 2} \to 0$), while the third gives the one associated to the anti-quark gluon splitting ($\vec{A}_{\bar{q}g}^{\; 2} \to 0$). \\

The calculation technique of the divergent contributions is very similar for the different contributions afflicted by collinear divergences. For this reason, we present in the following sections the explicit extraction of the soft contribution and the explicit computation of one collinearly-divergent term. The others are obtained similarly.

\subsection{Fragmentation from quark}

In this section, we explicitly calculate the divergent contributions in the case of quark fragmentation. The evident symmetry with the case of anti-quarks allows us to provide the formulas for this second fragmentation mechanism as well. 

\subsubsection{Collinear contributions: $q$-$g$ splitting}
\label{sec:Collinearcontributionsqgsplitting}

\begin{figure}[h]
\begin{picture}(430,100)
\put(0,0){\includegraphics[scale=0.30]{images/FF_Single_hadron_NLO_Real_1.eps}}
\put(212,34){$-$}
\put(240,0){\includegraphics[scale=0.30]{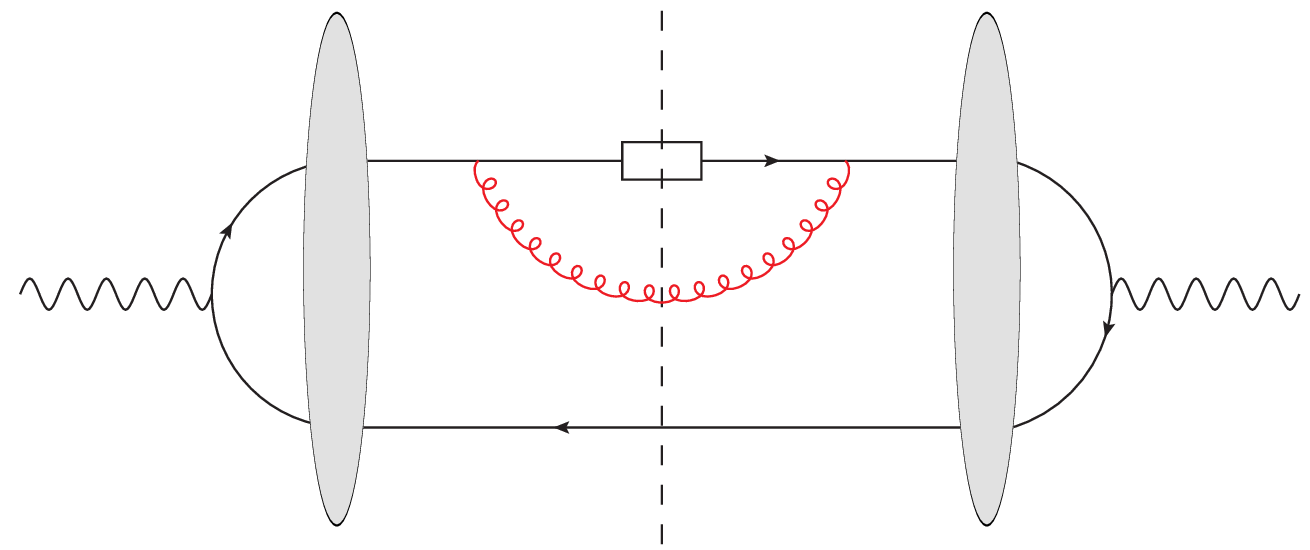}}
\end{picture} 
 \caption{The contribution containing the collinear-divergent part associated with the quark $\rightarrow$ quark $+$ gluon splitting, with the quark fragmenting into the identified hadronic state. The soft part (right diagram with the tiny red gluon) is subtracted from the total contribution (left diagram) in order to have the pure collinear divergence accompanied by finite terms.}
\label{fig:qgCollinear_Conribution}
\end{figure}
The first contribution we calculate is shown in Fig.~\ref{fig:qgCollinear_Conribution}. The left-hand side diagram in Fig.~\ref{fig:qgCollinear_Conribution} contains both soft and collinear divergences, as well as finite contributions. As mentioned before, soft contributions are treated separately, and therefore we subtract them from this contribution. The calculation of this contribution for the different cross sections $(LL,TL,TT)$ is practically identical, apart from the corrective factors, $\delta_{JI}$, which undergo trivial transformations but do not play any deep role. For simplicity of notation, we show the calculation in the $LL$ case and then give the final result in a completely general form valid for all different cases. \\

\begin{figure}
\begin{picture}(430,100)
\put(100,20){\includegraphics[scale=0.45]{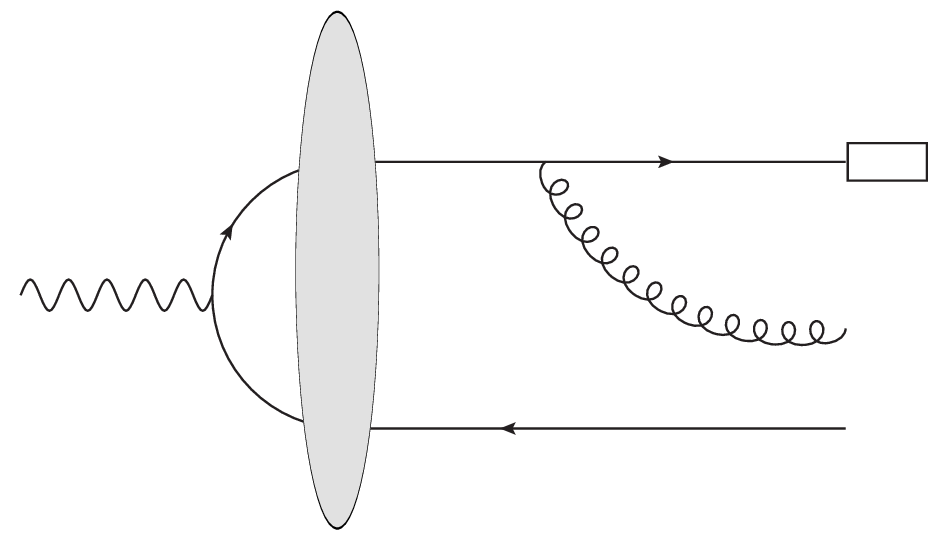}}
\put(192,110){$x_q$}
\put(230,110){$x_q' = \beta x_q$}
\put(250,75){$x_g = (1-\beta) x_q$}
\end{picture} 
 \caption{Illustration of the change of variables in Eq.~(\ref{Eq:Longitudinal_Trasformation}).}
 \label{Fig:ChangeOfVar}
\end{figure}
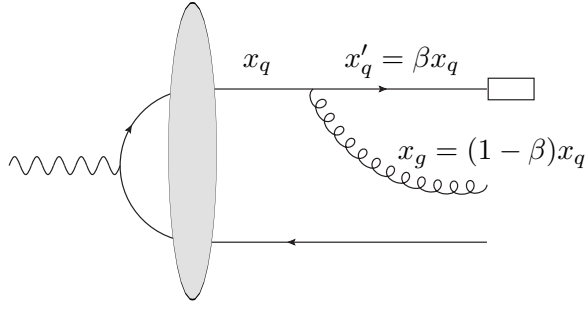

The term shown in Fig.~\ref{fig:qgCollinear_Conribution} corresponds to the first term in Eq.~(\ref{eq: div real impact factor}). Performing the convolution as in Eq.~(\ref{eq: coll facto}) and using the explicit form of the hard cross-section in (\ref{eq:PartonicDipxDip})\footnote{Keeping only the first term in the squared impact factor.}, we get  
\begin{align}
    & \frac{d \sigma_{ L L}^{q \rightarrow h}}{d x_{h} d^d p_{h \perp} } \bigg |_{(1)} \hspace{- 0.25 cm} = \frac{\alpha_s C_F}{\mu^{2\epsilon}} \frac{2 \alpha_{\mathrm{em}}Q^2}{(2\pi)^{4d}N_c} \sum_q Q_q^2 \int_{x_h}^1 \frac{d x_q'}{x_q'} D_{q}^{h} \left(\frac{x_h}{ x_q'}, \mu_{F} \right) \left( \frac{x_q'}{x_h} \right)^d \int d^d p_{\bar{q}\perp} \nonumber \\ 
    & \times \int_{\alpha}^{1-x_q'} \frac{d x_g}{x_g} \int \frac{  d^d p_{g\perp}}{ (2\pi)^d} \int d^d p_{1\perp} d^d p_{2\perp} \mathbf{F} \left(\frac{p_{12\perp}}{2}\right) \int d^d p_{1'\perp} d^d p_{2'\perp} \mathbf{F}^*\left(\frac{p_{1'2'\perp}}{2}\right) \nonumber \\ 
    & \times \delta \left( \frac{x_q'}{x_h} \vec{p}_h -p_{1\perp} + p_{\bar{q}2\perp} + p_{g\perp} \right) \delta \left( \frac{x_q'}{x_h} \vec{p}_h -p_{1'\perp} + p_{\bar{q}2'\perp} + p_{g\perp} \right) \nonumber \\ 
    & \times \frac{ \left( d x_g^2 + 4 x_q' (x_q' + x_g) \right) (x_q'+x_g)^2 (1-x_{q}'-x_g)^2 }{(x_q')^2\left( (x_q'+x_g) (1-x_{q}'-x_g) Q^2 + \vec{p}_{\bar{q}2}^{\,2} \right) \left( (x_q'+x_g) (1-x_{q}'-x_g) Q^2 + \vec{p}_{\bar{q}2'}^{\,2} \right) \left( \vec{p}_g - \frac{x_g }{x_h} \vec{p}_h \right)^2 } \; . 
    \label{Eq:CrossqgSplittingPreTrasf}
\end{align}
The cross section in Eq.~(\ref{Eq:CrossqgSplittingPreTrasf}) is expressed in terms of the fractions of longitudinal momenta of the initial photon carried by the quark and gluon produced after the splitting. However, to observe the cancellation of collinear divergences between this contribution and the counterterms coming from the renormalization of FF, it is necessary to perform the change of variables
\begin{equation}
   x_q' = \beta x_q \; , \hspace{0.5 cm}  x_g = (1-\beta) x_q \; ,
   \label{Eq:Longitudinal_Trasformation}
\end{equation}
where $x_q$ is the fraction of longitudinal momenta of the initial photon carried by the quark before the splitting, while $\beta$ is the fraction of longitudinal momenta of the initial quark carried by the final quark (see Fig. \ref{Fig:ChangeOfVar}). Then, the longitudinal integrations become
\begin{equation}
    \int_{x_h}^1 d x_q' \int_{\alpha}^{1-x_q'} d x_g \; \; \big[ ..... \big] = \int_{x_h}^1 d x_q \; x_q \int^{1-\frac{\alpha}{x_q}}_{\frac{x_h}{x_q}} d \beta \; \; \big[ ..... \big]_{\substack{  x_g = (1-\beta) x_q  \\ x_q' = \beta x_q }} \; ,
\end{equation}
where $\big[ ..... \big]$ represents the whole integrand function in Eq.~(\ref{Eq:CrossqgSplittingPreTrasf}) and the factor $x_q$ in the right hand side comes the Jacobian of the transformation in (\ref{Eq:Longitudinal_Trasformation}). After a bit of algebra, we end up with

\begin{align}
    & \frac{d \sigma_{ L L}^{q \rightarrow h}}{d x_{h} d^d p_{h \perp} } \bigg |_{(1)} \hspace{- 0.25 cm} = \frac{2 \alpha_s C_F}{\mu^{2\epsilon}} \frac{2 \alpha_{\mathrm{em}} Q^2}{(2\pi)^{4d} N_c \, x_h^d} \sum_q Q_q^2 \int_{x_h}^1 d x_q \, x_q^{1+d} (1-x_q)^2 \int^{1-\frac{\alpha}{x_q}}_{\frac{x_h}{x_q}} \frac{d \beta}{\beta}  \nonumber \\*
    & \times \frac{(1+\beta^2 + \epsilon (1-\beta)^2)}{\beta^{2-d} (1-\beta)}  D_{q}^{h} \left(\frac{x_h}{ \beta x_q}, \mu_{F} \right) \int d^d p_{\bar{q}\perp} \int \frac{  d^d p_{g\perp}}{ (2\pi)^d} \int d^d p_{2\perp} \frac{ \mathbf{F} \left(\frac{\beta x_q}{2 x_h} \vec{p}_{h} + \frac{\vec{p}_{\bar{q}}}{2} + \frac{\vec{p}_g}{2} - \vec{p}_2 \right)}{\left( x_q (1-x_{q}) Q^2 + \vec{p}_{\bar{q}2}^{\,2} \right)} \nonumber \\ 
    & \times \int d^d p_{2'\perp} \frac{ \mathbf{F}^{*} \left(\frac{\beta x_q}{2 x_h} \vec{p}_{h} + \frac{\vec{p}_{\bar{q}}}{2} + \frac{\vec{p}_g}{2} - \vec{p}_{2'}\right)}{\left( x_q (1-x_{q}) Q^2 + \vec{p}_{\bar{q}2'}^{\,2} \right)}  \frac{1}{ \left( \vec{p}_g - \frac{(1-\beta) x_q}{x_h} \vec{p}_h \right)^2 } \,. 
\label{Eq:Collqg_Intermidiate}
\end{align}
Then, it is convenient to express the functions $\mathbf{F}, \mathbf{F}^*$ in terms of their Fourier transforms Eq.~\eqref{eq:FTF}, i.e. 
\begin{equation*}
    \mathbf{F} \left(\frac{\beta x_q}{2 x_h} \vec{p}_{h} + \frac{\vec{p}_{\bar{q}}}{2} + \frac{\vec{p}_g}{2} - \vec{p}_2 \right) = \int d^d \vec{z}_1 \, F ( \vec{z}_1 ) \, e^{- i \vec{z}_1 \cdot \left( \frac{\beta x_q}{2 x_h} \vec{p}_{h} + \frac{\vec{p}_{\bar{q}}}{2} + \frac{\vec{p}_g}{2} - \vec{p}_2 \right) } 
\end{equation*}
and
\begin{equation*}
    \mathbf{F}^{*} \left(\frac{\beta x_q}{2 x_h} \vec{p}_{h} + \frac{\vec{p}_{\bar{q}}}{2} + \frac{\vec{p}_g}{2} - \vec{p}_{2'}\right) = \int d^d \vec{z}_2 \, F^{*}( \vec{z}_2 ) \, e^{i \vec{z}_2 \cdot \left( \frac{ \beta x_q}{2 x_h} \vec{p}_{h} + \frac{\vec{p}_{\bar{q}}}{2} + \frac{\vec{p}_g}{2} - \vec{p}_{2'} \right) }  \; ,
\end{equation*}
in order to integrate over $\vec{p}_g$ by using 
\begin{equation}
    \mu^{- 2 \epsilon} \int \frac{d^d \vec{p}_g }{(2 \pi)^d} e^{- i \left( \frac{\vec{z}_{1} - \vec{z}_2}{2} \right) \cdot \vec{p}_g } \frac{1}{\vec{p}_g^{\; 2}} = \frac{1}{4 \pi} \frac{1}{\hat{\epsilon}} \frac{\Gamma (1+\epsilon)}{\Gamma(1-\epsilon)} \left( \frac{ \vec{z}_{12}^{\; 2} \mu^2 }{16} \right)^{-\epsilon} \; .
\end{equation}
Finally, we get\footnote{Color transparency prevents the function $F(z_i)$ to have any singularity in the small dipole size limit, i.e. $z_i \sim 0$. Thus the large $\vec{p}_g$ region in Eq.~(\ref{Eq:Collqg_Intermidiate}) does not lead to divergences.}
\begin{align}
    & \frac{d \sigma_{ L L}^{q \rightarrow h}}{d x_{h} d^d p_{h \perp} } \bigg |_{(1)} \hspace{- 0.45 cm} = \frac{2 \alpha_{\mathrm{em}} Q^2}{(2\pi)^{4d} N_c \, x_h^d} \sum_q Q_q^2 \int_{x_h}^1 \hspace{- 0.1 cm} d x_q x_q^{1+d} (1-x_q)^2 \int^{1-\frac{\alpha}{x_q}}_{\frac{x_h}{x_q}} \frac{d \beta}{\beta} \frac{(1+\beta^2 + \epsilon (1-\beta)^2)}{\beta^{2-d} (1-\beta)}  \nonumber \\ 
    & \times D_{q}^{h} \left(\frac{x_h}{ \beta x_q}, \mu_{F} \right) \frac{\alpha_s C_F}{2 \pi} \int d^d p_{\bar{q}\perp} \int d^d p_{2\perp} \frac{ \displaystyle \int d^d \vec{z}_1 F( \vec{z}_1 ) e^{- i \vec{z_1} \cdot \left( \frac{x_q}{2 x_h} \vec{p}_{h} + \frac{\vec{p}_{\bar{q}}}{2} - \vec{p}_2 \right) } }{\left( x_q (1-x_{q}) Q^2 + \vec{p}_{\bar{q}2}^{\,2} \right)} \nonumber \\ 
    & \times \int d^d p_{2'\perp} \frac{ \displaystyle \int d^d \vec{z}_2 F^{*}( \vec{z}_2 ) e^{i \vec{z_2} \cdot \left( \frac{x_q}{2 x_h} \vec{p}_{h} + \frac{\vec{p}_{\bar{q}}}{2} - \vec{p}_{2'} \right)}}{\left( x_q (1-x_{q}) Q^2 + \vec{p}_{\bar{q}2'}^{\,2} \right)} \frac{1}{\hat{\epsilon}} \frac{\Gamma (1+\epsilon)}{\Gamma(1-\epsilon)} \left( \frac{ \vec{z}_{12}^{\; 2} \mu^2 }{16 \beta^2 } \right)^{-\epsilon}  \,. 
    \label{Eq:CollqgSoftCollFin}
\end{align}
We can also write 
\begin{align}
  \frac{1}{\hat{\epsilon}} \frac{\Gamma (1+\epsilon)}{\Gamma(1-\epsilon)} \left( \frac{ \vec{z}_{12}^{\; 2} \mu^2 }{16 \beta^2 } \right)^{-\epsilon} & =  \frac{1}{\hat{\epsilon}} \left( \frac{\vec{z}_{12}^{\; 2}  \mu^2 }{16 e^{-2 \gamma_E} \beta^2 } \right)^{-\epsilon} + o \left( \epsilon \right) \nonumber \\
  &= \frac{1}{\hat{\epsilon}} \left( \frac{ \vec{z}_{12}^{\; 2}  \mu^2 }{16 e^{-2 \gamma_E} } \right)^{-\epsilon} + 2 \ln \beta + o \left( \epsilon \right) \; .
    \label{Eq:CollqgPartialExpEps}
\end{align}
We can collect the term proportional to $\ln \beta$ and the third term in the square bracket $(1+\beta^2 + \epsilon (1-\beta)^2)$ to obtain a first finite term. As mentioned above, adapting the result to the $TL$ and $TT$ cases is simple, so we present the result for this first term in the completely general form
\begin{align}
& \frac{d \sigma_{ J I}^{q \rightarrow h}}{d x_{h} d^d p_{h \perp} } \bigg |_{\text{coll}(qg),\text{fin}1} \hspace{- 0.25 cm} = \frac{2 \alpha_{\mathrm{em}} Q^2 }{(2\pi)^{4d}N_c \; x_h^d } \sum_{q}  Q_q^2 \int_{x_{h}}^1 \hspace{- 0.25 cm} d x_q \;  x_q^{1+d} (1-x_{q})^2 f_{JI} \nonumber \\
& \times \frac{\alpha_s}{2 \pi} C_F \int_{\frac{x_h}{x_q}}^{1} \frac{d \beta}{\beta}  \left[ (1-\beta) + 2 \frac{1+\beta^2}{(1-\beta)} \ln \beta \right] D_{q}^{h} \left(\frac{x_h}{ \beta x_q}, \mu_{F} \right) \; . 
\label{eq:CollConqgFin1}
\end{align}
Coming back to the $LL$ case, from Eq.~(\ref{Eq:CollqgSoftCollFin}) the remaining part is
\begin{align}
    & \frac{d \sigma_{ L L}^{q \rightarrow h}}{d x_{h} d^d p_{h \perp} } \bigg |_{(1), \text{ rest}} \hspace{- 0.45 cm} = \frac{2 \alpha_{\mathrm{em}} Q^2}{(2\pi)^{4d} N_c \, x_h^d} \sum_q Q_q^2 \int_{x_h}^1 \hspace{- 0.1 cm} d x_q x_q^{1+d} (1-x_q)^2 \int^{1-\frac{\alpha}{x_q}}_{\frac{x_h}{x_q}} \frac{d \beta}{\beta} \frac{(1+\beta^2)}{ (1-\beta)}  \nonumber \\ 
    & \times D_{q}^{h} \left(\frac{x_h}{ \beta x_q}, \mu_{F} \right) \frac{\alpha_s C_F}{2 \pi} \int d^d p_{\bar{q}\perp} \int d^d p_{2\perp} \frac{ \displaystyle \int d^d \vec{z}_1 F( \vec{z}_1 ) e^{- i \vec{z_1} \cdot \left( \frac{x_q}{2 x_h} \vec{p}_{h} + \frac{\vec{p}_{\bar{q}}}{2} - \vec{p}_2 \right) } }{ x_q (1-x_{q}) Q^2 + \vec{p}_{\bar{q}2}^{\,2} } \nonumber \\ 
    & \times \int d^d p_{2'\perp} \frac{ \displaystyle \int d^d \vec{z}_2 F^{*}( \vec{z}_2 ) e^{i \vec{z_2} \cdot \left( \frac{x_q}{2 x_h} \vec{p}_{h} + \frac{\vec{p}_{\bar{q}}}{2}  - \vec{p}_{2^{\prime} }\right)}}{ x_q (1-x_{q}) Q^2 + \vec{p}_{\bar{q}2'}^{\,2} } \frac{1}{\hat{\epsilon}} \left( \frac{\vec{z}_{12}^{\; 2} \mu^2 }{16 e^{-2 \gamma_{E}}} \right)^{-\epsilon}  \, . 
\end{align}
It is now necessary to isolate and remove the soft contribution. To do this, we perform the following manipulation:
\begin{align}
    & \int_{\frac{x_h}{x_q}}^{1-\frac{\alpha}{x_q}} d \beta \frac{f(\beta)}{(1-\beta)} = \int_{\frac{x_h}{x_q}}^1 d \beta \frac{f(\beta)-f(1)}{(1-\beta)} + \int_{\frac{x_h}{x_q}}^{1-\frac{\alpha}{x_q}} d \beta \frac{f(1)}{(1-\beta)} \nonumber \\ 
    & = \int_{\frac{x_h}{x_q}}^1 d \beta \frac{f(\beta)-f(1)}{(1-\beta)} - \int_0^{\frac{x_h}{x_q}} d \beta \frac{f(1)}{(1-\beta)}  + \int_{\frac{x_h}{x_q}}^{1-\frac{\alpha}{x_q}} d \beta \frac{f(1)}{(1-\beta)} + \int_0^{\frac{x_h}{x_q}} d \beta \frac{f(1)}{(1-\beta)} \nonumber \\ 
    & = \int_{\frac{x_h}{x_q}}^1 d \beta \frac{f(\beta)}{(1-\beta)_+}  + \int_{\frac{x_h}{x_q}}^{1-\frac{\alpha}{x_q}} d \beta \frac{f(1)}{(1-\beta)} + \int_0^{\frac{x_h}{x_q}} d \beta \frac{f(1)}{(1-\beta)} \; ,
    \label{Eq:PlusPreManipulation}
\end{align}
where the $+$-prescription was introduced in (\ref{eq: plus prescription}). Among the three terms in the last equality~(\ref{Eq:PlusPreManipulation}), the second is the soft contribution. Indeed, it contains the singularity when $\beta \rightarrow 1 \; (x_g \rightarrow 0)$, regularized by the cut-off $\alpha$.\footnote{This term also includes the soft and collinear divergence, because this latter has been extracted regardless of the value of $\beta$ when integrating over the gluon transverse momenta.} As mentioned previously, this term is subtracted and added back when we extract the soft contribution that comes from the sum of all diagrams. We return to this point in the section~\ref{Sec:Soft_contribution}, where we demonstrate that the term removed here coincides with the one considered later. 
The rest leads us to
\begin{align}
& \frac{d \sigma_{ L L}^{q \rightarrow h}}{d x_{h} d^d p_{h \perp} } \bigg |_{\text{coll}(qg), \text{ rest} } \hspace{- 0.25 cm} = \frac{2 \alpha_{\mathrm{em}} Q^2 }{(2\pi)^{4d}N_c \; x_h^d } \sum_{q}  Q_q^2 \int_{x_{h}}^1 \hspace{- 0.25 cm} d x_q \;  x_q^{1+d} (1-x_{q})^2 \hspace{-0.15 cm} \int \hspace{-0.10 cm} d^{d} p_{ \bar{q} \perp} \hspace{-0.10 cm} \int d^{d} p_{2 \perp} \hspace{-0.10 cm} \int d^{d} p_{2' \perp} \nonumber \\ 
& \times \int d^d \vec{z}_1 \frac{ e^{ -i \vec{z}_1 \cdot \left( \frac{x_q}{2x_{h}}  \vec{p}_{h} + \frac{1}{2} \vec{p}_{ \bar{q} } - \vec{p}_{2} \right) } F \left( \vec{z}_1 \right)}{ \vec{p}_{\bar{q} 2}^{\; 2} + x_q (1-x_{q}) Q^{2}} \int d^d \vec{z}_2 \frac{ e^{ i \vec{z}_2 \cdot \left( \frac{x_q}{2x_{h}}  \vec{p}_{h} + \frac{1}{2} \vec{p}_{ \bar{q} } - \vec{p}_{2'} \right) } F^* \left( \vec{z}_2 \right)}{ \vec{p}_{\bar{q} 2'}^{\; 2} + x_q (1-x_{q}) Q^{2}} \; \; \frac{\alpha_s}{2 \pi} C_F \; \nonumber \\
& \times  \frac{1}{\hat{\epsilon}} \left( \frac{\vec{z}_{12}^{\; 2} \mu^2 }{16 e^{-2 \gamma_{E}}} \right)^{-\epsilon} \left[ \int_{\frac{x_h}{x_q}}^{1} \hspace{- 0.1 cm} \frac{d \beta}{\beta}  \frac{1+\beta^2}{(1-\beta)_{+}} D_{q}^{h} \left(\frac{x_h}{ \beta x_q}, \mu_{F} \right) - 2 \ln \left( 1 - \frac{x_h}{x_q} \right) D_{q}^{h} \left(\frac{x_h}{x_q}, \mu_{F} \right) \right] \; .
\label{Eq:CollqgCollFin}
\end{align}
Performing the expansion 
\begin{equation}
    \frac{1}{\hat{\epsilon}} \left( \frac{\vec{z}_{12}^{\; 2} \mu^2 }{16 e^{-2 \gamma_{E}}} \right)^{-\epsilon} =  \frac{1}{\hat{\epsilon}} - \ln \left( \frac{\vec{z}_{12}^{\; 2} \mu^2}{16 e^{-2 \gamma_{E}}} \right) + o \left( \epsilon \right) \; ,
\end{equation}
we can further separate the contribution in Eq.~(\ref{Eq:CollqgCollFin}) into a divergent (div) and finite (fin) part. Moreover, as before, it is simple to include corrective factors, $\delta_{JI}$, in~(\ref{deltaTT}) to get the general result valid also in the $TL$ and $TT$ case. Indeed, the divergent part associated to this contribution reads
\begin{align}
& \frac{d \sigma_{ J I}^{q \rightarrow h}}{d x_{h} d^d p_{h \perp} } \bigg |_{\text{coll}(qg),\text{div}} \hspace{- 0.25 cm} = \frac{2 \alpha_{\mathrm{em}} Q^2 }{(2\pi)^{4d}N_c \; x_h^d } \sum_{q}  Q_q^2 \int_{x_{h}}^1 \hspace{- 0.25 cm} d x_q \;  x_q^{1+d} (1-x_{q})^2 f_{JI}  \nonumber \\ 
& \times \frac{\alpha_s}{2 \pi} \frac{1}{\hat{\epsilon}} \left[ \int_{\frac{x_h}{x_q}}^{1} \hspace{- 0.1 cm} \frac{d \beta}{\beta}  \frac{1+\beta^2}{(1-\beta)_{+}} C_F D_{q}^{h} \left(\frac{x_h}{ \beta x_q}, \mu_{F} \right) - 2 C_F \ln \left( 1 - \frac{x_h}{x_q} \right) D_{q}^{h} \left(\frac{x_h}{x_q}, \mu_{F} \right) \right] \; ,
\label{eq:CollConqgDiv}
\end{align}
while the second finite part is 
\begin{align}
& \frac{d \sigma_{ J I}^{q \rightarrow h}}{d x_{h} d^d p_{h \perp} } \bigg |_{\text{coll}(qg),\text{fin}2} \hspace{- 0.25 cm} = - \frac{2 \alpha_{\mathrm{em}} Q^2 }{(2\pi)^{4d}N_c \; x_h^d } \sum_{q}  Q_q^2 \int_{x_{h}}^1 \hspace{- 0.25 cm} d x_q \;  x_q^{1+d} (1-x_{q})^2 \hspace{-0.15 cm} \int \hspace{-0.10 cm} d^{d} p_{ \bar{q} \perp} \hspace{-0.10 cm} \int d^{d} p_{2 \perp} \hspace{-0.10 cm} \int d^{d} p_{2' \perp} \nonumber \\
& \times \int d^d \vec{z}_1 \frac{ e^{ -i \vec{z}_1 \cdot \left( \frac{x_q}{2x_{h}}  \vec{p}_{h} + \frac{1}{2} \vec{p}_{ \bar{q} } - \vec{p}_{2} \right) } F \left( \vec{z}_1 \right)}{ \vec{p}_{\bar{q} 2}^{\; 2} + x_q (1-x_{q}) Q^{2}} \int d^d \vec{z}_2 \frac{ e^{ i \vec{z}_2 \cdot \left( \frac{x_q}{2x_{h}}  \vec{p}_{h} + \frac{1}{2} \vec{p}_{ \bar{q} } - \vec{p}_{2'} \right) } F^{*} \left( \vec{z}_2 \right)}{ \vec{p}_{\bar{q} 2'}^{\; 2} + x_q (1-x_{q}) Q^{2}} \; \delta_{JI} \; \frac{\alpha_s}{2 \pi} C_F \; \nonumber \\
& \times \ln \left( \frac{\vec{z}_{12}^{\; 2} \mu^2}{16 e^{-2 \gamma_{E}}} \right) \left[ \int_{\frac{x_h}{x_q}}^{1} \hspace{- 0.1 cm} \frac{d \beta}{\beta}  \frac{1+\beta^2}{(1-\beta)_{+}} D_{q}^{h} \left(\frac{x_h}{ \beta x_q}, \mu_{F} \right) - 2 \ln \left( 1 - \frac{x_h}{x_q} \right) D_{q}^{h} \left(\frac{x_h}{x_q}, \mu_{F} \right) \right] \; .
\label{eq:CollConqgFin2}
\end{align}

The corresponding contributions in the case of fragmentation from anti-quark are obtained by including a minus sign in the argument of the function \textbf{F}($F$), extending the sum over $q$ to the five anti-quark flavor species and performing the $(q \leftrightarrow \bar{q})$ relabelling.

\subsubsection{Collinear contributions: $\bar{q}$-$g$ splitting}
\label{SubSec_Collqbarg}
\begin{figure}[h]
\begin{picture}(430,100)
\put(0,0){\includegraphics[scale=0.30]{images/FF_Single_hadron_NLO_Real_2.eps}}
\put(212,34){$-$}
\put(240,0){\includegraphics[scale=0.30]{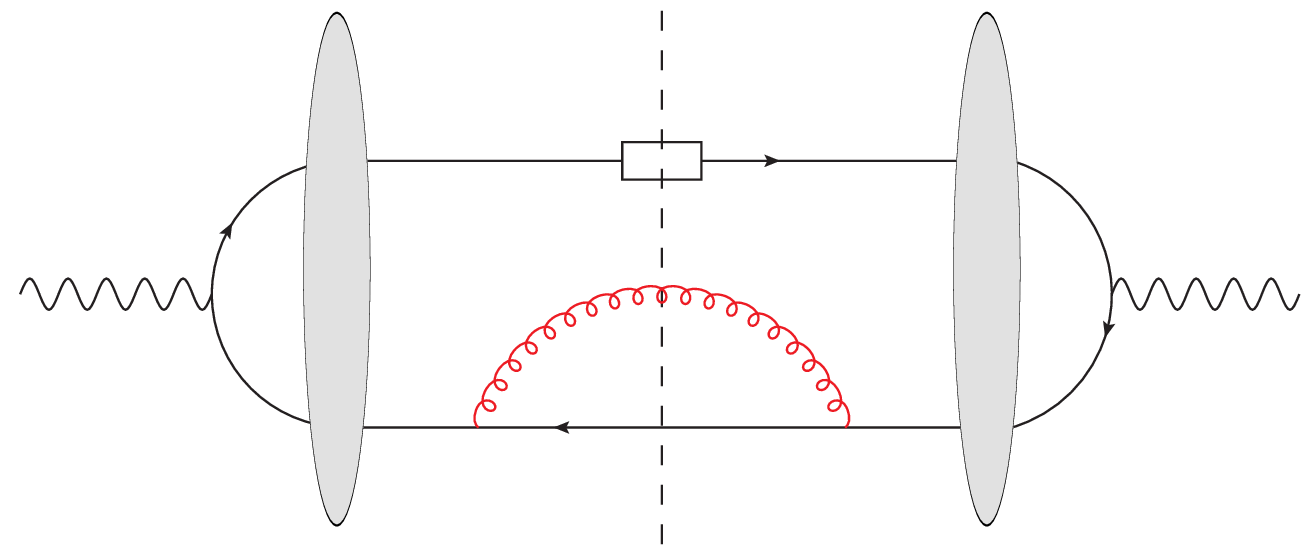}}
\end{picture} 
 \caption{The IR-divergent contribution associated with the anti-quark $\rightarrow$ anti-quark $+$ gluon splitting, but with the spectator quark fragmenting into the identified hadronic state. The soft part (right diagram with the tiny red gluon) is subtracted from the total contribution (left diagram).}
\label{fig:qbargCollinear_Conribution}
\end{figure}
There is another divergent contribution to consider in the case of quark fragmentation, it is shown in Fig.~\ref{fig:qbargCollinear_Conribution}. This contribution arise from the fact that the right diagram in Fig.~\ref{fig:qbargCollinear_Conribution} has a singular behaviour of the type
\begin{equation}
   \frac{1}{\epsilon} \int_{\alpha}^{1-x_q} \frac{d x_g }{x_g} \frac{\left( 4 (1-x_{q}-x_g)^2 + 4 (1-x_q-x_g) x_g + d x_g^2 \right)}{(1-x_{q}-x_g)^2} \; .
   \label{Anom_Soft}
\end{equation}
The term proportional to $ ( \frac{1}{\epsilon} \ln \alpha)$ is the soft term that we subtract (represented by the red gluon in Fig.~\ref{fig:qbargCollinear_Conribution}). Despite this subtraction there is still a singularity of the type $1/\epsilon$ (see Eq.~(\ref{Anom_Soft})) that needs to be extracted. This singularity is not associated with a divergence of the type $\ln \alpha$ and does not combine naturally with the other four diagrams that we consider when calculating the soft contribution. This divergence is of collinear nature. \\

From a technical point of view, the calculation is similar to the one of the previous section, except for the fact that the fragmenting particle is not involved in the splitting, and it is therefore convenient to integrate directly over $x_g$ without carrying out a transformation of the type~(\ref{Eq:Longitudinal_Trasformation}). We obtain two contributions, the first of which, containing the divergent part, reads
\begin{align}
& \frac{d \sigma_{ J I}^{q \rightarrow h}}{d x_{h} d^d p_{h \perp} } \bigg |_{\text{coll}(\bar{q} g)} \hspace{- 0.25 cm} = \frac{2 \alpha_{\mathrm{em}} Q^2 }{(2\pi)^{4d}N_c \; x_h^d } \sum_{q}  Q_q^2 \int_{x_{h}}^1 \hspace{- 0.25 cm} d x_q \;  x_q^{1+d} (1-x_{q})^2 f_{JI} D_{q}^{h} \left(\frac{x_h}{ x_q}, \mu_{F} \right) \nonumber \\ 
& \times \frac{\alpha_s}{2 \pi} C_F \left[ - \frac{3}{2} \frac{1}{\hat{\epsilon}} + 3 - 4 \zeta (2) \right] \equiv \frac{d \sigma_{ J I}^{q \rightarrow h}}{d x_{h} d^d p_{h \perp} } \bigg |_{\text{coll}(\bar{q} g),\text{div}} + \frac{d \sigma_{ J I}^{q \rightarrow h}}{d x_{h} d^d p_{h \perp} } \bigg |_{\text{coll}(\bar{q} g),\text{fin}1} \; ,
\label{eq:CollConqbargDiv} 
\end{align}
where the term labelled with "fin1" contains the terms without $1/ \epsilon$. The second is
\begin{align}
& \frac{d \sigma_{ J I}^{q \rightarrow h}}{d x_{h} d^d p_{h \perp} } \bigg |_{\text{coll}(\bar{q}g),\text{fin}2} \hspace{- 0.25 cm} = \frac{2 \alpha_{\mathrm{em}} Q^2 }{(2\pi)^{4d}N_c \; x_h^d } \sum_{q}  Q_q^2 \int_{x_{h}}^1 \hspace{- 0.25 cm} d x_q \;  x_q^{1+d} (1-x_{q})^2 \int \hspace{-0.10 cm} d^{d} p_{ \bar{q} \perp} \hspace{-0.15 cm} \int d^{d} p_{2 \perp} \hspace{-0.10 cm} \int d^{d} p_{2' \perp} \nonumber \\ 
& \times  \int d^d \vec{z}_1 \frac{ e^{ -i \vec{z}_1 \cdot \left( \frac{x_q}{2x_{h}}  \vec{p}_{h} + \frac{1}{2} \vec{p}_{ \bar{q} } - \vec{p}_{2} \right) } F \left( \vec{z}_1 \right)}{ \vec{p}_{\bar{q} 2}^{\; 2} + x_q (1-x_{q}) Q^{2}} \int d^d \vec{z}_2 \frac{ e^{ i \vec{z}_2 \cdot \left( \frac{x_q}{2x_{h}}  \vec{p}_{h} + \frac{1}{2} \vec{p}_{ \bar{q} } - \vec{p}_{2'} \right) } F^{*} \left( \vec{z}_2 \right)}{ \vec{p}_{\bar{q} 2'}^{\; 2} + x_q (1-x_{q}) Q^{2}} \delta_{JI} \; \frac{\alpha_s}{2 \pi} C_F \; \nonumber \\
& \times  \frac{3}{2}  \ln \left( \frac{\vec{z}_{12}^{\; 2} \; \mu^2 }{16 e^{-2 \gamma_{E}}} \right) D_{q}^{h} \left(\frac{x_h}{ x_q}, \mu_{F} \right) \; .
\label{eq:CollConqbargFin2}
\end{align}

The corresponding contributions in the case of fragmentation from anti-quark are obtained by including a minus sign in the argument of the function \textbf{F}($F$), extending the sum over $q$ to the five anti-quark flavor species and performing the $(q \leftrightarrow \bar{q})$ relabelling.

\subsubsection{Soft contribution}
\label{Sec:Soft_contribution}

\begin{figure}[h]
\begin{picture}(430,170)
\put(20,0){\includegraphics[scale=0.28]{images/FF_Single_hadron_NLO_Real_Soft_2.eps}}
\put(215,132){$+$}
\put(240,0){\includegraphics[scale=0.28]{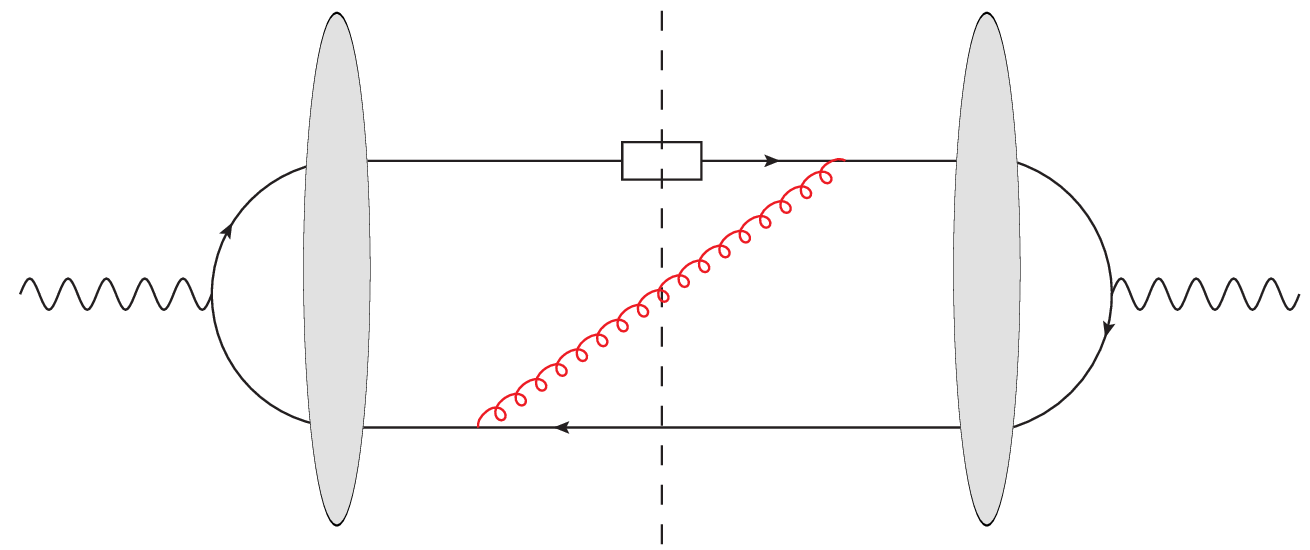}}
\put(20,100){\includegraphics[scale=0.28]{images/FF_Single_hadron_NLO_Real_Soft_1.eps}}
\put(0,32){$+$}
\put(215,32){$+$}
\put(240,100){\includegraphics[scale=0.28]{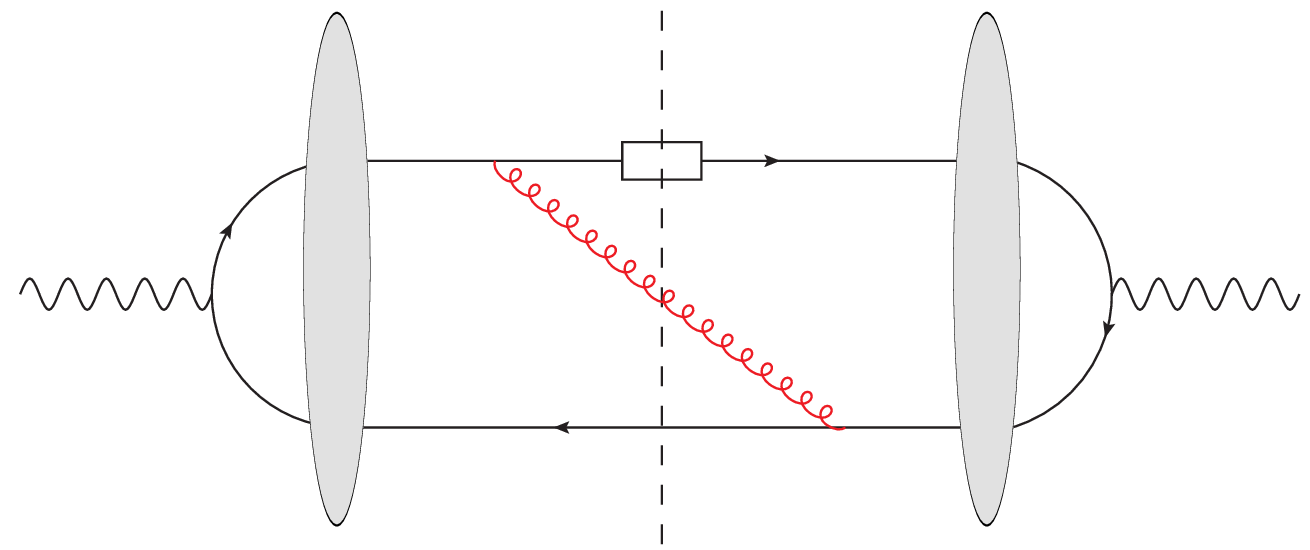}}
\end{picture} 
 \caption{The full soft contribution of real corrections. The tiny red line denotes a soft gluon.}
\label{fig:Soft_Conribution}
\end{figure}

In this section we deal with the associated soft divergences, working in a completely general way with respect to the different cross sections. \\

The four soft-divergent diagrams are shown in Fig.~\ref{fig:Soft_Conribution}. We start from Eq.~(\ref{eq:PartonicDipxDip}) and we make the rescaling $\vec{p}_g = x_g \vec{p}_g^{\; '}$. This parameterization is important because in the soft limit we want all components of the gluon momentum to vanish linearly. The aforementioned rescaling allows us to work in terms of a new non-vanishing transverse component $\vec{p}_g^{\; '}$ and $x_g$ which becomes the only variable in terms of which the soft limit is defined. Relabelling $\vec{p}_g^{\; '}$ as $\vec{p}_g$ after the substitution and setting $x_g$ to zero where possible, we then find  
\begin{gather}
\frac{d \hat{\sigma}_{3JI} }{d x_q \; d^d p_{q\perp} } \bigg |_{\text{Soft}} \hspace{-0.3 cm} = \frac{\alpha_s C_F}{\mu^{2\epsilon}} \frac{8 \alpha_{\mathrm{em}} Q^2 Q_q^2  }{(2\pi)^{4d}N_c}   \int d^d p_{\bar{q}\perp} \frac{d^d p_{g\perp}}{ (2\pi)^d} \hspace{-0.1 cm} \int d^d p_{2\perp}  \frac{\mathbf{F}\left(\frac{x_q'}{2x_{h}}  p_{h \perp} + \frac{1}{2} p_{ \bar{q} \perp} +  \frac{1}{2} x_g p_{ g \perp} - p_{2 \perp} \right)}{ \vec{p}_{\bar{q} 2}^{\; 2} + x_q' (1-x_{q}') Q^{2}}   \nonumber \\ 
\times  \int d^d p_{2'\perp} \frac{\mathbf{F}^{*} \left(\frac{x_q'}{2x_{h}}  p_{h \perp} + \frac{1}{2} p_{ \bar{q} \perp} +  \frac{1}{2} x_g p_{ g \perp} - p_{2' \perp} \right)}{ \vec{p}_{\bar{q} 2'}^{\; 2} + x_q' (1-x_{q}') Q^{2}} \delta_{JI} \int_{\alpha}^{1-x_q'} \hspace{-0.2 cm} \frac{d x_g }{x_g^{1-2 \epsilon}} \frac{ x_q'^{\; 2} (1-x_q')^2 }{\left( \vec{p}_g - \frac{\vec{p}_q}{x_q'} \right)^2 \left( \vec{p}_g - \frac{\vec{p}_{\bar{q}}}{1-x_{q}'} \right)^2 } \nonumber \\ \times  \left \{ \left( \vec{p}_g - \frac{\vec{p}_q}{x_q'} \right)^2 + \left( \vec{p}_g - \frac{\vec{p}_{\bar{q}}}{1-x_{q}'} \right)^2 - 2 \left( \vec{p}_g - \frac{\vec{p}_q}{x_q'} \right) \left( \vec{p}_g - \frac{\vec{p}_{\bar{q}}}{1-x_{q}'} \right)  \right \} \,. 
\label{Eq:SoftComb1}
\end{gather}
In Eq.~(\ref{Eq:SoftComb1}), we kept the factor $\frac{1}{2} x_g p_{g \perp}$, vanishing in the soft limit, in the argument of $\mathbf{F}$ and $\mathbf{F}^{*}$. The reason is that, before proceeding with the explicit computation, we want to briefly come back to our discussion on the soft subtraction and explain why the second term in Eq.~(\ref{Eq:PlusPreManipulation}) coincides with the first term in the last line of Eq.~(\ref{Eq:SoftComb1}). Introducing the Fourier transforms of $\mathbf{F}$ and $\mathbf{F}^{*}$ as was done after Eq.~(\ref{Eq:Collqg_Intermidiate}) and repeating all steps up to Eq.~(\ref{Eq:CollqgSoftCollFin}), the factor $x_g^{-1+2 \epsilon}$ generates an additional factor $(1-\beta)^{2\epsilon}$ that then is compensated by an analogous factor when the integral 
\begin{equation}
    \mu^{- 2 \epsilon} \int \frac{d^d \vec{p}_g }{(2 \pi)^d} e^{- i (1-\beta) \left( \frac{\vec{z}_{1} - \vec{z}_2}{2} \right) \cdot \vec{p}_g } \frac{1}{\vec{p}_g^{\; 2}} = \frac{1}{4 \pi} \frac{1}{\hat{\epsilon}} \frac{\Gamma (1+\epsilon)}{\Gamma(1-\epsilon)} \left( \frac{ \vec{z}_{12}^{\; 2} \mu^2 (1-\beta)^2 }{16} \right)^{-\epsilon} \; 
\end{equation}
is performed, leaving us with exactly the same contribution coming from the second term in Eq.~(\ref{Eq:PlusPreManipulation})\footnote{The exponential factor generated by the shift in $\vec{p}_g$ can be immediately set to one taking $x_g \rightarrow 0$.}. It is important to note that, to treat this single denominator term correctly and simply, it is important to keep the factor $x_g p_{g \perp}/2$, in the argument of $\mathbf{F}$ and $\mathbf{F}^{*}$, different from zero. This problem is related to the UV-behavior of the tadpole integral that is generated by taking the soft limit too naively in this single term. Nonetheless, this problem disappears in the combination of the four contributions in Fig.~\ref{fig:Soft_Conribution}, where the generated integral is
\begin{equation}
    \int \frac{d^d p_{g\perp}}{ (2\pi)^d} \frac{ \left( \frac{\vec{p}_q}{x_q'} - \frac{\vec{p}_{\bar{q}}}{1-x_{q}'} \right)^2 }{\left( \vec{p}_g - \frac{\vec{p}_q}{x_q'} \right)^2 \left( \vec{p}_g - \frac{\vec{p}_{\bar{q}}}{1-x_{q}'} \right)^2 } \; ,
\end{equation}
which contains the correct collinear singularities\footnote{We remind that the superposition of soft and collinear singularities is included in this term that we added and subtracted.} and it is well-behaving at large value $p_{g \perp}$. In this combination, we can conveniently set $x_g \rightarrow 0$ in the argument of $\mathbf{F}$ and $\mathbf{F}^{*}$ and obtain 
\begin{align}
& \frac{d \hat{\sigma}_{3JI} }{d x_q \; d^d p_{q\perp} } \bigg |_{\text{Soft}} \hspace{-0.3 cm} = \frac{\alpha_s C_F}{\mu^{2\epsilon}} \frac{8 \alpha_{\mathrm{em}} Q^2 Q_q^2  }{(2\pi)^{4d}N_c}   \int d^d p_{\bar{q}\perp} \hspace{-0.1 cm} \int d^d p_{2\perp} \frac{\mathbf{F}\left(\frac{x_q'}{2x_{h}}  p_{h \perp} + \frac{1}{2} p_{ \bar{q} \perp} - p_{2 \perp} \right)}{ \vec{p}_{\bar{q} 2}^{\; 2} + x_q' (1-x_{q}') Q^{2}} x_q'^{\; 2} (1-x_q')^2 \nonumber \\ 
& \times \int d^d p_{2'\perp} \frac{\mathbf{F}^{*} \left(\frac{x_q'}{2x_{h}}  p_{h \perp} + \frac{1}{2} p_{ \bar{q} \perp} - p_{2' \perp} \right)}{ \vec{p}_{\bar{q} 2'}^{\; 2} + x_q' (1-x_{q}') Q^{2}} \delta_{JI} \int_{\alpha}^{1-x_q'} \hspace{-0.2 cm} \frac{d x_g }{x_g^{1-2 \epsilon}} \int \frac{d^d p_{g\perp}}{ (2\pi)^d} \frac{ \left( \frac{\vec{p}_q}{x_q'} - \frac{\vec{p}_{\bar{q}}}{1-x_{q}'} \right)^2 }{\left( \vec{p}_g - \frac{\vec{p}_q}{x_q'} \right)^2 \left( \vec{p}_g - \frac{\vec{p}_{\bar{q}}}{1-x_{q}'} \right)^2 } \,. 
\label{Eq:SoftComb2} 
\end{align} 
The integration over the transverse momentum $\vec{p}_g$ is simple and gives
\begin{equation}
    I_{\text{T}} = \int \frac{d^d p_{g\perp}}{ (2\pi)^d} \frac{ \left( \frac{\vec{p}_q}{x_q'} - \frac{\vec{p}_{\bar{q}}}{1-x_{q}'} \right)^2 }{\left( \vec{p}_g - \frac{\vec{p}_q}{x_q'} \right)^2 \left( \vec{p}_g - \frac{\vec{p}_{\bar{q}}}{1-x_q'} \right)^2 } = \frac{\Gamma^2 (\epsilon) \Gamma (1-\epsilon)}{ (4 \pi)^{1+\epsilon} \Gamma (2 \epsilon)} \left[  \left( \frac{\vec{p}_q}{x_q'} - \frac{\vec{p}_{\bar{q}}}{1-x_q'} \right)^2 \right]^{\epsilon} \; .
\end{equation}
The next step is to perform the convolution between the hard cross section and the FF as in Eq.~(\ref{eq: coll facto}); this leads to
\begin{align}
& \frac{d \sigma_{ J I}^{q \rightarrow h}}{d x_{h} d^d p_{h \perp} } \bigg |_{\text{Soft}} \hspace{- 0.25 cm} = \frac{8 \alpha_{\mathrm{em}} Q^2 }{(2\pi)^{4d}N_c x_h^d } \sum_q Q_q^2 \int_{x_h}^1 d x_q' (x_q')^{1+d} (1-x_q')^2  D_q^h \left( \frac{x_h}{x_q'} , \mu_F \right) \int \hspace{-0.1 cm} d^d p_{2\perp} \hspace{-0.1 cm} \int \hspace{-0.1 cm} d^d p_{2'\perp} \nonumber \\
& \times \frac{\alpha_s C_F}{\mu^{2\epsilon}} \int \hspace{-0.1 cm} d^d p_{\bar{q}\perp} \hspace{-0.1 cm} \frac{\mathbf{F}\left(\frac{x_q'}{2x_{h}}  p_{h \perp} + \frac{1}{2} p_{ \bar{q} \perp} - p_{2 \perp} \right)}{ \vec{p}_{\bar{q} 2}^{\; 2} + x_q' (1-x_{q}') Q^{2}} \frac{\mathbf{F}^{*} \left(\frac{x_q'}{2x_{h}}  p_{h \perp} + \frac{1}{2} p_{ \bar{q} \perp} - p_{2' \perp} \right)}{ \vec{p}_{\bar{q} 2'}^{\; 2} + x_q' (1-x_{q}') Q^{2}} \delta_{JI} \int_{\alpha}^{1-x_q'} \frac{d x_g }{x_g^{1-2 \epsilon}} \; I_{\text{T}} \,. 
\end{align}

Before proceeding with the longitudinal integration, an observation is necessary. During the calculation we came across integrals in both variables $x_q'$ and $x_g$, which we treated slightly differently. In particular, in the calculation of the virtual contributions (see section \ref{SubSec:DivVirt}) and in the collinear contribution due to the $\bar{q}g$ splitting (see section \ref{SubSec_Collqbarg}), we integrated directly over $x_g$, while, in the contribution due to the splitting $qg$ we first carried out the change of variables in Eq.~(\ref{Eq:Longitudinal_Trasformation}), to obtain a form of divergences similar to that present in the counter-terms. Since the integration over the final variable $x_q$ is never done explicitly, this can make it difficult to observe the cancellation at the integrand level. We clarify this statement with a toy example. Suppose to have the integral
\begin{gather}
\int_{x_h}^1 d x_q' \int_{\alpha}^{1-x_q'} d x_g \frac{1}{x_g} \; .
\end{gather}
If we integrate over $x_g$ and then rename $x_q'$ as $x_q$, we get
\begin{gather}
I_1 = \int_{x_h}^1 d x_q \ln \left( \frac{1-x_q}{\alpha} \right) \; .   
\end{gather}
From the other side, if we perform the change of variables in~(\ref{Eq:Longitudinal_Trasformation}) and then integrate over $\beta$, we get
\begin{gather}
I_2 = \int_{x_h}^1 d x_q \ln \left( \frac{x_q-x_h}{\alpha} \right) \; .   
\end{gather}
The difference between $I_1$ and $I_2$ is obviously zero since they are the same integral, however, the cancellation is only seen by integrating over $x_q$,
\begin{gather}
I_1 - I_2 = \int_{x_h}^1 d x_q \ln \left( \frac{1-x_q}{x_q-x_h} \right) = 0 \; ,
\end{gather}
i.e. not at the level of the integrand. This problem can be overcome by treating the soft contribution in a symmetrical way with respect to the two different procedures. That is, separating the soft cross-section into two equal parts and treating them according to the two different ways explained before. We also observe that, in the contribution in which the transformation~(\ref{Eq:Longitudinal_Trasformation}) is carried out, since we are in the soft limit $\beta$ can be set to 1 everywhere, except in the term $(1-\beta)^{-1+2 \epsilon}$, which is clearly singular. \\

Proceeding as described above, the final form of the soft contribution is
\begin{align}
& \frac{d \sigma_{ J I}^{q \rightarrow h}}{d x_{h} d^d p_{h \perp} } \bigg |_{\text{Soft}} \hspace{- 0.25 cm} = \frac{2 \alpha_{\mathrm{em}} Q^2 }{(2\pi)^{4d}N_c \; x_h^d } \sum_{q}  Q_q^2 \int_{x_{h}}^1 \hspace{- 0.25 cm} d x_q \;  x_q^{1+d} (1-x_{q})^2 \int d^{d} p_{ \bar{q} \perp} \int d^{d} p_{2 \perp} \int d^{d} p_{2' \perp} \nonumber \\ 
& \times \frac{\mathbf{F}\left(\frac{x_q}{2x_{h}}  p_{h \perp} + \frac{1}{2} p_{ \bar{q} \perp} - p_{2 \perp} \right)}{ \vec{p}_{\bar{q} 2}^{\; 2} + x_q (1-x_{q}) Q^{2}} \; \frac{\mathbf{F}^{*} \left(\frac{x_q}{2x_{h}}  p_{h \perp} + \frac{1}{2} p_{ \bar{q} \perp} - p_{2' \perp} \right)}{ \vec{p}_{\bar{q} 2'}^{\; 2} + x_q (1-x_{q}) Q^{2}} D_q^h \left( \frac{x_h}{x_q} , \mu_F \right) \delta_{JI} \nonumber \\ 
& \times \frac{\alpha_s}{2 \pi} C_F  \left( \frac{1}{\hat{\epsilon}} A_{\text{Soft,div}} + A_{\text{Soft,fin}} \right) \equiv \frac{d \sigma_{ J I}^{q \rightarrow h}}{d x_{h} d^d p_{h \perp} } \bigg |_{\text{Soft,div}} + \frac{d \sigma_{ J I}^{q \rightarrow h}}{d x_{h} d^d p_{h \perp} } \bigg |_{\text{Soft,fin}} \; .
\label{eq:Softpart}
\end{align}
where
\begin{align}
A_{\text{Soft,div}} &  = - 4 \ln \alpha + 2 \ln x_q (1-x_q) + 2 \ln \left( 1 - \frac{x_h}{x_q} \right) \nonumber \\ 
& + 4 \epsilon \ln \alpha \ln \left( \frac{x_h^2 (1-x_q)^2 \mu^2}{ \left( x_h \vec{p}_{\bar{q}} - (1-x_q) \vec{p}_h \right)^2} \right) - \epsilon \ln^2 \alpha^2  \; ,  \\
A_{\text{Soft,fin}} &  = - 2   \ln \left( x_q (1-x_q) \left( 1 - \frac{x_h}{x_q} \right) \right) \ln \left( \frac{x_h^2 (1-x_q)^2 \mu^2}{  \left( x_h \vec{p}_{\bar{q}} - (1-x_q) \vec{p}_h \right)^2} \right) \nonumber \\ 
& + 2 \ln^2 (1-x_q) + 2 \ln^2 \left( x_q \left(1- \frac{x_h}{x_q} \right)  \right) \; ,
\end{align}
and the functions $\delta_{JI}$ are defined in~(\ref{deltaTT}). \\

The corresponding contributions in the case of fragmentation from anti-quark are obtained by including a minus sign in the argument of the function \textbf{F}, extending the sum over $q$ to the five anti-quark flavor species and performing the $(q \leftrightarrow \bar{q})$ relabelling (also inside the functions $A_{{\rm{Soft}},{\rm{div}}}$, $A_{{\rm{Soft}},{\rm{fin}}}$, $\delta_{JI}$). 

\subsection{Cancellation of divergences in the quark fragmentation case}
We can now show the cancellation of divergences in the quark fragmentation channel. First, we combine the divergent virtual, Eq.~(\ref{eq:SVpart}), and soft, Eq.~(\ref{eq:Softpart}), contributions, 
\begin{align}
   &  \frac{d \sigma_{ J I}^{q \rightarrow h}}{d x_{h} d^d p_{h \perp} } \bigg |_{S_V\text{,div}} \hspace{- 0.25 cm} + \frac{d \sigma_{ J I}^{q \rightarrow h}}{d x_{h} d^d p_{h \perp} } \bigg |_{\text{Soft,div}} = \frac{2 \alpha_{\mathrm{em}} Q^2 }{(2\pi)^{4d}N_c \; x_h^d }  \nonumber \\
   & \times \sum_{q}  Q_q^2  \int_{x_{h}}^1  d x_q \;  x_q^{1+d} (1-x_{q})^2 f_{JI} D_{q}^{h} \left(\frac{x_h}{ x_q}, \mu_{F} \right) \frac{\alpha_s}{2 \pi} C_F  \frac{1}{\hat{\epsilon}} \left[3 + 2 \ln \left( 1 - \frac{x_h}{x_q} \right) \right] \; 
\label{Eq:DivCanc1}
\end{align}
and we observe the full cancellation of $\alpha$-divergent terms. Then, we sum the divergent term proportional to the $P_{qq}$ in Eq.~(\ref{eq:FFNLOCounterTerm}) (see also Eq.~(\ref{eq:FFNLOCounterTerm2})) with Eq.~(\ref{eq:CollConqgDiv}) and the divergent contributions in (\ref{eq:CollConqbargDiv}) and find
\begin{align}
  & \hspace{0.15 cm} \frac{d \sigma_{J I}^{q \rightarrow h}}{d x_{h} d^d p_{h \perp} } \bigg |_{\text{ct, div, }P_{qq}} \hspace{- 0.25 cm} + \frac{d \sigma_{ J I}^{q \rightarrow h}}{d x_{h} d^d p_{h \perp} } \bigg |_{\text{coll}(q g),\text{div}} \hspace{- 0.25 cm}  + \hspace{0.15 cm} \frac{d \sigma_{ J I}^{q \rightarrow h}}{d x_{h} d^d p_{h \perp} } \bigg |_{\text{coll}(\bar{q} g),\text{div}} \hspace{- 0.25 cm}  = \frac{2 \alpha_{\mathrm{em}} Q^2 }{(2\pi)^{4d}N_c \; x_h^d }  \nonumber \\
  & \times \sum_{q}  Q_q^2  \int_{x_{h}}^1  d x_q \;  x_q^{1+d} (1-x_{q})^2 f_{JI} D_{q}^{h} \left(\frac{x_h}{ x_q}, \mu_{F} \right) \frac{\alpha_s}{2 \pi} C_F  \frac{1}{\hat{\epsilon}} \left[-3 - 2 \ln \left( 1 - \frac{x_h}{x_q} \right) \right] \; .
  \label{Eq:DivCanc2}
\end{align}
The two contributions in Eqs.~(\ref{Eq:DivCanc1}, \ref{Eq:DivCanc2}) cancel each other, giving a full cancellation in the quark fragmentation case. The cancellation in the case of anti-quark fragmentation takes place in the same way.

\subsection{Fragmentation from gluon}

Finally, we can have a contribution coming from the fragmentation of a gluon. As already mentioned, the two divergent contributions are diagrams $(1)$ and $(3)$ of Fig.~\ref{fig:GluonInANutshell}. The contribution of the diagram $(3)$ is easy to derive once that of the diagram $(1)$ has been calculated. 

\subsubsection{Collinear contributions: $q$-$g$ splitting}

\begin{figure}[h]
\begin{picture}(430,100)
\put(120,0){\includegraphics[scale=0.30]{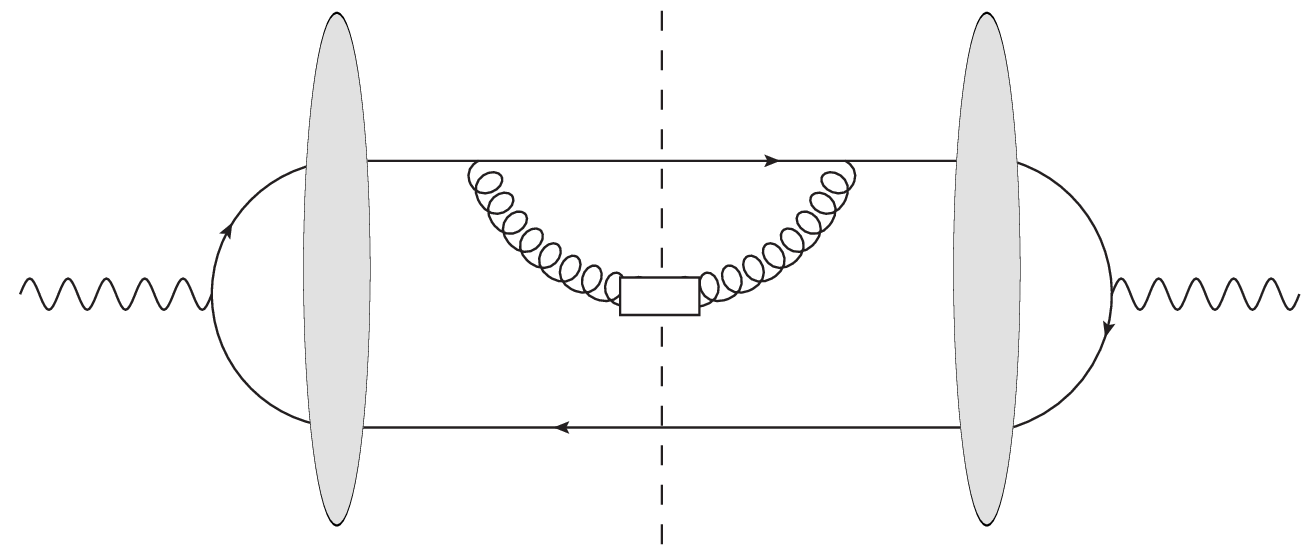}}
\end{picture} 
 \caption{The IR-divergent contribution associated with the quark $\rightarrow$ quark $+$ gluon splitting, with the gluon fragmenting into the identified hadronic state.}
\label{fig:CollinearGluon_Contribution}
\end{figure}

The strategy of the computation is identical to that of section~\ref{sec:Collinearcontributionsqgsplitting}, but considerably much simpler because there are no soft divergences involved. This time the correct change of variables to make is 
\begin{equation}
   x_g = \beta x_q \; , \hspace{0.5 cm}  x_q' = (1-\beta) x_q \; .
   \label{Eq:Longitudinal_Trasformation_Gluon}
\end{equation}
The divergent part associated to this contribution reads
\begin{align}
& \frac{d \sigma_{ J I}^{g \rightarrow h}}{d x_{h} d^d p_{h \perp} } \bigg |_{\text{coll}(qg),\text{div}} \hspace{- 0.25 cm} = \frac{2 \alpha_{\mathrm{em}} Q^2 }{(2\pi)^{4d}N_c \; x_h^d } \sum_{q}  Q_q^2 \int_{x_{h}}^1 \hspace{- 0.25 cm} d x_q \;  x_q^{1+d} (1-x_{q})^2 f_{JI}  \nonumber \\ 
& \times \frac{\alpha_s}{2 \pi} \frac{1}{\hat{\epsilon}} \int_{\frac{x_h}{x_q}}^{1} \hspace{- 0.1 cm} \frac{d \beta}{\beta}  \frac{1+(1-\beta)^2}{ \beta} C_F  D_{g}^{h} \left(\frac{x_h}{ \beta x_q}, \mu_{F} \right) \; ,
\label{eq:CollConqgDivGluon}
\end{align}
while the finite part reads
\begin{align}
& \frac{d \sigma_{ J I}^{g \rightarrow h}}{d x_{h} d^d p_{h \perp} } \bigg |_{\text{coll}(qg),\text{fin}} \hspace{- 0.25 cm} = \frac{2 \alpha_{\mathrm{em}} Q^2 }{(2\pi)^{4d}N_c \; x_h^d } \sum_{q}  Q_q^2 \int_{x_{h}}^1 \hspace{- 0.25 cm} d x_q \;  x_q^{1+d} (1-x_{q})^2 \hspace{-0.15 cm} \int \hspace{-0.10 cm} d^{d} p_{ \bar{q} \perp} \hspace{-0.10 cm} \int d^{d} p_{2 \perp} \hspace{-0.10 cm} \int d^{d} p_{2' \perp} \nonumber \\ 
& \times \int d^d \vec{z}_1 \frac{ e^{ -i \vec{z}_1 \cdot \left( \frac{x_q}{2x_{h}}  \vec{p}_{h} + \frac{1}{2} \vec{p}_{ \bar{q} } - \vec{p}_{2} \right) } F \left( \vec{z}_1 \right)}{ \vec{p}_{\bar{q} 2}^{\; 2} + x_q (1-x_{q}) Q^{2}} \int d^d \vec{z}_2 \frac{ e^{ i \vec{z}_2 \cdot \left( \frac{x_q}{2x_{h}}  \vec{p}_{h} + \frac{1}{2} \vec{p}_{ \bar{q} } - \vec{p}_{2'} \right) } F^{*} \left( \vec{z}_2 \right)}{ \vec{p}_{\bar{q} 2'}^{\; 2} + x_q (1-x_{q}) Q^{2}} \; \delta_{JI} \; \frac{\alpha_s}{2 \pi} C_F \; \nonumber \\
& \times \int_{ \frac{x_h}{x_q}}^1 d \beta \left[ 1 - \frac{1+(1-\beta)^2}{\beta^2} \ln \left( \frac{\vec{z}_{12}^{\; 2} \; \mu^2 }{16 \beta^2 e^{-2 \gamma_{E}}} \right) \right] D_{g}^{h} \left(\frac{x_h}{ \beta x_q}, \mu_{F} \right) \; .
\label{eq:CollConqgFinGluon}
\end{align}
The divergent contribution, Eq.~(\ref{eq:CollConqgDivGluon}), exactly cancels the divergent term proportional to the $P_{gq}$ in Eq.~(\ref{eq:FFNLOCounterTerm}) (see also Eq.~(\ref{eq:FFNLOCounterTerm2})). This completes our proof of the cancellation of divergences. \\

The corresponding contributions in the case of gluon emission from anti-quark (see diagram (3) in Fig.~\ref{fig:GluonInANutshell}) are obtained by including a minus sign in the argument of the function \textbf{F}($F$), extending the sum over $q$ to the five anti-quark flavor species and performing the $(q \leftrightarrow \bar{q})$ relabelling. Obviously, the divergences that appear in this case cancel out that proportional to the $P_{gq}$ in the renormalization of the FF of the anti-quarks. 

\section{NLO cross-section: Finite part of real corrections}

The finite contributions to the real corrections are obtained by convolving the hard cross sections (calculated from the squared impact factors in Appendix~\ref{sec: appendixC}) with FFs as in Eq.~(\ref{eq: coll facto}).

\subsection{Fragmentation from quark}

\subsubsection{Finite remainder in the $\tilde{\Phi}_3^{\alpha} \tilde{\Phi}_3^{\beta *}$ part}

\begin{figure}[h]
\begin{picture}(430,170)
\put(20,0){\includegraphics[scale=0.28]{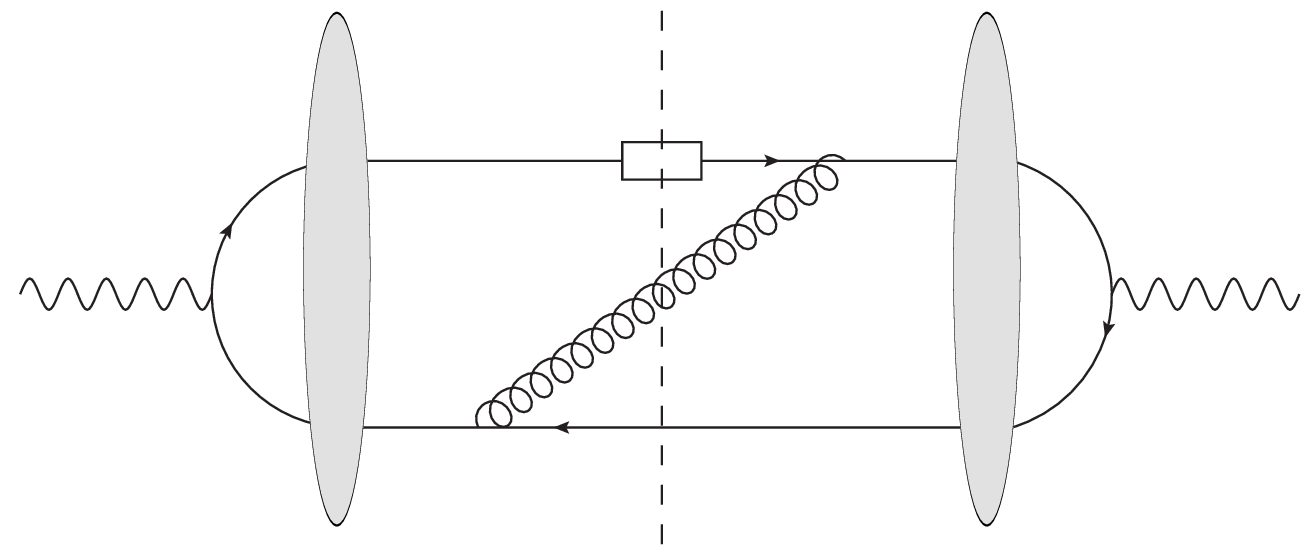}}
\put(215,132){$-$}
\put(240,0){\includegraphics[scale=0.28]{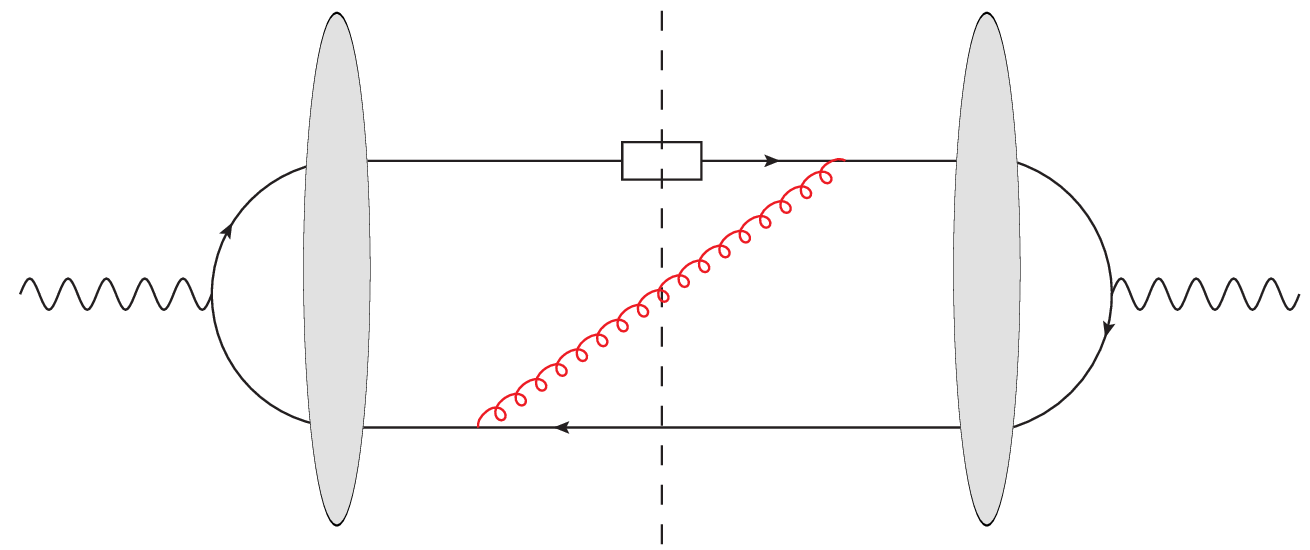}}
\put(20,100){\includegraphics[scale=0.28]{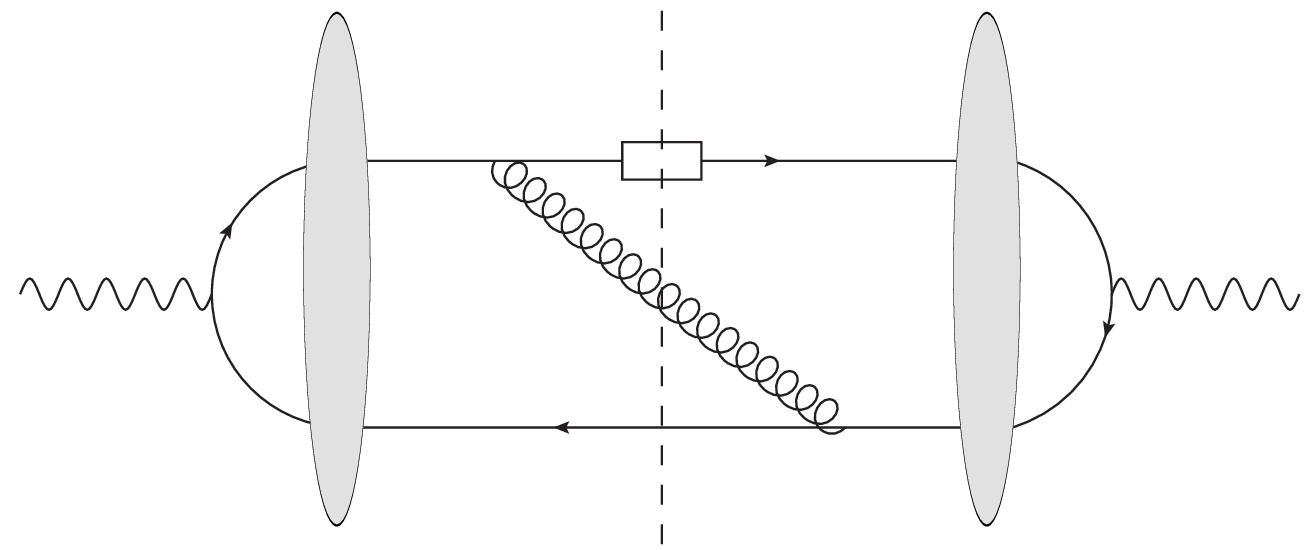}}
\put(0,32){$+$}
\put(215,32){$-$}
\put(240,100){\includegraphics[scale=0.28]{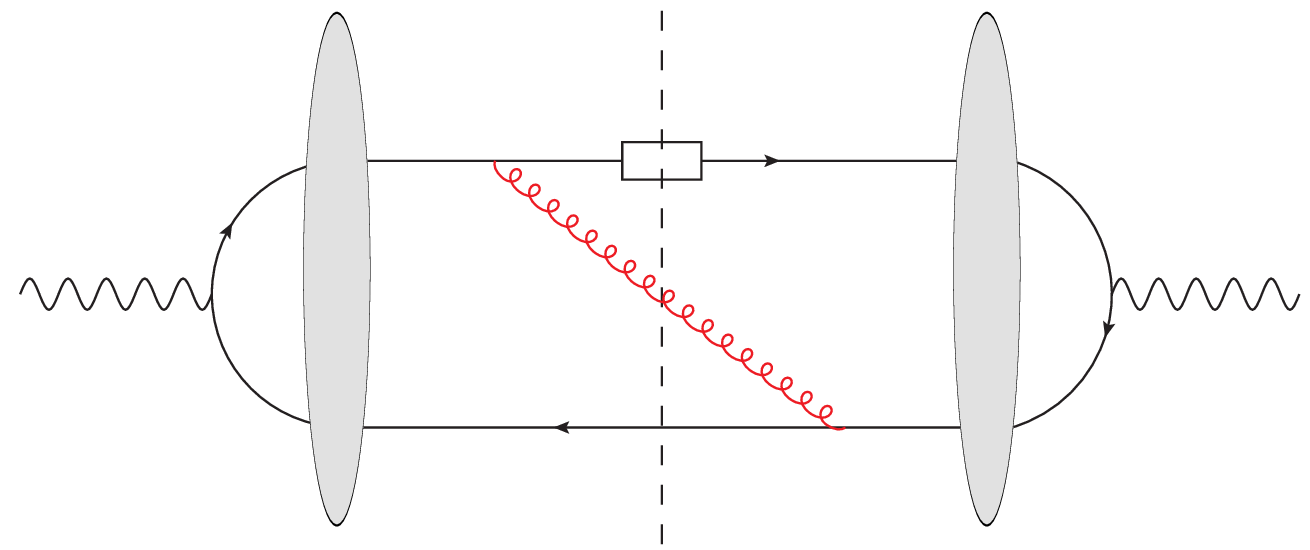}}
\end{picture} 
 \caption{The finite remainder in the $\tilde{\Phi}_3^{\alpha} \tilde{\Phi}_3^{\beta *}$ part, in the case of quark fragmentation. It is basically constituted by the two non-collinearly divergent contributions from which the soft part (diagrams with the tiny red gluon) is subtracted.}
\label{fig:Finit_SoftSub_Conribution}
\end{figure}

The subtraction of the soft contribution leave the finite remainder shown in Fig.~\ref{fig:Finit_SoftSub_Conribution}. This contribution is
\begin{align}
    & \frac{d \sigma_{ J I}^{q \rightarrow h}}{d x_{h} d^2 p_{h \perp} } \bigg |_{\text{real, fin. sub.}} \hspace{- 0.25 cm} = \frac{ \alpha_{em}}{(2 \pi)^8 N_c \, x_h^2} \sum_q Q_q^2 \int_{x_h}^{1} d x_q D_q^h \left( \frac{x_h}{x_q}, \mu_F \right) \int \; d^{2}\vec{p}_{1} \; d^2 \vec{p}_{2} \; d^{2} \vec{p}_{1^{\prime}} \; d^2 \vec{p}_{2^\prime} \nonumber \\ 
    & \times  \frac{1}{(p_{\gamma}^+)^2} \int d^{2} \vec{p}_{\bar{q}} \int \frac{d^{2} \vec{p}_{g}}{(2 \pi)^2} \int_0^{1-x_q} d x_g \frac{ x_g }{ (1-x_q-x_g)} \mathbf{F}\left( \frac{\vec{p}_{12}}{2}\right)\mathbf{F}^\ast\left( \frac{\vec{p}_{1^\prime 2^\prime }}{2}\right) \varepsilon_{I \alpha } \varepsilon_{J \beta}^\ast \frac{\alpha_s}{4} C_F \nonumber \\ 
    & \hspace{-0.2 cm} \times \hspace{-0.1 cm} \left \{ \hspace{-0.1 cm} \left[ \hat{\Phi}_3^{\alpha} \left( \vec{p}_1, \vec{p}_2 \right) \hat{\Phi}_3^{\beta \ast} \left( \vec{p}_{1'}, \vec{p}_{2'} \right) \delta \left( \frac{x_q}{x_h} \vec{p}_h - \vec{p}_1 + \vec{p}_{\bar{q} 2} + \vec{p}_g \right) \delta \left( \frac{x_q}{x_h} \vec{p}_h - \vec{p}_{1'} + \vec{p}_{\bar{q} 2'} + \vec{p}_g  \right) \right]_{(\vec{p}_g \rightarrow x_g \vec{p}_g )} \right. \nonumber \\ 
    & \left. \hspace{-0.3 cm} - \left[ \hat{\Phi}_3^{\alpha} \left( \vec{p}_1, \vec{p}_2 \right) \hat{\Phi}_3^{\beta \ast} \left( \vec{p}_{1'}, \vec{p}_{2'} \right) \delta \left( \frac{x_q}{x_h} \vec{p}_h - \vec{p}_1 + \vec{p}_{\bar{q} 2} + \vec{p}_g \right) \delta \left( \frac{x_q}{x_h} \vec{p}_h - \vec{p}_{1'} + \vec{p}_{\bar{q} 2'} + \vec{p}_g  \right) \right]_{\left( \substack{\vec{p}_g \rightarrow x_g \vec{p}_g  \vspace{0.05 cm} \\ x_g \sim 0 } \right)} \right \} ,
\label{Eq:FiniteRemainderOfSoftSub}
\end{align}
where 
\begin{gather}
\hat{\Phi}_3^{+} \left( \vec{p}_1, \vec{p}_2 \right) \hat{\Phi}_3^{+ \ast} \left( \vec{p}_{1'}, \vec{p}_{2'} \right) = \frac{8 x_q x_{\bar{q}}(p_\gamma^+)^4 \left( d x_g^2 - 2 x_g - 4 x_q x_{\bar{q}} \right) \left(x_q \vec{p}_g - x_g \vec{p}_q \right) \cdot \left( x_{\bar{q}} \vec{p}_g - x_g \vec{p}_{\bar{q}} \right)}{\left(Q^2 + \frac{\vec{p}_{\bar{q}2'}^{\; 2}}{x_{\bar{q}}(1-x_{\bar{q}})} \right) \left(Q^2 + \frac{\vec{p}_{q1}^{\;2}}{x_q (1-x_q)} \right)  \left(x_q \vec{p}_g - x_g \vec{p}_q \right)^2  \left( x_{\bar{q}} \vec{p}_g - x_g \vec{p}_{\bar{q}} \right)^2 } \nonumber \\ - \frac{8 x_q x_{\bar{q}} (p_\gamma^+)^4 \left(2 x_g -d x_g^2 + 4 x_q x_{\bar{q}} \right) \left(x_q  \vec{p}_g - x_g \vec{p}_q \right) \cdot \left( x_{\bar{q}} \vec{p}_g - x_g \vec{p}_{\bar{q}} \right)}{\left(Q^2 + \frac{\vec{p}_{q1'}^{\; 2}}{x_q (1-x_q)} \right) \left(Q^2 + \frac{\vec{p}_{\bar{q}2}^{\; 2}}{x_{\bar{q}} (1-x_{\bar{q}})} \right)  \left(x_q \vec{p}_g - x_g \vec{p}_q \right)^2  \left( x_{\bar{q}} \vec{p}_g - x_g \vec{p}_{\bar{q}} \right)^2 } \; \; ,
\label{HatPhi1}
\end{gather}
\begin{gather}
 \hat{\Phi}_3^{+} \left( \vec{p}_{1},\vec{p}_{2} \right) \hat{\Phi}_3^{i \ast} \left(\vec{p}_{1'},\vec{p}_{2'}\right) = \frac{4 x_q\left(p_\gamma^{+}\right)^3}{\left(x_q +x_g\right)\left(Q^2+\frac{\vec{p}_{\bar{q} 2'}^{\,2}}{x_{\bar{q}} \left(1-x_{\bar{q}}\right)}\right)\left(Q^2+\frac{\vec{p}_{q1}^{\; 2}}{x_q\left(1-x_q\right)}\right)} \nonumber \\ \times \left( \frac{ \left(x_q p_{g\perp} - x_g p_{q\perp}\right)_\mu \left(x_{\bar{q}} p_{g\perp} - x_g p_{\bar{q}\perp}\right)_\nu}{\left(x_q \vec{p}_g - x_g \vec{p}_q\right)^2 \left(x_{\bar{q}} \vec{p}_g - x_g \vec{p}_{\bar{q}}\right)^2}\right) \left[x_g\left(4 x_{\bar{q}} +x_g d-2\right)\left(p_{\bar{q} 2' \perp}^\mu g_{\perp}^{i \nu}-p_{\bar{q} 2' \perp}^\nu g_{\perp}^{\mu i}\right) \nonumber \right. \\  \left. -\left(2 x_{\bar{q}}-1\right)\left(4 x_q x_{\bar{q}}+x_g\left(2-x_g d\right)\right) g_{\perp}^{\mu \nu} p_{\bar{q} 2' \perp}^i\right] + (q \leftrightarrow \bar{q}) \; ,
\label{HatPhi2}
\end{gather}
\begin{align}
&  \hat{\Phi}_3^i\left(\vec{p}_1,\vec{p}_2\right) \hat{\Phi}_3^{k \ast}\left(\vec{p}_{1'},\vec{p}_{2'}\right) =\frac{-2\left(p_\gamma^{+}\right)^2}{\left(x_q+x_g\right)\left(x_{\bar{q}}+x_g\right)\left(Q^2+\frac{\vec{p}_{\bar{q} 2}^{\; 2}}{x_{\bar{q}} \left(1-x_{\bar{q}}\right)}\right)\left(Q^2+\frac{\vec{p}_{q1^{\prime}}^{\; 2}}{x_q \left(1-x_q \right)}\right)} \nonumber \\
& \times  \left(\frac{ \left(x_q p_{g\perp} - x_g p_{q\perp}\right)_\mu \left(x_{\bar{q}} p_{g\perp} - x_g p_{\bar{q}\perp}\right)_\nu}{\left(x_q \vec{p}_g - x_g \vec{p}_q\right)^2  \left(x_{\bar{q}} \vec{p}_g - x_g \vec{p}_{\bar{q}}\right)^2}\right) \left\{ x_g \left(\left(d-4\right)\right) x_g -2) \left[p_{q 1^{\prime} \perp}^\nu \left(p_{\bar{q} 2 \perp}^\mu g_{\perp}^{i k}+p_{\bar{q} 2 \perp}^k g_{\perp}^{\mu i}\right) \right. \right. \nonumber \\
& \left. + \; g_{\perp}^{\mu \nu}\left(\left(\vec{p}_{q 1^{\prime}} \cdot \vec{p}_{\bar{q} 2}\right) g_{\perp}^{i k}+p_{q 1^{\prime} \perp}^i p_{\bar{q} 2 \perp}^k\right) -g_{\perp}^{\nu k} p_{q 1^{\prime} \perp}^i p_{\bar{q} 2 \perp}^\mu -g_{\perp}^{\mu i} g_{\perp}^{\nu k}\left(\vec{p}_{q 1^{\prime}} \cdot \vec{p}_{\bar{q} 2}\right) \right] -g_{\perp}^{\mu \nu} \nonumber \\
& \times  \left[ \left(2x_q -1 \right) \left(2 x_{\bar{q}} - 1\right) p_{q1'\perp}^k p_{\bar{q}2\perp}^i \left( 4 x_q x_{\bar{q}} + x_g (2 - x_g d)\right)  + 4 x_q x_{\bar{q}} ((\vec{p}_{q1'} \cdot \vec{p}_{\bar{q}2})g_\perp^{ik} + p_{q1'\perp}^i p_{\bar{q}2\perp}^k  )\right] \nonumber \\
& + \left( p_{q1'\perp}^\mu p_{\bar{q}2\perp}^\nu g_\perp^{ik} - p_{q1'\perp}^\mu p_{\bar{q}2\perp}^k g_\perp^{\nu i } - p_{q1'\perp}^i p_{\bar{q}2\perp}^\nu g_\perp^{\mu k } - g_\perp^{\mu k } g_\perp^{\nu i } (\vec{p}_{q1'} \cdot \vec{p}_{\bar{q}2} ) \right) \nonumber \\ 
& \times x_g ((d-4)x_g + 2) + x_g (2x_{\bar{q}} - 1 ) (x_g d + 4 x_q -2 ) \left( g_\perp^{\mu k } p_{q1'\perp}^\nu - g_\perp^{\nu k} p_{q1'\perp}^\mu \right) p_{\bar{q}2\perp}^i \nonumber \\
& \left.  + \,  x_g (2 x_q -1 ) p_{q1'\perp}^k (4 x_{\bar{q}} + x_g d -2) \left( g_\perp^{\nu i } p_{\bar{q}2\perp}^\mu -g_\perp^{\mu i } p_{\bar{q}2\perp}^\nu \right) \right\} + (q \leftrightarrow \bar{q}) \; .
\label{HatPhi3}
\end{align}
Here one has to fix $x_{\bar{q}} = 1 -x_q-x_g$ and $\vec{p}_q = \frac{x_q}{x_h} \vec{p}_h$. Eqs.~(\ref{HatPhi1}, \ref{HatPhi2}, \ref{HatPhi3}) contain the non-collinearly divergent terms of the corresponding $\tilde{\Phi}_3^{\alpha} \tilde{\Phi}_3^{\beta *}$ in Eqs.~(\ref{eq: div real impact factor}, \ref{Divergent_Part_impact_factor_TL}, \ref{Divergent_Part_impact_factor_TT}). In this case, just for simplicity of notation, we presented these contributions without renaming the two longitudinal fractions $x_q$ and $x_{\bar{q}}$ as done before (see text after Eq.~(\ref{eq:ShortHand})) when calculating the divergent contributions. \\

For clarity, the notation $x_g \sim 0$ in the second term of~\ref{Eq:FiniteRemainderOfSoftSub} indicates that, after extracting the singularity $1/x_g^2$ (i.e. after the rescaling $\vec{p}_g \rightarrow x_g \vec{p}_g$), throughout the remaining regular part, $x_g$ can be set to zero. Then the subtraction between the two terms will make the divergence of the type $1/x_g$ and this will be fully compensated by the factor $x_g$ in the numerator of the second line of the Eq.~(\ref{Eq:FiniteRemainderOfSoftSub}). \\

The corresponding anti-quark contribution is easily obtained by exchanging $x_q$ and $\vec{p}_{\bar{q}}$ with $x_{\bar{q}}$ and $\vec{p}_{q}$ in Eq.~(\ref{Eq:FiniteRemainderOfSoftSub}) and setting $x_q=1-x_{\bar{q}}-x_g$ and $\vec{p}_{\bar{q}} = \frac{x_{\bar{q}}}{x_h} \vec{p}_h$ in Eqs.~(\ref{HatPhi1}, \ref{HatPhi2}, \ref{HatPhi3}).

\subsubsection{Additional finite part of the dipole $\times$ dipole contribution}
\label{sec:RealDipxDip2}
It is important to note that the dipole $\times$ dipole contributions do not end with the diagrams shown in Fig.~\ref{fig:QuarkInANutshell}. Indeed, the dipole part of the impact factor can be expressed as
\begin{equation}
    \Phi_3^\alpha = \Tilde{\Phi}_3^\alpha + \Phi_4^\alpha |_{\vec{p}_3 = 0} ,
\end{equation}
where the second term corresponds to a contribution in which the gluon is emitted before the Shockwave, but passes through it without receiving any transverse kick ($\vec{p}_3=0$)\,. In considering the square of the impact factor we must also include all contributions involving $\Phi_4^{\alpha} |_{\vec{p}_3 = 0}$. \\

Hence, we have a second finite dipole $\times$ dipole contribution, which reads
\begin{align}
   &  \frac{d \sigma_{ J I}^{q \rightarrow h}}{d x_{h} d^2 p_{h \perp} } \bigg |_{\substack{\text{real, finite} \\ \text{dip. $\times$ dip.}}} \hspace{- 0.25 cm} = \frac{ \alpha_{em}}{(2 \pi)^8 N_c \, x_h^2} \sum_q Q_q^2 \int_{x_h}^{1} d x_q D_q^h \left( \frac{x_h}{x_q}, \mu_F \right) \; \int d^{2}\vec{p}_{1} \; d^2 \vec{p}_{2} \; d^{2} \vec{p}_{1^{\prime}} \; d^2 \vec{p}_{2^\prime} \nonumber \\ 
   & \times \int d^{2} \vec{p}_{\bar{q}}  \int \frac{d^{2} \vec{p}_{g}}{(2 \pi)^2} \int_0^{1-x_q} \frac{d x_g}{x_g} \frac{1}{ (1-x_q-x_g)} \delta \left( \frac{x_q}{x_h} \vec{p}_h - \vec{p}_{1} + \vec{p}_{\bar{q} 2} + \vec{p}_g  \right) \delta \left( \vec{p}_{1 1'} + \vec{p}_{2 2'} \right) \nonumber \\ 
   & \hspace{-0.2 cm} \times \mathbf{F}\left( \frac{\vec{p}_{12}}{2}\right)\mathbf{F}^\ast\left( \frac{\vec{p}_{1^\prime 2^\prime }}{2}\right) \varepsilon_{I \alpha } \varepsilon_{J \beta}^\ast \frac{\alpha_s C_F}{4 (p_{\gamma}^+)^2} \left[ \Phi_3^{\alpha} \left( \vec{p}_1, \vec{p}_2 \right) \Phi_3^{\beta \ast} \left( \vec{p}_{1'}, \vec{p}_{2'} \right) - \tilde{\Phi}_3^{\alpha} \left( \vec{p}_1, \vec{p}_2 \right) \tilde{\Phi}_3^{\beta \ast} \left( \vec{p}_{1'}, \vec{p}_{2'} \right) \right]  ,
    \label{Eq:QuarkFinDipxDip2}
\end{align}
where, one has to fix $x_{\bar{q}} = 1 -x_q-x_g$ and $\vec{p}_q = \frac{x_q}{x_h} \vec{p}_h$\footnote{In this case, in the term $\Phi_3^{\alpha} \Phi_3^{\beta \ast}$ in Eqs.~(\ref{eq: div real impact factor}, \ref{Divergent_Part_impact_factor_TL}, \ref{Divergent_Part_impact_factor_TT}) one has to use $x_q$ and $x_{\bar{q}}$ instead of $x_q'$ and $x_{\bar{q}}'$.}.

\subsubsection{Dipole $\times$ double dipole contribution}
\label{sec:RealDipxDouble}

\begin{figure}[h]
\begin{picture}(430,100)
\put(120,0){\includegraphics[scale=0.30]{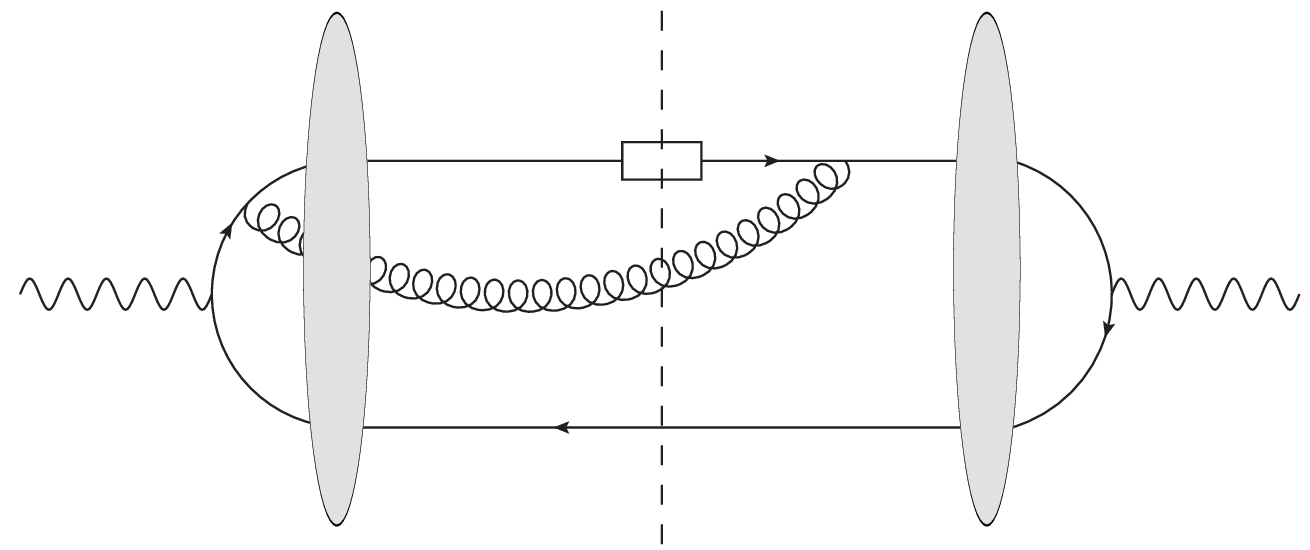}}
\end{picture} 
 \caption{An example of diagram contributing to the dipole $\times$ double dipole real part of the cross-section.}
\label{fig:Real_DipxDoubleDip}
\end{figure}

In the dipole $\times$ double dipole contribution, at cross section level, the gluon crosses at least once the shockwave\footnote{If it crosses twice, in one of the two cases it should not receive any transverse kick from the shockwave.} (see Fig.~\ref{fig:Real_DipxDoubleDip}). The dipole $\times$ double dipole contribution is
\begin{align}
   &  \frac{d \sigma_{ J I}^{q \rightarrow h}}{d x_{h} d^2  p_{h \perp} } \bigg |_{\text{real, dip. $\times$ d. dip.}} \hspace{- 0.25 cm} = \frac{ \alpha_{em}}{2 (2 \pi)^8 N_c \; x_h^2} \sum_q Q_q^2 \int_{x_h}^{1} d x_q D_q^h \left( \frac{x_h}{x_q}, \mu_F \right) \int d^{2}\vec{p}_{1} d^2 \vec{p}_{2}  d^{2} \vec{p}_{1^{\prime}} d^2 \vec{p}_{2^\prime} \nonumber \\ 
   &  \hspace{-0.15 cm} \times \int \hspace{-0.15 cm} d^{2} \vec{p}_{3} \; \frac{d^2 \vec{p}_{3^\prime}}{(2 \pi)^2} \int d^{2} \vec{p}_{\bar{q}}  \int \frac{d^{2} \vec{p}_{g}}{(2 \pi)^2} \int_0^{1-x_q} \frac{d x_g}{x_g} \frac{\delta \left( \vec{p}_{1 1'} + \vec{p}_{2 2'} + \vec{p}_{3 3'} \right)}{ (1-x_q-x_g)} \delta \left( \frac{x_q}{x_h} \vec{p}_h - \vec{p}_{1} + \vec{p}_{\bar{q} 2} + \vec{p}_{g3}  \right) \nonumber \\ 
   & \hspace{-0.2 cm} \times \varepsilon_{I \alpha } \varepsilon_{J \beta}^\ast \frac{\alpha_s }{4 (p_{\gamma}^+)^2} \left[ \Phi_3^\alpha(\vec{p}_{1}, \vec{p}_{2}) \Phi_4^{\beta \ast }(\vec{p}_{1^\prime}, \vec{p}_{2^\prime}, \vec{p}_{3^\prime}) \mathbf{F}\left(\frac{\vec{p}_{12}}{2}\right) \widetilde{\mathbf{F}}^{\ast} \left( \frac{\vec{p}_{1^\prime 2^\prime}}{2}, \vec{p}_{3^\prime} \right) \delta(\vec{p}_{3}) \right. \nonumber \\ 
   & +  \left. \Phi_4^\alpha (\vec{p}_{1}, \vec{p}_{2}, \vec{p}_{3}) \Phi_3^{\beta\ast} ( \vec{p}_{1'}, \vec{p}_{2'} ) \widetilde{\mathbf{F}}\left(\frac{\vec{p}_{12}}{2}, \vec{p}_{3}\right) \mathbf{F}^\ast\left(\frac{\vec{p}_{1^\prime 2^\prime}}{2} \right) \delta(\vec{p}_{3^\prime}) \right] \; .
    \label{Eq:QuarkFinDipxDouble}
\end{align}
The expressions for the interferences $\Phi_3^{\alpha} \Phi_4^{\beta \ast }, \Phi_4^\alpha \Phi_3^{\beta\ast}$, in the various cases, are given in appendix~\ref{sec: appendixC}. In those formulas one has to fix $x_{\bar{q}} = 1 -x_q-x_g$ and $\vec{p}_q = \frac{x_q}{x_h} \vec{p}_h$.

\subsubsection{Double dipole $\times$ double dipole contribution}
\label{sec:RealDoublexDouble}

\begin{figure}[h!]
\begin{picture}(430,100)
\put(120,0){\includegraphics[scale=0.30]{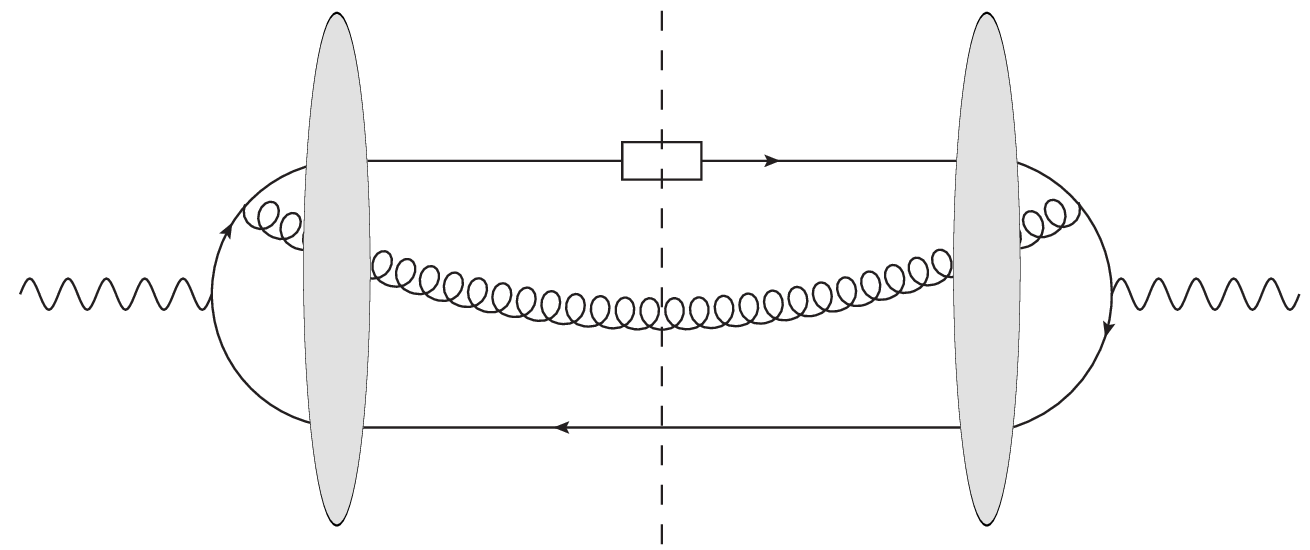}}
\end{picture} 
 \caption{An example of diagram contributing to the double dipole $\times$ double dipole real part of the cross-section.}
\label{fig:Real_DoubleDipxDoubleDip}
\end{figure}
In the double dipole $\times$ double dipole contribution, at cross section level, the gluon crosses twice the shockwave (see Fig.~\ref{fig:Real_DoubleDipxDoubleDip}). The double dipole $\times$ double dipole contribution is
\begin{align}
    & \frac{d \sigma_{ J I}^{q \rightarrow h}}{d x_{h} d^2 p_{h \perp} } \bigg |_{\substack{\text{real,} \\ \text{d. dip. $\times$ d. dip.}}} \hspace{- 0.35 cm} = \frac{ \alpha_{em}}{2 (2 \pi)^8 (N_c^2-1) \; x_h^2} \sum_q Q_q^2 \int_{x_h}^{1} \hspace{-0.2 cm} d x_q D_q^h \left( \frac{x_h}{x_q}, \mu_F \right) \hspace{-0.1 cm} \int  \hspace{-0.1 cm} d^{2}\vec{p}_{1} d^2 \vec{p}_{2}  d^{2} \vec{p}_{1^{\prime}} d^2 \vec{p}_{2^\prime} \nonumber \\
    & \hspace{-0.15 cm} \times  \int \hspace{-0.15 cm} \frac{d^{2} \vec{p}_{3}}{(2 \pi)^2} \; \frac{d^2 \vec{p}_{3^\prime}}{(2 \pi)^2} \int d^{2} \vec{p}_{\bar{q}}  \int \frac{d^{2} \vec{p}_{g}}{(2 \pi)^2} \int_0^{1-x_q} \hspace{-0.2 cm} d x_g \frac{\delta \left( \vec{p}_{1 1'} + \vec{p}_{2 2'} + \vec{p}_{3 3'} \right)}{x_g (1-x_q-x_g)} \delta \left( \frac{x_q}{x_h} \vec{p}_h - \vec{p}_{1} + \vec{p}_{\bar{q} 2} + \vec{p}_{g3}  \right) \nonumber \\ 
    & \hspace{-0.2 cm} \times \varepsilon_{I \alpha } \varepsilon_{J \beta}^\ast \frac{\alpha_s}{4 (p_{\gamma}^+)^2} \Phi_4^\alpha\left(\vec{p}_{1},\vec{p}_{2},\vec{p}_{3}\right) \Phi_4^{\beta \ast}(\vec{p}_{1^\prime}, \vec{p}_{2^\prime}, \vec{p}_{3^\prime}) \widetilde{\mathbf{F}}\left( \frac{\vec{p}_{12}}{2}, \vec{p}_{3} \right) \widetilde{\mathbf{F}}^\ast \left(\frac{\vec{p}_{1^\prime 2^\prime}}{2}, \vec{p}_{3^\prime} \right) \; .
    \label{Eq:QuarkFinDoublexDouble}
\end{align}
The expressions for the interferences $\Phi_4^{\alpha} \Phi_4^{\beta \ast }$, in the various cases, are given in appendix~\ref{sec: appendixC}. In those formulas one has to fix $x_{\bar{q}} = 1 -x_q-x_g$ and $\vec{p}_q = \frac{x_q}{x_h} \vec{p}_h$. \\

The corresponding anti-quark contributions, of subsections \ref{sec:RealDipxDip2}, \ref{sec:RealDipxDouble}, \ref{sec:RealDoublexDouble}, are easily obtained by exchanging $x_q$ and $\vec{p}_{\bar{q}}$ with $x_{\bar{q}}$ and $\vec{p}_{q}$ in Eqs.~(\ref{Eq:QuarkFinDipxDip2}, \ref{Eq:QuarkFinDipxDouble}, \ref{Eq:QuarkFinDoublexDouble}) and setting $x_q=1-x_{\bar{q}}$ and $\vec{p}_{\bar{q}} = \frac{x_{\bar{q}}}{x_h} \vec{p}_h$ in the various interferences of impact factors present in the formulas of appendix~\ref{sec: appendixC}.

\subsection{Fragmentation from gluon}
The finite contributions in the case of gluon fragmentation are obtained in a similar way to what is shown in the case of quark fragmentation. 

\subsubsection{Finite remainder in the $\tilde{\Phi}_3$ part}

\begin{figure}[h]
\begin{picture}(430,100)
\put(0,0){\includegraphics[scale=0.30]{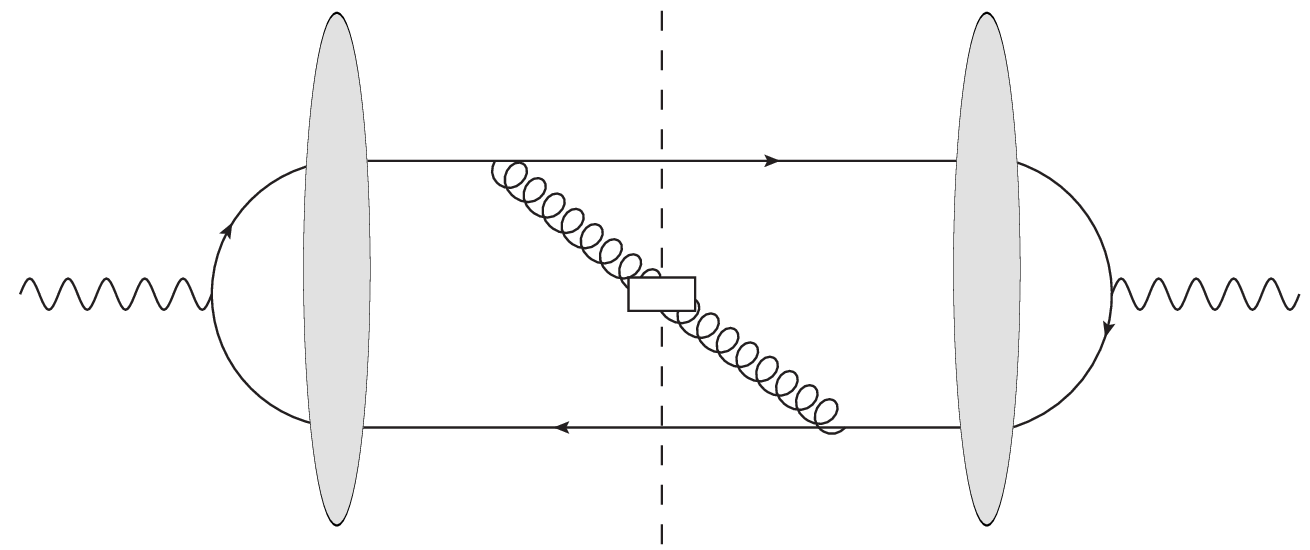}}
\put(212,34){$+$}
\put(240,0){\includegraphics[scale=0.30]{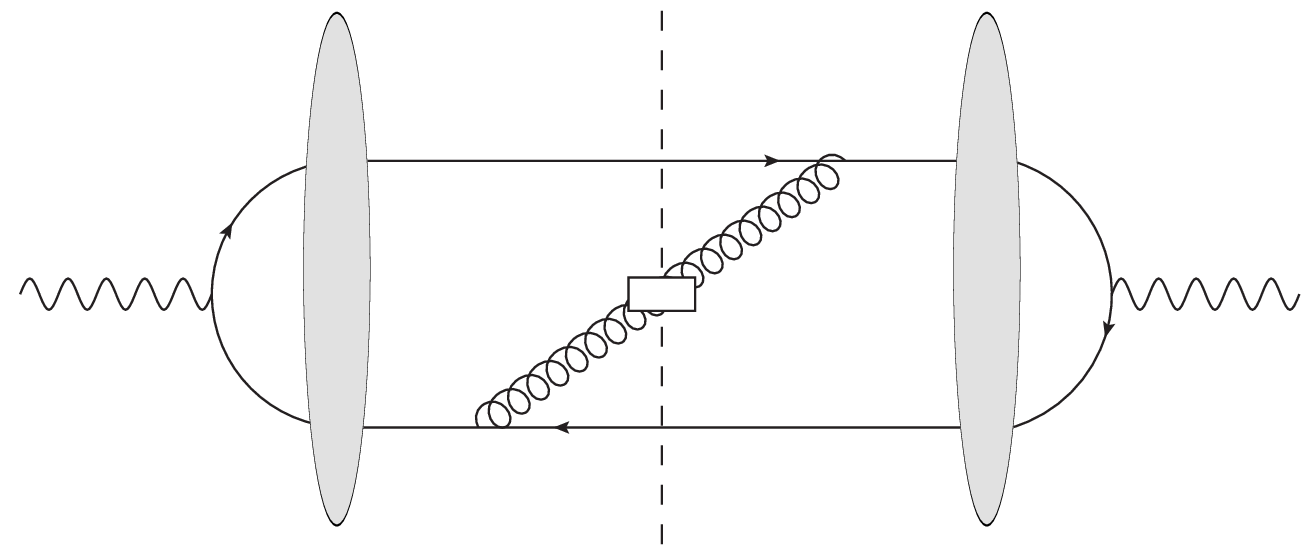}}
\end{picture} 
 \caption{The non IR-divergent contributions of the $\tilde{\Phi}_3$ part, in the case of gluon fragmentation.}
\label{fig:GluonFin_Contribution}
\end{figure}
First of all, we have to consider the two IR-finite diagrams in which the gluon does not cross the shockwave (see Fig. \ref{fig:GluonFin_Contribution}). This contribution reads
\begin{align}
    & \frac{d \sigma_{ J I}^{g \rightarrow h}}{d x_{h} d^2 p_{h \perp} } \bigg |_{\text{dip. $\times$ dip., 1}} \hspace{- 0.15 cm} = \frac{ \alpha_{em}}{(2 \pi)^8 N_c \; x_h^2} \sum_q Q_q^2 \int_{x_h}^{1} d x_g D_g^h \left( \frac{x_h}{x_g}, \mu_F \right) \int \; d^{2}\vec{p}_{1} \; d^2 \vec{p}_{2} \; d^{2} \vec{p}_{1^{\prime}} \; d^2 \vec{p}_{2^\prime} \nonumber \\ 
    & \times  \int d^{2} \vec{p}_{\bar{q}} \int \frac{d^{2} \vec{p}_{q}}{(2 \pi)^2} \int_0^{1-x_g} d x_q \frac{1}{x_q (1-x_q-x_g)} \mathbf{F}\left( \frac{\vec{p}_{12}}{2}\right)\mathbf{F}^\ast\left( \frac{\vec{p}_{1^\prime 2^\prime }}{2}\right) \varepsilon_{I \alpha } \varepsilon_{J \beta}^\ast \frac{\alpha_s C_F}{4 (p_{\gamma}^+)^2} \nonumber \\
    & \times \delta \left( \frac{x_g}{x_h} \vec{p}_h + \vec{p}_{q 1} + \vec{p}_{\bar{q} 2} \right) \delta \left( \vec{p}_{1 1'} + \vec{p}_{2 2'} \right) \hat{\Phi}_3^{\alpha} \left( \vec{p}_1, \vec{p}_2 \right) \hat{\Phi}_3^{\beta \ast} \left( \vec{p}_{1'}, \vec{p}_{2'} \right) \; .
    \label{Eq:FinRemDipxDipGluon}
\end{align}

\subsubsection{Additional finite part of the dipole $\times$ dipole contribution}

In the dipole $\times$ dipole contribution, we also have a contribution analogous to the one presented in the section~\ref{sec:RealDipxDip2}.
Therefore, the second finite dipole $\times$ dipole contribution is
\begin{align}
   &  \frac{d \sigma_{ J I}^{g \rightarrow h}}{d x_{h} d^2 p_{h \perp} } \bigg |_{\text{dip. $\times$ dip., 2}} \hspace{- 0.15 cm} = \frac{ \alpha_{em}}{(2 \pi)^8 N_c \; x_h^2} \sum_q Q_q^2 \int_{x_h}^{1} d x_g D_g^h \left( \frac{x_h}{x_g}, \mu_F \right) \int \; d^{2}\vec{p}_{1} \; d^2 \vec{p}_{2} \; d^{2} \vec{p}_{1^{\prime}} \; d^2 \vec{p}_{2^\prime} \nonumber \\ 
   & \times  \int d^{2} \vec{p}_{\bar{q}} \int \frac{d^{2} \vec{p}_{q}}{(2 \pi)^2} \int_0^{1-x_g} d x_q \frac{1}{x_q (1-x_q-x_g)} \mathbf{F}\left( \frac{\vec{p}_{12}}{2}\right)\mathbf{F}^\ast\left( \frac{\vec{p}_{1^\prime 2^\prime }}{2}\right) \varepsilon_{I \alpha } \varepsilon_{J \beta}^\ast \frac{\alpha_s C_F}{4 (p_{\gamma}^+)^2} \nonumber \\ & \times \delta \left( \frac{x_g}{x_h} \vec{p}_h + \vec{p}_{q 1} + \vec{p}_{\bar{q} 2} \right) \delta \left( \vec{p}_{1 1'} + \vec{p}_{2 2'} \right) \left[  \Phi_3^{\alpha} \left( \vec{p}_1, \vec{p}_2 \right) \Phi_3^{\beta \ast} \left( \vec{p}_{1'}, \vec{p}_{2'} \right) -  \tilde{\Phi}_3^{\alpha} \left( \vec{p}_1, \vec{p}_2 \right) \tilde{\Phi}_3^{\beta \ast} \left( \vec{p}_{1'}, \vec{p}_{2'} \right) \right] .
   \label{Eq:FinRemDipxDipGluon2}
\end{align}

\subsubsection{Dipole $\times$ double dipole contribution}

\begin{figure}[h]
\centering
 \includegraphics[scale=0.35]{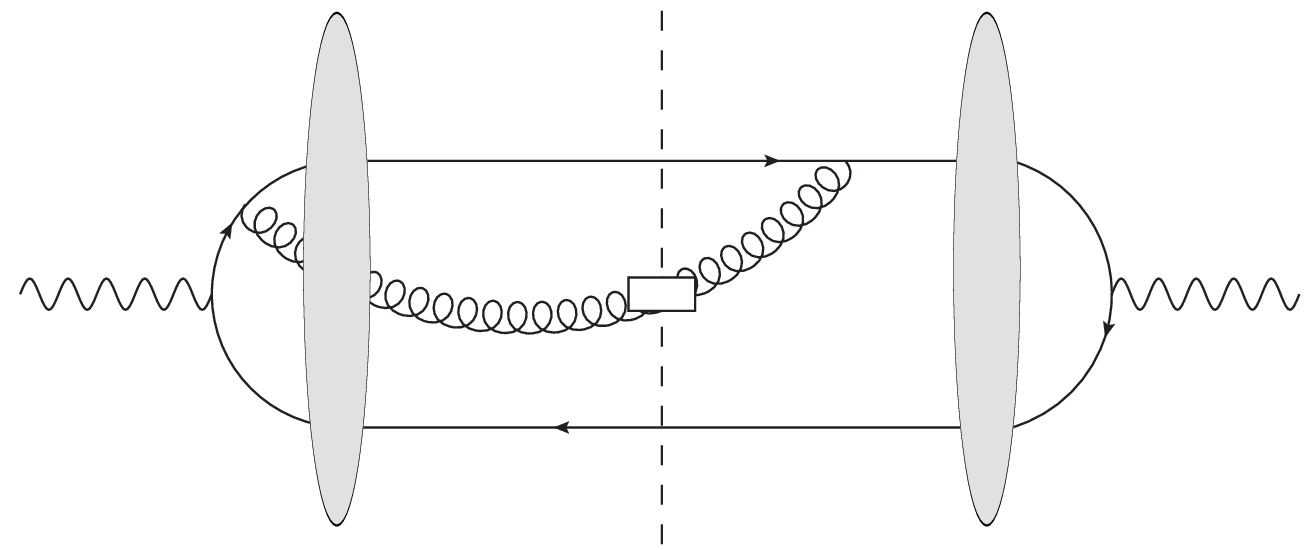}
 \caption{An example of diagram which contributes to the dipole $\times$ double dipole part of real contributions, in the gluon fragmentation case.}
    \label{fig:NLO_Real_DipxDouble_Dip}
\end{figure}
An example of dipole $\times$ double dipole contribution, in the gluon fragmentation case, is shown in Fig.~\ref{fig:NLO_Real_DipxDouble_Dip}. The complete dipole $\times$ double dipole contribution to the cross-section is
\begin{align}
    & \frac{d \sigma_{ J I}^{g \rightarrow h}}{d x_{h} d^2 p_{h \perp} } \bigg |_{\text{dip. $\times$ d. dip.}} \hspace{- 0.15 cm} = \frac{ \alpha_{em}}{2 (2 \pi)^8 N_c \; x_h^2} \sum_q Q_q^2 \int_{x_h}^{1} d x_g D_g^h \left( \frac{x_h}{x_g}, \mu_F \right) \int \; d^{2}\vec{p}_{1} \; d^2 \vec{p}_{2} \; d^{2} \vec{p}_{1^{\prime}} \; d^2 \vec{p}_{2^\prime} \nonumber \\
    & \times \int d^{2} \vec{p}_{3} \; d^2 \vec{p}_{3^\prime}  \int d^{2} \vec{p}_{\bar{q}} \int \frac{d^{2} \vec{p}_{q}}{(2 \pi)^2} \int_0^{1-x_g} \frac{d x_q}{x_q} \frac{\delta \left( \vec{p}_{1 1'} + \vec{p}_{2 2'} + \vec{p}_{3 3'} \right)}{(1-x_q-x_g)} \delta \left( \frac{x_g}{x_h} \vec{p}_h + \vec{p}_{q 1} + \vec{p}_{\bar{q} 2} - \vec{p}_{3}  \right) \nonumber \\ 
    & \hspace{-0.2 cm} \times \varepsilon_{I \alpha } \varepsilon_{J \beta}^\ast \frac{\alpha_s }{4 (p_{\gamma}^+)^2} \left[ \Phi_3^\alpha(\vec{p}_{1}, \vec{p}_{2}) \Phi_4^{\beta \ast }(\vec{p}_{1^\prime}, \vec{p}_{2^\prime}, \vec{p}_{3^\prime}) \mathbf{F}\left(\frac{\vec{p}_{12}}{2}\right) \widetilde{\mathbf{F}}^{\ast} \left( \frac{\vec{p}_{1^\prime 2^\prime}}{2}, \vec{p}_{3^\prime} \right) \delta(\vec{p}_{3}) \right. \nonumber \\
    & +  \left. \Phi_4^\alpha (\vec{p}_{1}, \vec{p}_{2}, \vec{p}_{3}) \Phi_3^{\beta\ast} ( \vec{p}_{1'}, \vec{p}_{2'} ) \widetilde{\mathbf{F}}\left(\frac{\vec{p}_{12}}{2}, \vec{p}_{3}\right) \mathbf{F}^\ast\left(\frac{\vec{p}_{1^\prime 2^\prime}}{2} \right) \delta(\vec{p}_{3^\prime}) \right] \; .
    \label{Eq:FinRemDipxDoubleGluon}
\end{align}

\subsubsection{Double dipole $\times$ double dipole contribution}

\begin{figure}[h]
\centering
 \includegraphics[scale=0.35]{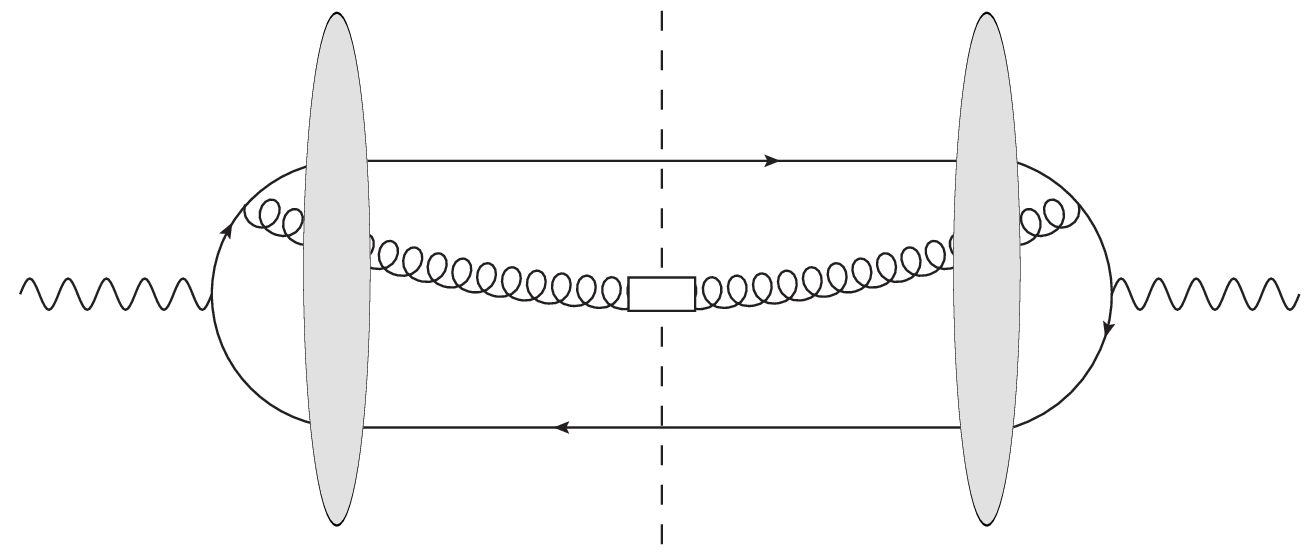}
 \caption{An example of diagram which contributes to the double dipole $\times$ double dipole part of real contributions, in the gluon fragmentation case.}
    \label{fig:NLO_Real_Double_DipxDouble_Dip}
\end{figure}

The last contribution to be taken into account is the double dipole $\times$ double dipole one (see Fig.~\ref{fig:NLO_Real_Double_DipxDouble_Dip}), which reads
\begin{align}
    & \frac{d \sigma_{ J I}^{g \rightarrow h}}{d x_{h} d^2 p_{h \perp} } \bigg |_{ \text{d. dip. $\times$ d. dip.}} \hspace{- 0.35 cm} = \frac{ \alpha_{em}}{2(2 \pi)^8 (N_c^2-1) \; x_h^2} \sum_q Q_q^2 \int_{x_h}^{1} \hspace{-0.2 cm} d x_g D_g^h \left( \frac{x_h}{x_g}, \mu_F \right) \hspace{-0.1 cm} \int  \hspace{-0.1 cm} d^{2} \vec{p}_{1} d^2 \vec{p}_{2}  d^{2} \vec{p}_{1^{\prime}} d^2 \vec{p}_{2^\prime} \nonumber \\ 
    & \times \hspace{-0.15 cm} \int \hspace{-0.15 cm} \frac{d^{2} \vec{p}_{3}}{(2 \pi)^2} \; \frac{d^2 \vec{p}_{3^\prime}}{(2 \pi)^2} \int d^{2} \vec{p}_{\bar{q}}  \int \frac{d^{2} \vec{p}_{q}}{(2 \pi)^2} \int_0^{1-x_g} \hspace{-0.2 cm} d x_q \frac{\delta \left( \vec{p}_{1 1'} + \vec{p}_{2 2'} + \vec{p}_{3 3'} \right)}{x_q (1-x_q-x_g)} \delta \left( \frac{x_g}{x_h} \vec{p}_h + \vec{p}_{q1} + \vec{p}_{\bar{q} 2} - \vec{p}_{3}  \right) \nonumber \\ 
    & \hspace{-0.2 cm} \times \varepsilon_{I \alpha } \varepsilon_{J \beta}^\ast \frac{\alpha_s}{4 (p_{\gamma}^+)^2} \Phi_4^\alpha\left(\vec{p}_{1},\vec{p}_{2},\vec{p}_{3}\right) \Phi_4^{\beta \ast}(\vec{p}_{1^\prime}, \vec{p}_{2^\prime}, \vec{p}_{3^\prime}) \widetilde{\mathbf{F}}\left( \frac{\vec{p}_{12}}{2}, \vec{p}_{3} \right) \widetilde{\mathbf{F}}^\ast \left(\frac{\vec{p}_{1^\prime 2^\prime}}{2}, \vec{p}_{3^\prime} \right) \; .
    \label{Eq:FinRemDoublexDoubleGluon}
\end{align}

\newpage

\section{Summary}
For the reader's convenience, in this section we give a brief summary in the form of a table of the finite terms with reference to the corresponding equations. \\ 

\begin{tabularx}{0.9\textwidth} { 
  | >{\raggedright\arraybackslash}X 
  | >{\centering\arraybackslash}X 
  | >{\raggedleft\arraybackslash}X | }
 \hline
 \textbf{Fragmenting parton} & \textbf{Contribution} & \textbf{Equation(s)} \\
 \hline
Quark$^{\dagger}$\footnote{For all contributions denoted with $^{\dagger}$, the corresponding contribution in which the quark is replaced by the anti-quark should be produced accordingly to the explanation in the main text.} & LO & Eq.~(\ref{eq:LL-LO}) \\
\hline
Quark or gluon$^{\dagger}$ & NLO FFs counterterm & Eq.~(\ref{eq:FFNLOCounterTerm})*\footnote{For all contributions denoted with *, the equation we are referring to also contains a singular piece that cancels out and must be ignored here. Moreover, in the finite parts,
the limit $d \rightarrow 2$ is to be taken.} \\
\hline
Quark$^{\dagger}$ & Finite part of the singular virtual dipole $\times$ dipole & Eq.~(\ref{eq:SVpart})*  \\
\hline
Quark$^{\dagger}$ & Non-singular virtual dipole $\times$ dipole & Eqs.~(\ref{eq:VirtDipXDip}, \ref{eq:VirtDipXDipTL}, \ref{eq:VirtDipXDipTT})$^{\#}$\footnote{For all contributions denoted with $^{\#}$, the three equations refers to $LL,TL,TT$ cases respectively.}  \\
\hline
Quark$^{\dagger}$ & Virtual double dipole $\times$ dipole & \hspace{-0.2 cm} Eqs.~(\ref{eq:VirtDoubleDipXDipLL}, \ref{eq:VirtDoubleDipXDipTL}, \ref{eq:VirtDoubleDipXDipTT})$^{\#}$  \\
\hline
Quark$^{\dagger}$ & Dipole $\times$ dipole real collinear: $q$-$g$ & Eqs.~(\ref{eq:CollConqgFin1}, \ref{eq:CollConqgFin2})  \\
\hline
Quark$^{\dagger}$ & Dipole $\times$ dipole real collinear: $\bar{q}$-$g$ & Eqs.~(\ref{eq:CollConqbargDiv}*, \ref{eq:CollConqbargFin2})   \\
\hline
Quark$^{\dagger}$ &  Real soft & Eq.~(\ref{eq:Softpart})*  \\
\hline
Gluon$^{\dagger}$ & Dipole $\times$ dipole real collinear: $q$-$g$ & Eq.~(\ref{eq:CollConqgFinGluon})  \\
\hline
Quark$^{\dagger}$ & Finite remainder of the soft subtraction & Eq.~(\ref{Eq:FiniteRemainderOfSoftSub})  \\
\hline
Quark$^{\dagger}$ & Additional finite part of the real dipole $\times$ dipole & Eq.~(\ref{Eq:QuarkFinDipxDip2})  \\
\hline
Quark$^{\dagger}$ & Real dipole $\times$ double dipole & Eq.~(\ref{Eq:QuarkFinDipxDouble})  \\
\hline
Quark$^{\dagger}$ & Real double dipole $\times$ double dipole & Eq.~(\ref{Eq:QuarkFinDoublexDouble})  \\
\hline
Gluon & First finite part of the dipole $\times$ dipole & Eq.~(\ref{Eq:FinRemDipxDipGluon})  \\
\hline
Gluon & Additional finite part of the dipole $\times$ dipole & Eq.~(\ref{Eq:FinRemDipxDipGluon2})  \\
\hline
Gluon & Dipole $\times$ double dipole & Eq.~(\ref{Eq:FinRemDipxDoubleGluon})  \\
\hline
Gluon & Double dipole $\times$ double dipole & Eq.~(\ref{Eq:FinRemDoublexDoubleGluon})  \\
\hline
\end{tabularx}

\section{Conclusion}

In the present work, we have continued our study of diffractive processes in the saturation framework relying on the shockwave approach. In particular, we computed the cross section for the diffractive production of a hadron, at large $p_T$, in $\gamma^{(*)}$ nucleon/nucleus scattering, in rather general kinematics, which includes both lepto- and photoproduction. Our main result is the explicit proof of cancellation of any kind of divergences and the extraction of the finite remainder. \\

Diffractive productions are important channel to investigate the gluon tomography in the nucleon (see \cite{Marquet:2009ca,Hatta:2022lzj,Iancu:2021rup}) and the achievement of an appropriate level of precision calls for a full NLL description. This new class of processes provides an access to precision physics of gluon saturation dynamics, with very promising future phenomenological studies both at the LHC in UPC (in photoproduction) and at the future EIC (both in photoproduction and leptoproduction). It adds a new piece in the list of processes which are very promising to probe gluonic saturation in nucleons and nuclei at NLO, which includes inclusive DIS~\cite{Beuf:2022ndu}, inclusive photoproduction of dijets~\cite{Altinoluk:2020qet,Taels:2022tza}, photon-dijet production in DIS~\cite{Roy:2019hwr}, dijets in DIS~\cite{Caucal:2021ent,Caucal:2022ulg,Caucal:2023fsf}, single hadron~\cite{Bergabo:2022zhe}  and dihadrons production in DIS~\cite{Bergabo:2022tcu,Iancu:2022gpw}, diffractive exclusive dijets~\cite{Boussarie:2014lxa,Boussarie:2016ogo,Boussarie:2019ero} and exclusive light meson production~\cite{Boussarie:2016bkq,Mantysaari:2022bsp}, exclusive quarkonium production~\cite{Mantysaari:2021ryb,Mantysaari:2022kdm}, inclusive DDIS~\cite{Beuf:2022kyp}, diffractive di-hadron production~\cite{Fucilla:2022wcg}, forward production of a Drell-Yan pair and a jet~\cite{Taels:2023czt}. 

\acknowledgments

E.~L. and S.~W. thank Charlotte Van Hulse and Ronan McNulty for early discussions which motivated the present work.
We thank Renaud Boussarie, Michel Fontannaz, Saad Nabeebaccus, Maxim A.~Nefedov, Alessandro Papa and Farid Salazar for many useful discussions.

This  project  has  received  funding  from  the  European  Union’s  Horizon  2020  research  and  innovation program under grant agreement STRONG–2020 (WP 13 "NA-Small-x").
The work by M.~F. is supported by
Agence Nationale de la Recherche under the contract ANR-17-CE31-0019.
The  work of  L.~S. is  supported  by  the  grant  2019/33/B/ST2/02588  of  the  National  Science Center  in  Poland. L.~S. thanks the P2IO Laboratory
of Excellence (Programme Investissements d'Avenir ANR-10-LABEX-0038) and the P2I - Graduate School of Physics of Paris-Saclay University for support.
This work was also partly supported by the French CNRS via the GDR QCD.

\appendix

\section{Virtual corrections to $\gamma^{*} \rightarrow q \bar{q}$ impact factor}
\label{App_Virtual_Corrections}
\subsection{One-loop computation}
\begin{figure}
\centering
 \includegraphics[scale=0.28]{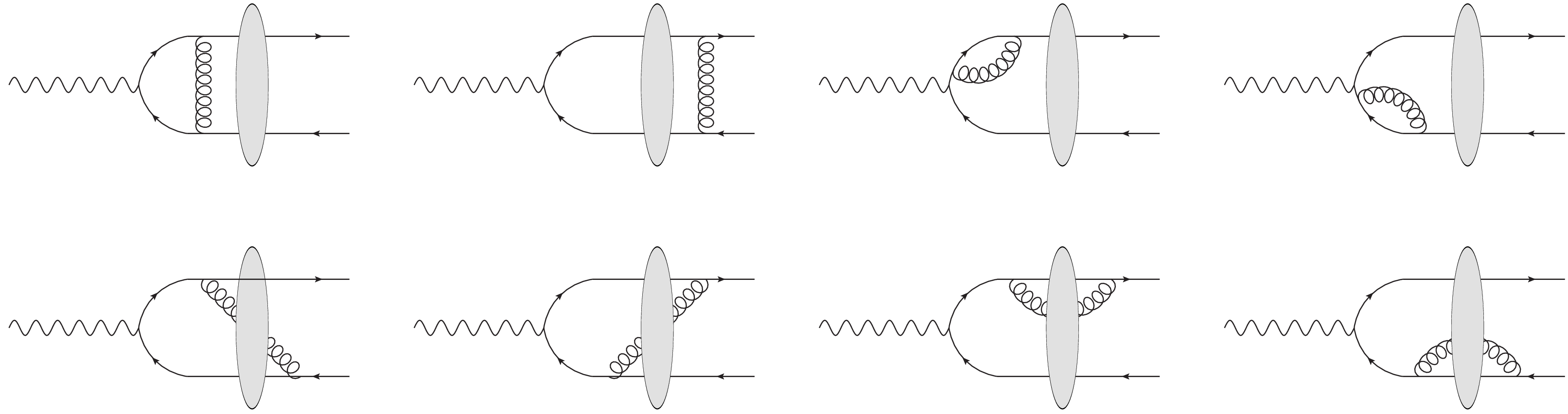}
 \caption{One loop diagrams for $\gamma^{*} \rightarrow q \bar{q}$.}
    \label{fig:OneLoopDipole}
\end{figure}

Eight one-loop diagrams represented in Fig.~\ref{fig:OneLoopDipole} contribute to the virtual reduced matrix element. Diagrams are of two kinds as the loop gluon may or may not cross the shockwave. In both cases, the singlet projector is the same as for the LO case $ \displaystyle \frac{\delta_{ln}}{\sqrt{N_c}}$. Diagrams of the same kind have the same colour factors and Wilson line operators. For diagrams where the gluon does not cross the shockwave (top four diagrams in Fig.~\ref{fig:OneLoopDipole}), the factor containing color factors and Wilson line operators is
\begin{equation}
     C_F N_c \; \mathcal{U}_{12} \; ,
\end{equation}
where $\mathcal{U}_{12}$ is the dipole operator in coordinate space and $C_F$ the Casimir of the fundamental representation. For diagrams where the gluon crosses the shockwave, we have
\begin{equation}
     - \frac{N_c^2}{2} (\mathcal{U}_{13} + \mathcal{U}_{32} -\mathcal{U}_{12} -\mathcal{U}_{13} \mathcal{U}_{32} ) - C_F N_c \; \mathcal{U}_{12} \; ,
\end{equation}
where $\mathcal{U}_{12}\mathcal{U}_{32} $ is the double dipole operator in coordinate space.
From those colours factors, the virtual reduced matrix element can be split into two terms
associated to the dipole or double dipole operator:
\begin{gather}
     T_1^\alpha =-\alpha_s \frac{N_c \Gamma(1-\epsilon)}{(4 \pi)^{1+\epsilon}} \int d^d \vec{p}_1 d^d \vec{p}_2\left\{\delta\left(\vec{p}_{q 1}+\vec{p}_{\bar{q} 2}\right)\left(\frac{N_c^2-1}{N_c}\right) \tilde{\mathcal{U}}_{12}\left(\vec{p}_1, \vec{p}_2\right) \Phi_1^\alpha\left(\vec{p}_1, \vec{p}_2\right)\right. \nonumber \\ \left.+N_c \int \frac{d^d \vec{p}_3}{(2 \pi)^d} \delta\left(\vec{p}_{q 1}+\vec{p}_{\bar{q} 2}-\vec{p}_3\right)\left[\tilde{\mathcal{U}}_{13}+\tilde{\mathcal{U}}_{32}-\tilde{\mathcal{U}}_{12}-\widetilde{\mathcal{U}_{13} \mathcal{U}_{32}}\right]\left(\vec{p}_1, \vec{p}_2, \vec{p}_3\right) \Phi_2^\alpha\left(\vec{p}_1, \vec{p}_2, \vec{p}_3\right)\right\} \; ,
     \label{Eq:App_TransMatrEl}
\end{gather}
where $\tilde{\mathcal{U}}_{12}$ and $\widetilde{\mathcal{U}_{13} \mathcal{U}_{32}}$ denote the Fourier transforms of  $\mathcal{U}_{12}$ and $\mathcal{U}_{13} \mathcal{U}_{32}$. The dipole impact factor has contributions from both types of diagrams, while the double dipole impact factor only gets contributions from diagrams where the gluon crosses the shockwave. \\

Five of the diagrams in Fig.~\ref{fig:OneLoopDipole} are independent, the computational steps are the same for each of them:
\begin{itemize}
    \item[\textbullet] Write the S-matrix element of the diagram as an integral over the coordinate of every vertex. Replace the building blocks by their expressions in mixed space representation, as a function of $x^+$ and $p^+$, $\vec{p}$ \; ; e.g., for a free quark propagator,
     \begin{gather}
     G_0(x) =\int \frac{d p^{+} d^d \vec{p}}{2 p^{+}(2 \pi)^{d+1}} \exp \left[-i x^{+} \frac{\vec{p}^{\; 2}-i \varepsilon}{2 p^{+}}-i p^{+} x^{-}+i \vec{p} \cdot \vec{x}\right] \nonumber \\  \times\left\{\left(p^{+} \gamma^{-}+\frac{\vec{p}^{\; 2}}{2 p^{+}} \gamma^{+}+p_{\perp}\right)\left(\theta\left(p^{+}\right) \theta\left(x^{+}\right)-\theta\left(-p^{+}\right) \theta\left(-x^{+}\right)\right)+i \delta\left(x^{+}\right) \gamma^{+}\right\}.
     \end{gather}
     \item[\textbullet] Integrate over the $-$ and transverse components of the vertex coordinates. This gives the explicit conservation of the longitudinal momenta, characteristic of the eikonal interactions. It also gives a delta function of the transverse momentum components, including the momentum kick $ \vec{p}_1, \vec{p}_2, \vec{p}_3$ from the $t$-channel shockwave to the $q, \bar{q}, g$ lines.
     \item[\textbullet] With those delta functions, the integrals over the momenta of the intermediate particles can be trivially done, apart from one momentum, which is always chosen to be the momentum of the loop gluon.
     \item[\textbullet] One then integrates over the $+$ component of the vertex coordinates, leading to an expression with only an integral over $l^+$ and $\vec{l}$. 
     \item[\textbullet] The last step is the integration over $\vec{l}$ and $l^+$. Integration over $\vec{l}$ is regularized by dimensional regularization and done using the Schwinger, or equivalently, the Feynman parametrization. The longitudinal $z=l/p_{\gamma}^+$ integral is regularized by an infrared cutoff $\alpha$. To extract the divergent part, the following $+$ prescription has been introduced, defined here as 
     \begin{equation}
         \int_{\alpha}^{z_0} dz \; \phi (z) = \int_{\alpha}^{z_0} dz \; \phi_0 (z) + \int_{\alpha}^{z_0} dz \left [ \phi (z) - \phi_0 (z) \right ] = \int_{\alpha}^{z_0} dz \; \phi_0 (z) + \int_{\alpha}^{z_0} dz \left [ \phi (z) \right ]_+ \; ,
     \end{equation}
     where $\phi_0 (z)$ is the divergent part of $\phi (z)$ when $z \rightarrow 0$. \vspace{0.5 cm}
\end{itemize}
\textbf{IR-singular part of the dipole contribution}: The divergences of soft and collinear nature are entirely contained in the dipole part of the impact factor, that can be written as
\begin{equation}
    \Phi_1^{\alpha} (\vec{p}_1, \vec{p}_2) = \frac{S_V}{2} \Phi_0^{\alpha} (\vec{p}_1, \vec{p}_2) + \Phi_{1R}^{\alpha} (\vec{p}_1, \vec{p}_2) \; ,
\end{equation}
where $\Phi_0^{\alpha} (\vec{p}_1, \vec{p}_2)$ is the leading order impact factor (Eqs.~(3.7) and (3.8) of Ref.~\cite{Boussarie:2016ogo}), $\Phi_{1R}^{\alpha} (\vec{p}_1, \vec{p}_2)$ the regular part of the one-loop dipole contribution (Eqs.~(3.43) and (3.44) of Ref.~\cite{Boussarie:2016ogo}) and 
\begin{equation}
   \frac{S_V}{2}=\left[\ln \left(\frac{x \bar{x}}{\alpha^2}\right)-\frac{3}{2}\right]\left[\ln \left(\frac{x \bar{x} \mu^2}{\left(x \vec{p}_{\bar{q}}-\bar{x} \vec{p}_q\right)^2}\right)-\frac{1}{\epsilon}\right]+i \pi \ln \left(\frac{x \bar{x}}{\alpha^2}\right)+\frac{1}{2} \ln ^2\left(\frac{x \bar{x}}{\alpha^2}\right)-\frac{\pi^2}{6}+3 \; . \vspace{0.2 cm}
\end{equation}

\subsection{B-JIMWLK dipole evolution} The rapidity divergences are fully contained into the double dipole part of the impact factor, $\Phi_2^{\alpha}$. These rapidity singular contributions have been extracted (in the form of $\ln \alpha$) and isolated in Eqs.~(3.18), (3.19), (3.20) and (3.21) of Ref~\cite{Boussarie:2016ogo}. They have to be re-absorbed into the renormalized Wilson operators with the help of BK equation. Using Eq.~(\ref{Eq:BKnonExp}), we get
\begin{gather} \tilde{\mathcal{U}}_{12}^\alpha =\tilde{\mathcal{U}}_{12}^\eta-\int_\alpha^{e^\eta} \frac{d \rho}{\rho} 2 \alpha_s N_c\left(\mu^2\right)^{1-\frac{d}{2}} \int \frac{d^d \vec{q}_1 d^d \vec{q}_2 d^d \vec{q}_3}{(2 \pi)^{2 d}} \delta\left(\vec{p}_1+\vec{p}_2-\vec{q}_1-\vec{q}_2-\vec{q}_3\right) \nonumber \\  \times\left[\tilde{\mathcal{U}}_{13}^\eta+\tilde{\mathcal{U}}_{32}^\eta-\tilde{\mathcal{U}}_{12}^\eta-\widetilde{\mathcal{U}_{13}^\eta \mathcal{U}_{32}^\eta}\right]\left(\vec{q}_1, \vec{q}_2, \vec{q}_3\right) \nonumber \\ \times\left\{\frac{2\left(\vec{p}_1-\vec{q}_1\right) \cdot\left(\vec{p}_2-\vec{q}_2\right)}{\left(\vec{p}_1-\vec{q}_1\right)^2\left(\vec{p}_2-\vec{q}_2\right)^2}+\frac{\pi^{\frac{d}{2}} \Gamma\left(1-\frac{d}{2}\right) \Gamma^2\left(\frac{d}{2}\right)}{\Gamma(d-1)}\left(\frac{\delta\left(\vec{p}_1-\vec{q}_1\right)}{\left[\left(\vec{p}_2-\vec{q}_2\right)^2\right]^{1-\frac{d}{2}}}+\frac{\delta\left(\vec{p}_2-\vec{q}_2\right)}{\left[\left(\vec{p}_1-\vec{q}_1\right)^2\right]^{1-\frac{d}{2}}}\right)\right\} .
\end{gather}
Using the above equation in the leading order impact factor to pass from the cutoff $\alpha$ to the rapidity divide $e^{\eta}$, we produce the next-to-leading order term 
\begin{gather}
    T^{\alpha}_{{\rm{BK}}} = -\alpha_s N_c^2 \frac{\Gamma\left(2-\frac{d}{2}\right)}{(4 \pi)^{1+\epsilon}} \int \frac{d^d \vec{p}_1 d^d \vec{p}_2 d^d \vec{p}_3}{(2 \pi)^d} \delta\left(\vec{p}_{q 1}+\vec{p}_{\bar{q} 2}-\vec{p}_3\right) \nonumber \\  \times \left[\tilde{\mathcal{U}}_{13}^\eta+\tilde{\mathcal{U}}_{32}^\eta-\tilde{\mathcal{U}}_{12}^\eta-\widetilde{\mathcal{U}_{13}^\eta \mathcal{U}_{32}^\eta}\right]\left(\vec{p}_1, \vec{p}_2, \vec{p}_3\right) \Phi^{\alpha}_{BK} \left(\vec{p}_1, \vec{p}_2, \vec{p}_3\right) \; ,
    \label{Eq:App_BKTransMatrEl}
\end{gather}
where
\begin{gather}
  \Phi^{\alpha}_{{\rm{BK}}} =  2 \ln \left(\frac{e^\eta}{\alpha}\right) \left\{ \frac{2   \mu^{-2\epsilon} (4 \pi)^{1+\epsilon}  }{\Gamma\left(2-\frac{d}{2}\right)} \int \frac{d^d \vec{p}}{(2 \pi)^d} \frac{\vec{p} \cdot\left(\vec{p}-\vec{p}_3\right)}{\vec{p}^{\; 2}\left(\vec{p}_3-\vec{p}\right)^2} \Phi_0^\alpha\left(\vec{p}+\vec{p}_1, \vec{p}_2+\vec{p}_3-\vec{p}\right)\right. \nonumber \\  \left.+ \left( \frac{1}{\epsilon} + \ln \left(\frac{\vec{p}_3^{\; 2}}{\mu^2}\right) \right) \left(\Phi_0^\alpha\left(\vec{p}_1, \vec{p}_2+\vec{p}_3\right)+\Phi_0^\alpha\left(\vec{p}_3+\vec{p}_1, \vec{p}_2\right) \right)\right\} \; .
\end{gather}
$\Phi^{\alpha}_{{\rm{BK}}}$ can be computed explicitly; the final expressions can be found in Eqs.~(3.25) and (3.26) of Ref.~\cite{Boussarie:2016ogo}. Comparing Eq.~(\ref{Eq:App_TransMatrEl}) and Eq.~(\ref{Eq:App_BKTransMatrEl}), we immediately realize that we can add $\Phi^{\alpha}_{{\rm{BK}}}$ to the double dipole contribution to get
\begin{equation}
    \Phi^{' \alpha}_{2} = \Phi^{\alpha}_{2} + \Phi^{\alpha}_{{\rm{BK}}} \; , 
\end{equation}
which is the rapidity-divergence free double-dipole contribution to the impact factor\footnote{There is a small subtlety, adding $\Phi^{\alpha}_{BK}$ to $\Phi^{\alpha}_{2}$, it seems that some singular $\displaystyle \frac{1}{\epsilon}$ pieces survive. These are artificial UV poles originate from the fact that when one transforms the Wilson line operator into its momentum space representation straightforwardly, the property of vanishing
when $r_3 = r_2$ or $r_3 = r_1$ is not taken into account. This property reveals in the convolution of the impact factor killing all the artificial singularities.}. The final expressions of $\Phi^{' + }_{2}$ and $\Phi^{' i}_{2}$ can be found in Eqs. (3.31) and (3.32) of Ref.~\cite{Boussarie:2016ogo}.

\section{LO impact factor squared }
\label{LO impact factor squared}
The impact factors in the LL, TL and TT cases are respectively given by
\begin{align}
\sum_{\lambda_q,\lambda_{\bar{q}}} \Phi_0^+(\vec{p}_1,\vec{p}_2)\Phi_0^{+*}(\vec{p}_{1'},\vec{p}_{2'}) = \frac{32 (p_\gamma^+)^4 x_q^3 x_{\bar{q}}^3 }{\left(\vec{p}_{q1}^2 + x_q x_{\bar{q}} Q^2 \right) \left(\vec{p}_{q1'}^2 + x_q x_{\bar{q}} Q^2 \right)} \; ,
\label{eq:LeadingCS_LL}
\end{align}

\begin{align}
\sum_{\lambda_q,\lambda_{\bar{q}}} \Phi_0^+(\vec{p}_1,\vec{p}_2) \Phi_0^{i*}(\vec{p}_{1'},\vec{p}_{2'}) = \frac{16 (p_\gamma^+)^3 x_q^2 x_{\bar{q}}^2 p_{q1\perp}^i(1-2x_q)}{\left(\vec{p}_{q1}^2 + x_q x_{\bar{q}} Q^2 \right) \left(\vec{p}_{q1'}^2 + x_q x_{\bar{q}} Q^2 \right)} \; ,
\label{eq:LeadingCS_TL}
\end{align}

\begin{align}
 \sum_{\lambda_q,\lambda_{\bar{q}}} \Phi_0^{i} (\vec{p}_1,\vec{p}_2)\Phi_0^{k*}(\vec{p}_{1'},\vec{p}_{2'}) = \frac{8 (p_\gamma^+)^2 x_q x_{\bar{q}} \left[(1-2x_q)^2 g_\perp^{ri} g_\perp^{lk} - g_\perp^{rk} g_\perp^{li}+ g_\perp^{rl} g_\perp^{rl}g_\perp^{ik}\right] p_{q1\perp r} p_{q1'\perp l}}{\left(\vec{p}_{q1}^2 + x_q x_{\bar{q}} Q^2 \right) \left(\vec{p}_{q1'}^2 + x_q x_{\bar{q}} Q^2 \right)} \; .
 \label{eq:LeadingCS_TT}
\end{align}
The LT case is immediately obtained from TL by complex conjugation and $1,2 \leftrightarrow 1',2'$ substitution.

\section{Finite parts of virtual corrections}
\label{AppendixB}

\subsection{Building blocks integrals}
\label{sec:building_block}

\begin{eqnarray}
I_{1}^{k}(\vec{q}_1,\, \vec{q}_2,\, \Delta_1,\, \Delta_2) & \equiv & \frac{1}{\pi}\int\frac{d^{d}\vec{l}\left(l_{\perp}^{k}\right)}{\left[(\vec{l}-\vec{q}_{1})^{2}+\Delta_{1}\right]\left[(\vec{l}-\vec{q}_{2})^{2}+\Delta_{2}\right]\vec{l}^{^{\, \, 2}}} \label{I1k}, \\
I_2(\vec{q}_1,\, \vec{q}_2,\, \Delta_1,\, \Delta_2) & \equiv & \frac{1}{\pi}\int \frac{d^d \vec{l}}{\left[ (\vec{l}-\vec{q}_1)^2+\Delta_1 \right] \left[ (\vec{l}-\vec{q}_2)^2 +\Delta_2 \right]} \label{I2}, \\
I_3^k(\vec{q}_1,\, \vec{q}_2,\, \Delta_1,\, \Delta_2) & \equiv & \frac{1}{\pi}\int \frac{d^d \vec{l}\left( l_\bot^k \right)}{\left[ (\vec{l}-\vec{q}_1)^2+\Delta_1 \right] \left[ (\vec{l}-\vec{q}_2)^2 +\Delta_2 \right]} \label{I3k}, \\
I^{jk}(\vec{q}_1,\, \vec{q}_2,\, \Delta_1,\, \Delta_2) & \equiv & \frac{1}{\pi}\int\frac{d^{d}\vec{l}\left( l_{\perp}^j l_{\perp}^{k}\right)}{\left[(\vec{l}-\vec{q}_{1})^{2}+\Delta_{1}\right]\left[(\vec{l}-\vec{q}_{2})^{2}+\Delta_{2}\right]\vec{l}^{^{\, \, 2}}} \label{Ijk} \, .
\end{eqnarray}
The arguments of these integrals will be different for each diagram so we will write them explicitly before giving the expression of each diagram, but we will omit them in the equations for the reader's convenience. \\
Explicit results for the first 3 integrals in (\ref{I1k}-\ref{Ijk}) are obtained by a straightforward Feynman parameter integration. We will express them using the following variables :

\begin{eqnarray}
\rho_{1} & \equiv & \frac{\left(\vec{q}_{12}^{\, \, 2}+\Delta_{12}\right)-\sqrt{\left(\vec{q}_{12}^{\, \, 2}+\Delta_{12}\right)^{2}+4\vec{q}_{12}^{\, \, 2}\Delta_{2}}}{2\vec{q}_{12}^{\, \, 2}} ,\\
\rho_{2} & \equiv & \frac{\left(\vec{q}_{12}^{\, \, 2}+\Delta_{12}\right)+\sqrt{\left(\vec{q}_{12}^{\, \, 2}+\Delta_{12}\right)^{2}+4\vec{q}_{12}^{\, \, 2}\Delta_{2}}}{2\vec{q}_{12}^{\, \, 2}} \, , \label{rhovar}
\end{eqnarray}
where $\Delta_{ij} = \Delta_i - \Delta_j$ . \\
One gets :

\begin{eqnarray}
I_{1}^{k} & = & \frac{q_{1\perp}^{k}}{2\left[\vec{q}_{12}^{\, \, 2}\left(\vec{q}_{1}^{\, \, 2}+\Delta_{1}\right)\left(\vec{q}_{2}^{\, \, 2}+\Delta_{2}\right)-\left(\vec{q}_{1}^{\, \, 2}-\vec{q}_{2}^{\, \, 2}+\Delta_{12}\right)\left(\vec{q}_{1}^{\, \, 2}\Delta_{2}-\vec{q}_{2}^{\, \, 2}\Delta_{1}\right)\right]}\\ \nonumber
 & \times & \left\{ \frac{\left(\vec{q}_{2}^{\, \, 2}+\Delta_{2}\right)\vec{q}_{12}^{\, \, 2}+\vec{q}_{2}^{\, \, 2}\left(\Delta_{1}+\Delta_{2}\right)+\Delta_{2}\left(\Delta_{21}-2\vec{q}_{1}^{\, \, 2}\right)}{\left(\rho_{1}-\rho_{2}\right)\vec{q}_{12}^{\, \, 2}}\ln\left[\left(\frac{-\rho_{1}}{1-\rho_{1}}\right)\left(\frac{1-\rho_{2}}{-\rho_{2}}\right)\right]\right.\\ \nonumber
 & \times & \left.\left(\vec{q}_{2}^{\, \, 2}+\Delta_{2}\right)\ln\left[\frac{\Delta_{2}\left(\vec{q}_{1}^{\, \, 2}+\Delta_{1}\right)^{2}}{\Delta_{1}\left(\vec{q}_{2}^{\, \, 2}+\Delta_{2}\right)^{2}}\right]+\left(1\leftrightarrow2\right)\right\} \, ,
\end{eqnarray}

\begin{eqnarray}
I_{2} & = & \frac{1}{\vec{q}_{12}^{\, \, 2}\left(\rho_{1}-\rho_{2}\right)}\ln\left[\left(\frac{-\rho_{1}}{1-\rho_{1}}\right)\left(\frac{1-\rho_{2}}{-\rho_{2}}\right)\right] \, ,
\end{eqnarray}
and

\begin{eqnarray}
I_{3}^{k} & = & \frac{\left(\vec{q}_{12}^{\, \, 2}+\Delta_{12}\right)q_{1}^{k}+\left(\vec{q}_{21}^{\, \, 2}+\Delta_{21}\right)q_{2}^{k}}{2\left(\rho_{1}-\rho_{2}\right)(\vec{q}_{12}^{\, \, 2})^2}\ln\left[\left(\frac{-\rho_{1}}{1-\rho_{1}}\right)\left(\frac{1-\rho_{2}}{-\rho_{2}}\right)\right] \nonumber \\ &-& \frac{q_{12}^{k}}{2\vec{q}_{12}^{\, \, 2}}\ln\left(\frac{\Delta_{1}}{\Delta_{2}}\right) \, .
\end{eqnarray}
Please note that in some cases the real part of $\Delta_1$ or $\Delta_2$ will be negative so the previous results can acquire an imaginary part from the imaginary part $\pm \, i0$ of the arguments. \\ 
The last integral in (\ref{Ijk}) can be expressed in terms of the other ones by writing 
\begin{equation}
I^{jk} = I_{11}\left(q_{1\perp}^{j}q_{1\perp}^{k}\right)+I_{12}\left(q_{1\perp}^{j}q_{2\perp}^{k}+q_{2\perp}^{j}q_{1\perp}^{k}\right)+I_{22}\left(q_{2\perp}^{j}q_{2\perp}^{k}\right) \, ,
\end{equation}
with
\begin{align}
I_{11} & = -\frac{1}{2}\frac{\left[\vec{q}_{2}^{\, \, 2}q_{1\perp k}-\left(\vec{q}_{1}\cdot\vec{q}_{2}\right)q_{2\perp k}\right]}{\left[\vec{q}_{1}^{\, \, 2}\vec{q}_{2}^{\, \, 2}-\left(\vec{q}_{1} \cdot \vec{q}_{2}\right)^{2}\right]^{2}} \\ \nonumber
& \hspace{-0.25 cm} \times  \left[\left(\frac{\vec{q}_{1}\cdot\vec{q}_{2}}{\vec{q}_{1}^{\, \, 2}}\right)\ln\left(\frac{\vec{q}_{1}^{\, \, 2}+\Delta_{1}}{\Delta_{1}}\right)q_{1\perp}^{k}+\left(\vec{q}_{2}\cdot\vec{q}_{21}\right)I_{3}^{k}+\left\{ \vec{q}_{2}^{\, \, 2}\left(\vec{q}_{1}\cdot\vec{q}_{12}\right)+\Delta_{1}\vec{q}_{2}^{\, \, 2}-\Delta_{2}\left(\vec{q}_{1}\cdot\vec{q}_{2}\right)\right\} I_{1}^{k}\right]\\
I_{12} & = \frac{-1}{4\left[\vec{q}_{1}^{\, \, 2} \vec{q}_{2}^{\, \, 2} -\left(\vec{q}_1 \cdot \vec{q}_2\right)^2\right]} \ln \left(\frac{\vec{q}_{1}^{\, \, 2}+\Delta_1}{\Delta_1}\right)  \nonumber \\
&+ \frac{\vec{q}_{2}^{\, \, 2} \left(\vec{q}_1 \cdot \vec{q}_2\right)}{2\left[\vec{q}_{1}^{\, \, 2} \vec{q}_{2}^{\, \, 2} -\left(\vec{q}_1 \cdot \vec{q}_2\right)^2\right]^2}\left[\left(\vec{q}_{1}^{\, \, 2} +\Delta_1\right)\left(q_{1 \perp k} I_1^k\right)+\left(q_{1 \perp k} I_3^k\right)\right]  \nonumber\\
& -\frac{\left(\vec{q}_{1}^{\, \, 2} \vec{q}_{2}^{\, \, 2} \right)+\left(\vec{q}_1 \cdot \vec{q}_2\right)^2}{4\left[\vec{q}_{1}^{\, \, 2} \vec{q}_{2}^{\, \, 2} -\left(\vec{q}_1 \cdot \vec{q}_2\right)^2\right]^2}\left[\left(\vec{q}_{2}^{\, \, 2} +\Delta_2\right)\left(q_{1 \perp k} I_1^k\right)+\left(q_{1 \perp k} I_3^k\right)\right]+(1 \leftrightarrow 2), \\
I_{22}&  = I_{11}|_{1 \leftrightarrow 2} \, .
\end{align}

In what follows, for the $\phi$ function, $x=x_q$, $\bar{x} = x_{\bar{q}}$.

\subsection{$\phi_4$}

The arguments in the integrals of \ref{sec:building_block} are 
\begin{eqnarray*}
\vec{q}_{1} & = & \vec{p}_{1}-\left(\frac{x-z}{x}\right)\vec{p}_{q}, \quad \, \, \, \, \,
\vec{q}_{2}  =  \left(\frac{x-z}{x}\right)\left(x\vec{p}_{\bar{q}}-\bar{x}\vec{p}_{q}\right) \, ,\\
\Delta_{1} & = & \left(x-z\right)\left(\bar{x}+z\right)Q^{2}, \quad
\Delta_{2}  =  -\frac{x\left(\bar{x}+z\right)}{\bar{x}\left(x-z\right)}\vec{q}^{2}-i0\,.
\end{eqnarray*}
Let us write the impact factors in terms of these variables. \\They read: \vspace{0.2 cm} \\
(longitudinal NLO) $\times$ (longitudinal LO) contribution :
\begin{equation}
\left(\phi_{4}\right)_{LL}=-\frac{4(x-z)(\bar{x}+z)}{z}[-\bar{x}(x-z)(z+1)I_{2}+q_{2\bot k}(2x^{2}-(2x-z)(z+1))I_{1}^{k}] \, ,
\end{equation}
(longitudinal NLO) $\times$ (transverse LO) contribution :
\begin{equation}
\left(\phi_{4}\right)_{LT}^{j}=(1-2x)p_{q1^{\prime}}{}_{\bot}^{j}\left(\phi_{4}\right)_{LL}-4(x-z)(\bar{x}+z)(1-2x+z)[(\vec{q}\cdot\vec{p}_{q1^{\prime}})g_{\bot k}^{j}+q_{2\bot}^{j}p_{q1^{\prime}\bot k}]I_{1}^{k} \, ,
\end{equation}
(transverse NLO) $\times$ (longitudinal LO) contribution :
\begin{align}
\left(\phi_{4}\right)_{TL}^{i} & =2\{[(x-\bar{x}-z)q_{2\bot}^{i}q_{1\bot k}+(-8x\bar{x}-6xz+2z^{2}+3z+1)q_{1\perp}^{i}q_{2\bot k}]I_{1}^{k}\nonumber \\
 & -2[4x^{2}-x(3z+5)+(z+1)^{2}]q_{2\bot k}I^{ik}+(x-\bar{x}-z)\left(\vec{q}_{2}\cdot\vec{q}_{1}\right)I^{i}\nonumber \\
 & +I_{2}[(x-\bar{x}-z)q_{2\bot}^{i}+\bar{x}(2(x-z)^{2}-5x+3z+1)q_{1\perp}^{i}]\nonumber \\
 & -\bar{x}[2(x-z)^{2}-5x+3z+1]I_{3}^{i}\nonumber \\
 & +\frac{x\bar{x}(1-2x)}{z}[2q_{2\bot k}I^{ik}+I_{3}^{i}-q_{1\perp}^{i}(2q_{2\bot k}I_{1}^{k}+I_{2})]\} \, ,
\end{align}
(transverse NLO) $\times$ (transverse LO) contribution :
\begin{eqnarray} \nonumber
\left(\phi_{4}\right)_{TT}^{ij} & = & \left[(x-\bar{x}-2z)(x-\bar{x}-z)(\vec{q}_{2}\cdot\vec{p}_{q1^{\prime}})q_{1\perp}^{i}+(z+1)(\left(\vec{q}_{1}\cdot\vec{q}_{2}\right)p_{q1^{\prime}\perp}^{i}-(\vec{q}_{1}\cdot\vec{p}_{q1^{\prime}})q_{2\bot}^{i})\right]I_{1}^{j}\\ \nonumber
 &+& 2\bar{x}[q_{2\bot k}-(x-z)q_{1\perp k}](p_{q1^{\prime}\bot}^{i}I^{jk}-g_{\bot}^{ij}p_{q1^{\prime}\bot l}I^{kl}) \\ \nonumber
 &+& 2(x-z)[(2\bar{x}+z)(\vec{q}_{2}\cdot\vec{p}_{q1^{\prime}})-\bar{x}(\vec{q}_{1}\cdot\vec{p}_{q1^{\prime}})]I^{ij}\\ \nonumber
 &+& [(1-z)((\vec{q}_{1}\cdot\vec{p}_{q1^{\prime}})q_{2\bot}^{j}-(\vec{q}_{2}\cdot\vec{p}_{q1^{\prime}})q_{1\perp}^{j})-(1-2x)(\bar{x}-x+z)\left(\vec{q}_{1}\cdot\vec{q}_{2}\right)p_{q1^{\prime}\perp}^{j}]I_{1}^{i}\\ \nonumber
 &-& 2\left[(x-z)(\bar{x}q_{1\perp}^{j}-(2\bar{x}+z)q_{2\bot}^{j})p_{q1^{\prime}\perp k} \right. \\ \nonumber 
 &+& \left. (1-2x)\left(4x^{2}-(3z+5)x+(z+1)^{2}\right)q_{2\bot k}p_{q1^{\prime}}{}_{\bot}^{j}\right]I^{ik}\\ \nonumber
 &-& \bar{x}\left(\bar{x}-x\right)\left(2(x-z)^{2}-5x+3z+1\right)p_{q1^{\prime}\perp}^{j}I_{3}^{i} \\ \nonumber
 &+& \bar{x}\left(\bar{x}+z\right)(p_{q1^{\prime}\perp}^{i}I_{3}^{j}-g_{\bot}^{ij}p_{q1^{\prime}\perp k}I_{3}^{k})\\ \nonumber
 &+& I_{2}\left[g_{\bot}^{ij}\left((1-z)(\vec{q}_{2}\cdot\vec{p}_{q1^{\prime}})-\bar{x}(1+x-z)(\vec{q}_{1}\cdot\vec{p}_{q1^{\prime}})\right) \right. \\ \nonumber 
 &+& \left.((1-z)q_{2\bot}^{j}-\bar{x}(1+x-z)q_{1\perp}^{j})p_{q1^{\prime}}{}_{\bot}^{i}\right.\\ \nonumber
 &-& \left.(\bar{x}-x)\left((\bar{x}-x+z)q_{2\bot}^{i}-\bar{x}\left(2(x-z)^{2}-5x+3z+1\right)q_{1\perp}^{i}\right)p_{q1^{\prime}}{}_{\bot}^{j}\right]\\ \nonumber
 &+& I_{1}^{k}\left[g_{\bot}^{ij}\left((x-\bar{x}+z)(\vec{q}_{1}\cdot\vec{p}_{q1^{\prime}})q_{2\bot k}+(1-z)(\vec{q}_{2}\cdot\vec{p}_{q1^{\prime}})q_{1\bot k}-(z+1)\left(\vec{q}_{1}\cdot\vec{q}_{2}\right)p_{q1^{\prime}}{}_{\bot k}\right)\right.\\ \nonumber
 &+& q_{1\perp}^{j}((x-\bar{x}+z)q_{2\bot k}p_{q1^{\prime}\perp}^{i}-(z+1)q_{2\bot}^{i}p_{q1^{\prime}}{}_{\bot k})\\ \nonumber
 &+& q_{2\bot}^{j}((x-\bar{x}-2z)(x-\bar{x}-z)q_{1\perp}^{i}p_{q1^{\prime}\perp k}+(1-z)q_{1\perp k}p_{q1^{\prime}}{}_{\bot}^{i})\\ \nonumber
&-&\left.(1-2x)((1-2x+z)q_{2\bot}^{i}q_{1\perp k}-(2z^{2}+3z-x(8\bar{x}+6z)+1)q_{1\perp}^{i}q_{2\bot k})p_{q1^{\prime}}{}_{\bot}^{j}\right]\\ \nonumber
 &+& \frac{x\bar{x}}{z}\left[(x-\bar{x})^{2}p_{q1^{\prime}\perp}^{j}(2q_{2\bot k}I^{ik}+I_{3}^{i}-q_{1\perp}^{i}(I_{2}+2q_{2\bot k}I_{1}^{k}))\right.\\ \nonumber
 &+& p_{q1^{\prime}\perp}^{i}(q_{1\perp}^{j}(I_{2}+2q_{2\bot k}I_{1}^{k})-2q_{2\bot k}I^{jk}-I_{3}^{j})\\
 &+& \left.g_{\bot}^{ij}((\vec{q}_{1}\cdot\vec{p}_{q1^{\prime}})(I_{2}+2q_{2\bot k}I_{1}^{k})+p_{q1^{\prime}\perp k}(2q_{2\bot l}I^{kl}+I_{3}^{k}))\right]\, .
\end{eqnarray}

\subsection{$\phi_5$}

Here the integrals from \ref{sec:building_block} will have the following arguments :

\begin{equation}
\vec{q}_1 = \left( \frac{x-z}{x} \right) \vec{p}_3 -\frac{z}{x}\vec{p}_1, \quad \vec{q}_2 = \vec{p}_{q1} - \frac{z}{x}\vec{p}_q \, ,\label{var1D5}
\end{equation}
\begin{equation}
 \Delta_1 = \frac{z(x-z)}{x^2\bar{x}} (\vec{p}_{\bar{q}2}^{\, \, 2}+ x\bar{x}Q^2), \quad  \Delta_2 = (x-z)(\bar{x}+z)Q^2 \label{var2D5}\, ,
\end{equation}
With such variables, it is easy to see that the argument in the square roots in (\ref{rhovar}) are full squares.
In terms of the variables in (\ref{var1D5}), the impact factors read: \vspace{0.2 cm} \\
(longitudinal NLO) $\times$ (longitudinal LO) : 
\begin{equation}
\left(\phi_{5}\right)_{LL}=\frac{4(x-z)(-2x(\bar{x}+z)+z^{2}+z)}{xz}\left[\bar{x}(x-z)I_{2}-\left(zq_{1\perp k}-x\left(\bar{x}+z\right)q_{2\bot k}\right)I_{1}^{k}\right] \, ,
\end{equation}
(longitudinal NLO) $\times$ (transverse LO) : 
\begin{align}
\left(\phi_{5}\right)_{LT}^{j} & =(\bar{x}-x)p_{q1^{\prime}\bot}^{j}\left(\phi_{5}\right)_{LL} \\ \nonumber &+\frac{4(x-z)(x-\bar{x}-z)}{x}\left(zq_{1\perp}^{k}-x(\bar{x}+z)q_{2\perp}^{k}\right)p_{q1^{\prime}\perp l}\left(g_{\perp k}^{j}I_{1}^{l} - g_{\perp k}^{l} I_{1}^{j}\right)\, ,
\end{align}
(transverse NLO) $\times$ (longitudinal LO) : 
\begin{align}
\left(\phi_{5}\right)_{TL}^{i} & =2\left[(x-\bar{x}-z)\left(\vec{q}_{1}\cdot\vec{q}_{2}\right)-\bar{x}(x-z)^{2}Q^{2}+(\frac{z}{x}-x)\vec{q}_{1}^{\,\,2}\right]I_{1}^{i}\nonumber \\
 & +\frac{2}{x}\left[xq_{2\bot k}(-8x\bar{x}-6xz+2z^{2}+3z+1)+2q_{1\bot k}(2xz-2x^{2}+x-z^{2})\right]q_{1\perp}^{i}I_{1}^{k}\nonumber \\
 & +2q_{2\bot}^{i}q_{1\perp k}(x-\bar{x}-z)I_{1}^{k}+2\frac{\bar{x}}{x}(x(8x-3)-6xz+2z^{2}+z)I_{1}^{i}\nonumber \\
 & +\frac{2}{x}\left[xq_{2\bot}^{i}(x-\bar{x}-z)+q_{1\perp}^{i}(8x^{3}-6x^{2}(z+2)+x(z+3)(2z+1)-2z^{2})\right]I_{2}\nonumber \\
 & -\frac{4}{x}\left[(x-z)(\bar{x}+z)q_{1\perp k}+x(4x^{2}-x(3z+5)+(z+1)^{2})q_{2\perp k}\right]I^{ik}\nonumber \\
 & -\frac{4}{z}x\bar{x}(x-\bar{x})\left[q_{2\perp k}I^{ik}+I_{3}^{i}-q_{1\perp}^{i}\left(q_{2\perp k}I_{1}^{k}+I_{2}\right)\right] \, ,
\end{align}
(transverse NLO) $\times$ (transverse LO) : 

\begin{align*}
& \left(\phi_{5}\right)_{TT}^{ij}  =  -2(x-z)\left[\frac{z}{x}(\vec{q}_{1}\cdot\vec{p}_{q1^{\prime}})-(2\bar{x}+z)(\vec{q}_{2}\cdot\vec{p}_{q1^{\prime}})\right]I^{ij}\\ \nonumber
 & +  \left[-\bar{x}(x-z)^{2}Q^{2}p_{q1^{\prime}\perp}^{i}+(\bar{x}-x+2z)(\bar{x}-x+z)(\vec{q}_{2}\cdot\vec{p}_{q1^{\prime}})q_{1\perp}^{i}\right.\\
 & - \left.(\vec{q}_{1}\cdot\vec{p}_{q1^{\prime}})((z+1)q_{2\bot}^{i}-2\frac{z}{x}(2x-z)q_{1\perp}^{i}) \right. \\ \nonumber
 &+ \left. ((z+1)\left(\vec{q}_{1}\cdot\vec{q}_{2}\right)-\left(x+\frac{z}{x}\right)\vec{r}^{\,\,2})p_{q1^{\prime}\bot}^{i}\right]I_{1}^{j}\\ \nonumber
 & -  2\frac{\bar{x}}{x}(xq_{2\perp k}+(x-z)q_{1\perp k})\left(g_{\bot}^{ij}p_{q1^{\prime}\perp l}I^{kl}-p_{q1^{\prime}}{}_{\bot}^{i}I^{jk}\right)\\ \nonumber
 & + \left[\bar{x}\left(x-\bar{x}\right)(x-z)^{2}Q^{2}p_{q1^{\prime}\perp}^{j}-(z-1)(\vec{q}_{1}\cdot\vec{p}_{q1^{\prime}})q_{2\bot}^{j}\right.\\  \nonumber
 & +  \left.(z-1)(\vec{q}_{2}\cdot\vec{p}_{q1^{\prime}})q_{1\perp}^{j}+\frac{x-\bar{x}}{x}\left((x^{2}-z)\vec{q}_{1}^{\,\,2}+x(\bar{x}-x+z)(\vec{q}_{1}\cdot\vec{q}_{2})\right)p_{q1^{\prime}\perp}^{j}\right]I_{1}^{i}\\ \nonumber
 & +  2\left[\frac{x-\bar{x}}{x}\left(x(4x^{2}-(3z+5)x+(z+1)^{2})q_{2\bot k}+(x-z)(\bar{x}+z)q_{1\perp k}\right)p_{q1^{\prime}\perp}^{j}\right.\\
 & -  \left.\frac{x-z}{x}\left(x(2x-z-2)q_{2\bot}^{j}+zq_{1\perp}^{j}\right)p_{q1^{\prime}\perp k}\right]I^{ik} \\ \nonumber 
 & +  \frac{\bar{x}\left(\bar{x}-x\right)}{x}\left(2z^{2}-6xz+z+x(8x-3)\right)p_{q1^{\prime}\perp}^{j}I_{3}^{i}\\ \nonumber
 & + \left[(x-\bar{x})\left((\bar{x}-x+z)q_{2\bot}^{i}+\left(6(z+2)x-8x^{2}-(z+3)(2z+1)+2\frac{z^{2}}{x}\right)q_{1\perp}^{i}\right)p_{q1^{\prime}\perp}^{j}\right.\\ \nonumber
 & +  \left.(1-z)(g_{\bot}^{ij}(\vec{q}_{2}\cdot\vec{p}_{q1^{\prime}})+q_{2\bot}^{j}p_{q1^{\prime}\perp}^{i})+(2x+z-3)(g_{\bot}^{ij}(\vec{q}_{1}\cdot\vec{p}_{q1^{\prime}})+q_{1\perp}^{j}p_{q1^{\prime}\perp}^{i})\right]I_{2}\\ \nonumber
 & +  \left(3\bar{x}+z-\frac{z}{x}\right)p_{q1^{\prime}\perp}^{i}I_{3}^{j}-\frac{\bar{x}}{x}(3x-z)g_{\bot}^{ij}p_{q1^{\prime}\perp k}I_{3}^{k}\\
 & +  \left[(x-\bar{x})p_{q1^{\prime}\perp}^{j}\left\{ (\bar{x}-x+z)q_{2\bot}^{i}q_{1\perp k}-(2z^{2}-6xz+3z-8x\bar{x}+1)q_{2\perp k}q_{1\perp}^{i}\right.\right.\\ \nonumber
 & -  \left. 2(\bar{x}-x+2z-\frac{z^{2}}{x})q_{1\perp k}q_{1\perp}^{i}\right\} +\bar{x}(x-z)^{2}Q^{2}g_{\bot}^{ij}p_{q1^{\prime}\perp k} \\ \nonumber 
 &+ (1-z)q_{1\perp k}(g_{\bot}^{ij}(\vec{q}_{2}\cdot\vec{p}_{q1^{\prime}})+q_{2\bot}^{j}p_{q1^{\prime}\perp}^{i})\\ \nonumber
 & +  \left((x-\bar{x}+z)q_{2\bot k}-2q_{1\perp k}\right)(g_{\bot}^{ij}(\vec{q}_{1}\cdot\vec{p}_{q1^{\prime}})+q_{1\perp}^{j}p_{q1^{\prime}\perp}^{i}) \\ \nonumber 
 &+ g_{\bot}^{ij}\left(\left(x+\frac{z}{x}\right)\vec{q}_{1}^{\,\,2}-(z+1)(\vec{q}_{1}\cdot\vec{q}_{2})\right)p_{q1^{\prime}\perp k}\\ \nonumber
 & + \left.\left((x-\bar{x}-2z)(x-\bar{x}-z)q_{1\perp}^{i}q_{2\perp}^{j}-(z+1)q_{2\perp}^{i}q_{1\perp}^{j}+2(2x-z)\frac{z}{x}q_{1\perp}^{i}q_{1\perp}^{j}\right)p_{q1^{\prime}\perp k}\right]I_{1}^{k}\\ \nonumber
 & +  \frac{2x\bar{x}}{z}\left[(x-\bar{x})^{2}p_{q1^{\prime}\perp}^{j}(q_{2\bot k}I^{ik}+I_{3}^{i})-p_{q1^{\prime}\perp}^{i}(q_{2\bot k}I^{jk}+I_{3}^{j})+g_{\bot}^{ij}p_{q1^{\prime}\bot k}(q_{2\bot l}I^{kl}+I_{3}^{k})\right.\\ \nonumber
 & +  \left.(I_{2}+q_{2\perp k}I_{1}^{k})\left(g_{\bot}^{ij}(\vec{q}_{1}\cdot\vec{p}_{q1^{\prime}})+q_{1\perp}^{j}p_{q1^{\prime}\perp}^{i}-(1-2x)^{2}q_{1\perp}^{i}p_{q1^{\prime}\perp}^{j}\right)\right]. \numberthis
\end{align*}

\subsection{$\phi_6$}
We will use the variable
\begin{equation}
\vec{q}=\left( \frac{x-z}{x} \right) \vec{p}_{3}-\frac{z}{x}\vec{p}_1 \, .
\end{equation}
(longitudinal NLO) $\times $ (longitudinal NLO) :
\begin{equation}
(\phi_6)_{LL}=-4x\bar{x}^2 J_0 \, ,
\end{equation}
(longitudinal NLO) $\times $ (transverse NLO) :
\begin{equation}
(\phi_6)_{LT}^j = (1-2x)p_{q1^\prime\bot}^j(\phi_6)_{LL}\, ,
\end{equation}
(transverse NLO) $\times $ (longitudinal NLO) :
\begin{equation}
(\phi_6)_{TL}^i = 2\bar{x}\left[ (1-2x) p_{\bar{q}2\bot}^{i} J_0 - J_{1\bot}^i \right] \, ,
\end{equation}
(transverse NLO) $\times $ (transverse NLO) :
\begin{align}
(\phi_6)_{TT}^{ij} &  =\bar{x}\left[(x-\bar{x})^{2}p_{\bar{q}2\bot}^{i}%
p_{q1^{\prime}\bot}^{j}-g_{\bot}^{ij}(\vec{p}_{\bar{q}2}
\cdot\vec{p}_{q1^{\prime}})-p_{q1^{\prime}\bot}^{i}p_{\bar{q}2\bot}^{j}\right]J_0 \nonumber\\
&  +\bar{x} \left[(x-\bar{x})p_{q1^{\prime}\bot}^{j}g_{\bot k}^i - p_{q1^\prime\bot k}g_{\bot}^{ij}+p_{q1^{\prime
}\bot}^{i}g_{\bot k}^j\right]J_{1\bot}^k \, .
\end{align}
We introduced
\begin{align}
J_{1\bot}^{k}  &  =\frac{(x-z)^{2}}{x^{2}}\frac{q_{\bot}^{k}}{\vec
{q}^{\,\,2}}\ln\left(  \frac{\vec{p}_{\bar{q}2}^{\,\,2}+x\bar{x}Q^{2}}{\vec{p}_{\bar{q}2}^{\,\,2}+x\bar{x}Q^{2}+	\frac{x^2 \bar{x}}{z(x-z)}\vec{q}^{\,\,2}%
}\right)  ,\\ \nonumber
\mathrm{and} \\
J_0 &  =\frac{z}{x(\vec{p}_{\bar{q}2}^{\,\,2}+x\bar{x}Q^{2})}  -\frac{2x(x-z)+z^{2}}{xz(\vec{p}_{\bar{q}2}^{\,\,2}+x\bar{x}Q^{2})}
\ln\left(  \frac{x^2\bar{x}\mu^{2}}{z(x-z)(\vec{p}_{\bar{q}2}^{\,\,2}+x\bar{x}Q^{2})+x^{2}\bar{x}\vec{q}^{\,\,2}}\right)  .
\end{align}

\section{Finite part of the squared impact factors for real corrections  }
\label{sec: appendixC}
 \subsection{LL transition}
The double dipole $\times$ double dipole contribution is
\begin{align}
\label{phi4plus_squared}
\Phi &  _{4}^{+}(p_{1\bot},p_{2\bot},p_{3\bot})\Phi_{4}^{+*}(p_{1\bot}^{\prime
},p_{2\bot}^{\prime},p_{3\bot}^{\prime})=\frac{8p_{\gamma}^{+}{}^{4}%
}{z^{2}\left(  \frac{\vec{p}_{\bar{q}2^\prime}^{\, \, 2}%
}{x_{\bar{q}}\left( 1 - x_{\bar{q}} \right) }+Q^{2}\right)  \left( Q^2+\frac{\vec{p}_{q1^{\prime}}^{\,\,2}}{x_{q}%
} + \frac{\vec{p}_{\bar{q}2^\prime}^{\, \, 2}}{x_{\bar{q}}}+\frac
{\vec{p}_{g3^{\prime}}^{\,\,2}}{z}\right)  }\nonumber\\
\times &  \left[  \frac{x_{\bar{q}}\left(  dz^{2}+4x_{q}\left(  x_{q}%
+z\right)  \right)  \left(  x_{q}\vec{p}_{g3}-z\vec{p}_{q1})(x_{q}\vec
{p}_{g3^{\prime}}-z\vec{p}_{q1^{\prime}}\right)  }{x_{q}\left(  x_{q}%
+z\right)  ^{2}{}\left(  \frac{(\vec{p}_{g3}+\vec{p}_{q1}){}^{2}}{x_{\bar{q}%
}\left(  x_{q}+z\right)  }+Q^{2}\right)  \left(  \frac{(\vec{p}_{g3}+\vec
{p}_{q1}){}^{2}}{x_{\bar{q}}}+\frac{\vec{p}_{g3}^{\,\,2}}{z}+\frac{\vec{p}_{q1}%
^{\,\,2}}{x_{q}}+Q^{2}\right)  }\right. \nonumber\\
-  &  \left.  \frac{(4x_{q}x_{\bar{q}}+2z-dz^{2})(x_{\bar{q}}\vec{p}%
_{g3}-z\vec{p}_{\bar{q}2})(x_{q}\vec{p}_{g3^{\prime}}-z\vec{p}_{q1^{\prime}}%
)}{\left(  x_{\bar{q}}+z\right)  \left(  x_{q}+z\right)  \left(  \frac
{(\vec{p}_{\bar{q}2}+\vec{p}_{g3}){}^{2}}{x_{q}\left(  x_{\bar{q}}+z\right)
}+Q^{2}\right)  \left(  \frac{(\vec{p}_{\bar{q}2}+\vec{p}_{g3}){}^{2}}{x_{q}%
}+\frac{\vec{p}_{g3}^{\,\,2}}{z}+\frac{\vec{p}_{\bar{q}2}^{\,\,2}}{x_{\bar{q}}%
}+Q^{2}\right)  }\right]  +(q\leftrightarrow\bar{q}).
\end{align}
The interference term in the dipole $\times$ dipole contribution reads
{\allowdisplaybreaks
\begin{align*}
& \left( \tilde{\Phi}_3^+(\vec{p}_1, \vec{p}_2) \Phi_4^{+*}(\vec{p}_{1'}, \vec{p}_{2'},\vec{0}) +\Phi_4^+(\vec{p}_1, \vec{p}_2,\vec{0}) \tilde{\Phi}_3^{+*}(\vec{p}_{1'},\vec{p}_{2'})\right) \\
&  =\left[  \frac{8p_{\gamma}^{+}{}^{4}}{z\left(  x_{q}+z\right) \left(  \frac{\vec{p}{}_{\bar{q}2^{\prime}}^{\,\,2}}{x_{\bar{q}}\left(x_{q}+z\right)  }+Q^{2}\right)  \left(  \frac{\vec{p}{}_{q1^{\prime}}^{\,\,2}%
}{x_{q}}+\frac{\vec{p}{}_{\bar{q}2^{\prime}}^{\,\,2}}{x_{\bar{q}}}+\frac
{\vec{p}_{g}{}^{2}}{z}+Q^{2}\right)  }\right. \nonumber\\
&  \times\left\{  \frac{\left(  4x_{q}x_{\bar{q}}+z(2-dz)\right)  (\vec{p}_{g}%
-\frac{z}{x_{\bar{q}}}\vec{p}_{\bar{q}})(x_{q}\vec{p}_{g}-z\vec
{p}_{q1^{\prime}})}{(\vec{p}_{g}-\frac{z\vec{p}_{\bar{q}}}{x_{\bar{q}}}){}%
^{2}\left(  \frac{\vec{p}{}_{q1'}^{\,\,2}}{x_{q}\left(  x_{\bar{q}}+z\right)
}+Q^{2}\right)  }\right. \nonumber\\
&  -\left.  \left.  \frac{x_{\bar{q}}\left(  dz^{2}+4x_{q}\left(
x_{q}+z\right)  \right)  ({}\vec{p}_{g}-\frac{z}{x_{q}}\vec{p}_{q})(\vec{p}_{g}%
-\frac{z}{x_{q}}\vec{p}_{q1^{\prime}})}{(\vec{p}_{g}-\frac{z\vec{p}_{q}%
}{x_{q}}){}^{2}\left(  \frac{\vec{p}{}_{\bar{q}2}^{\,\,2}}{x_{\bar{q}}\left(
x_{q}+z\right)  }+Q^{2}\right)  }\right\}  +(q\leftrightarrow\bar{q})\right]
\nonumber\\
&  +(1\leftrightarrow1^{\prime},2\leftrightarrow2^{\prime}). \numberthis
\end{align*}}
The double dipole $\times$ dipole contribution has the form 

\begin{equation}
\Phi_{4}^{+}( \vec{p}_1, \vec{p}_2, \vec{p}_3 )\,\Phi_{3}^{+*}(\vec{p}_{1'}, \vec{p}_{2'})=\Phi_{4}^{+} ( \vec{p}_1, \vec{p}_2, \vec{p}_3 ) \Phi_{4}^{+*}(\vec{p}_{1'}, \vec{p}_{2'}, \vec{0})+\Phi_4^+(\vec{p}_1, \vec{p}_2, \vec{p}_3) \tilde{\Phi}_3^{+*}(\vec{p}_{1'}, \vec{p}_{2'}) ,
\end{equation}
where
\begin{align*}
\Phi_4^+(\vec{p}_1, \vec{p}_2, \vec{p}_3) & \tilde{\Phi}_3^{+*}(\vec{p}_{1'}, \vec{p}_{2'})  =\frac{8p_{\gamma}^{+}{}^{4}}{z\left(  x_{q}+z\right)  \left(
\frac{\vec{p}{}_{\bar{q}2}^{\,\,2}}{x_{\bar{q}}\left(  x_{q}+z\right)  }%
+Q^{2}\right)  \left(  \frac{\vec{p}{}_{q1}^{\,\,2}}{x_{q}}+\frac{\vec{p}%
{}_{\bar{q}2}^{\,\,2}}{x_{\bar{q}}}+\frac{\vec{p}_{g3}^{\,\,2}}{z}+Q^{2}\right)
}\nonumber\\
&  \times\left\{  \frac{\left(  4x_{q}x_{\bar{q}}+z(2-dz)\right)  (\vec{p}_{g}%
-\frac{z}{x_{\bar{q}}}\vec{p}_{\bar{q}})(x_{q}\vec{p}_{g3}-z\vec{p}_{q1}%
)}{(\vec{p}_{g}-\frac{z\vec{p}_{\bar{q}}}{x_{\bar{q}}}){}^{2}\left(
\frac{\vec{p}{}_{q1^{\prime}}^{\,\,2}}{x_{q}\left(  x_{\bar{q}}+z\right)
}+Q^{2}\right)  }\right. \nonumber\\
&  -\left.  \frac{x_{\bar{q}}\left(  dz^{2}+4x_{q}\left(  x_{q}+z\right)
\right)  (\vec{p}_{g}-\frac{z}{x_{q}}\vec{p}_{q})(\vec{p}_{g3}-\frac{z}{x_{q}}%
\vec{p}_{q1})}{(\vec{p}_{g}-\frac{z\vec{p}_{q}}{x_{q}}){}^{2}\left(
\frac{\vec{p}{}_{\bar{q}2^{\prime}}^{\,\,2}}{x_{\bar{q}}\left(  x_{q}%
+z\right)  }+Q^{2}\right)  }\right\}  +( q \leftrightarrow \bar{q} ). \numberthis[finite_double_dipole_dipole_LL]
\end{align*}
For the dipole $\times$ double dipole contribution, one just has to complex conjugate \eqref{eq:finite_double_dipole_dipole_LL} and also invert the name of the momenta i.e. $1',2' \leftrightarrow 1,2$. 

\subsection{LT/TL transition}
The double dipole $\times$ double dipole contribution is 
\begin{align*}
&  \Phi_{4}^{i}(p_{1\bot},p_{2\bot},p_{3\bot})\Phi_{4}^{+*}(p_{1\bot}^{\prime
},p_{2\bot}^{\prime},p_{3\bot}^{\prime}) \\
& =\frac{-4p_{\gamma}^{+}{}^{3}%
}{\left(  Q^{2}+\frac{\vec{p}{}_{g3}^{\,\,2}}{z}+\frac{\vec{p}{}_{q1}^{\,\,2}%
}{x_{q}}+\frac{\vec{p}{}_{{\bar{q}}2}^{\,\,2}}{x_{\bar{q}}}\right)  \left(
Q^{2}+\frac{\vec{p}{}_{g3^{\prime}}^{\,\,2}}{z}+\frac{\vec{p}{}_{q1^{\prime}%
}^{\,\,2}}{x_{q}}+\frac{\vec{p}{}_{{\bar{q}}2^{\prime}}^{\,\,2}}{x_{\bar{q}}%
}\right)  }\nonumber\\
& \hspace{-0.1 cm} \times \hspace{-0.1 cm} \left(  \frac{z\left(  (\vec{P} \hspace{-0.1 cm}  \cdot \hspace{-0.1 cm}  \vec{p}_{q1})G_{\bot}^{i} \hspace{-0.1 cm}  - \hspace{-0.1 cm} (\vec
{G} \hspace{-0.1 cm}  \cdot  \hspace{-0.1 cm} \vec{p}_{q1})P_{\bot}^{i}\right)  \left(  dz+4x_{q}-4\right)  -(\vec{G} \cdot 
\vec{P})p_{q1}^{i}{}_{\bot}\left(  2x_{q}-1\right)  \left(  4\left(
x_{q}-1\right)  x_{\bar{q}}-dz^{2}\right)  }{z^{2}x_{\bar{q}}\left(
z+x_{\bar{q}}\right)  {}^{3}\left(  Q^{2}+\frac{\vec{p}{}_{q1}^{\,\,2}}%
{x_{q}\left(  z+x_{\bar{q}}\right)  }\right)  \left(  Q^{2}+\frac{\vec{p}%
{}_{q1^{\prime}}^{\,\,2}}{x_{q}\left(  z+x_{\bar{q}}\right)  }\right)
}\right. \nonumber\\
 &  + \hspace{-0.05 cm}  \frac{z\left(  (\vec{P} \hspace{-0.1 cm} \cdot \hspace{-0.1 cm}  \vec{p}_{q1})H_{\bot}^{i} \hspace{-0.1 cm} -(\vec{H} \cdot \vec{p}_{q1})P_{\bot}^{i}\right)  \left(  dz+4x_{q}-2\right)  -(\vec{H} \cdot \vec{P}%
)p_{q1}^{i}{}_{\bot}\left(  2x_{q}-1\right)  \left(  z(2-dz)+4x_{q}x_{\bar{q}%
}\right)  }{z^{2}x_{q}\left(  z+x_{q}\right)  \left(  z+x_{\bar{q}}\right)
{}^{2}\left(  Q^{2}+\frac{\vec{p}{}_{\bar{q}2^{\prime}}^{\,\,2}}{\left(
z+x_{q}\right)  x_{\bar{q}}}\right)  \left(  Q^{2}+\frac{\vec{p}{}%
_{q1}^{\,\,2}}{x_{q}\left(  z+x_{\bar{q}}\right)  }\right)  }\nonumber\\
&  +   \left.  \frac{H_{\bot}^{i}\left(  z(zd+d-2)+x_{q}\left(  2-4x_{\bar{q}%
}\right)  \right)  x_{\bar{q}}}{z\left(  z+x_{q}\right)  {}^{2}\left(
z+x_{\bar{q}}\right)  \left(  Q^{2}+\frac{\vec{p}{}_{\bar{q}2^{\prime}%
}^{\,\,2}}{\left(  z+x_{q}\right)  x_{\bar{q}}}\right)  }\right)
+(q\leftrightarrow\bar{q}). \numberthis[phi_4_phi_4_LT]
\end{align*}
Here, 
\begin{equation}
G_{\bot}^{i}=x_{\bar{q}}p_{g3^{\prime}\bot}^{i}-zp_{\bar{q}2^{\prime}\bot}%
^{i},\quad H_{\bot}^{i}=x_{q}p_{g3^{\prime}\bot}^{i}-zp_{q1^{\prime}\bot}%
^{i},\quad P_{\bot}^{i}=x_{\bar{q}}p_{g3\bot}^{i}-zp_{\bar{q}2\bot}^{i}.
\end{equation}
The interference term in the dipole $\times$ dipole contribution reads
\begin{align*}
&  \left( \Phi_4^{i}(\vec{p}_{1}, \vec{p}_2, \vec{0}) \tilde{\Phi}_3^{+*}(\vec{p}_{1'}, \vec{p}_{2'}) + \tilde{\Phi}_3^{i}(\vec{p}_{1}, \vec{p}_{2}) \Phi_4^{+*} (\vec{p}_{1'}, \vec{p}_{2'}, \vec{0})\right) \\
& =4p_{\gamma}^{+}{}^{3}\left(  \frac{\Delta_{q}{}_{\bot}^{i}%
x_{q}x_{\bar{q}}\left(  dz^{2}+dz-2z+2x_{q}-4x_{q}x_{\bar{q}}\right)  }%
{\vec{\Delta}{}_{q}^{2}\left(  z+x_{q}\right)  {}^{2}\left(  z+x_{\bar{q}%
}\right)  \left(  Q^{2}+\frac{\vec{p}_{g}^{\,\,2}}{z}+\frac{\vec{p}{}%
_{q1}^{\,\,2}}{x_{q}}+\frac{\vec{p}{}_{\bar{q}2}^{\,\,2}}{x_{\bar{q}}}\right)
\left(  Q^{2}+\frac{\vec{p}{}_{\bar{q}2^{\prime}}^{\,\,2}}{\left(
z+x_{q}\right)  x_{\bar{q}}}\right)  }\right. \nonumber\\
&  - \frac{(\vec{J} \cdot \vec{\Delta}_{q})p_{\bar{q}2}^{i}{}_{\bot}\left(
dz^{2}+4x_{q}\left(  z+x_{q}\right)  \right)  \left(  1-2x_{\bar{q}}\right)
+z\left(  (\vec{J} \cdot \vec{p}_{\bar{q}2})\Delta_{q}^{i}{}_{\bot}-(\vec{p}_{\bar
{q}2} \cdot \vec{\Delta}_{q})J_{\bot}^{i}\right)  \left(  dz+4x_{\bar{q}}-4\right)
}{z\left(  z+x_{q}\right)  {}^{3}\vec{\Delta}{}_{q}^{2}\left(  Q^{2}%
+\frac{\vec{p}_{g}^{\,\,2}}{z}+\frac{\vec{p}{}_{q1^{\prime}}^{\,\,2}}{x_{q}%
}+\frac{\vec{p}{}_{\bar{q}2^{\prime}}^{\,\,2}}{x_{\bar{q}}}\right)  \left(
Q^{2}+\frac{\vec{p}{}_{\bar{q}2}^{\,\,2}}{\left(  z+x_{q}\right)  x_{\bar{q}}%
}\right)  \left(  Q^{2}+\frac{\vec{p}{}_{\bar{q}2^{\prime}}^{\,\,2}}{\left(
z+x_{q}\right)  x_{\bar{q}}}\right)  }\nonumber\\
&  -\frac{x_{q}\left(  z\left(  (\vec{K} \cdot \vec{p}_{\bar{q}2})\Delta_{q}^{i}%
{}_{\bot}-(\vec{p}_{\bar{q}2} \cdot \vec{\Delta}_{q})K_{\bot}^{i}\right)  \left(
dz+4x_{\bar{q}}-2\right)  +(\vec{K} \cdot \vec{\Delta}_{q})p_{\bar{q}2}^{i}{}_{\bot
}\left(  1-2x_{\bar{q}}\right)  \left(  z(dz-2)-4x_{q}x_{\bar{q}}\right)
\right)  }{z\left(  z+x_{q}\right)  {}^{2}x_{\bar{q}}\left(  z+x_{\bar{q}%
}\right)  \vec{\Delta}{}_{q}^{2}\left(  Q^{2}+\frac{\vec{p}_{g}^{\,\,2}}%
{z}+\frac{\vec{p}{}_{q1^{\prime}}^{\,\,2}}{x_{q}}+\frac{\vec{p}{}_{\bar
{q}2^{\prime}}^{\,\,2}}{x_{\bar{q}}}\right)  \left(  Q^{2}+\frac{\vec{p}%
{}_{\bar{q}2}^{\,\,2}}{\left(  z+x_{q}\right)  x_{\bar{q}}}\right)  \left(
Q^{2}+\frac{\vec{p}{}_{q1^{\prime}}^{\,\,2}}{x_{q}\left(  z+x_{\bar{q}%
}\right)  }\right)  }\nonumber\\
&  -\frac{z\left(  (\vec{p}_{q1} \cdot \vec{\Delta}_{q})X_{\bot}^{i}-(\vec{X} \cdot \vec{p}_{q1}) \Delta_{q}^{i}{}_{\bot}\right)  \left(  dz+4x_{q}-2\right)  +(\vec
{X} \cdot \vec{\Delta}_{q})p_{q1}^{i}{}_{\bot}\left(  1-2x_{q}\right)  \left(
z(dz-2)-4x_{q}x_{\bar{q}}\right)  }{z\vec{\Delta}{}_{q}^{2}\left(
z+x_{q}\right)  \left(  z+x_{\bar{q}}\right)  {}^{2}\left(  Q^{2}+\frac
{\vec{p}_{g}^{\,\,2}}{z}+\frac{\vec{p}{}_{q1}^{\,\,2}}{x_{q}}+\frac{\vec{p}%
{}_{\bar{q}2}^{\,\,2}}{x_{\bar{q}}}\right)  \left(  Q^{2}+\frac{\vec{p}%
{}_{\bar{q}2^{\prime}}^{\,\,2}}{\left(  z+x_{q}\right)  x_{\bar{q}}}\right)
\left(  Q^{2}+\frac{\vec{p}{}_{q1}^{\,\,2}}{x_{q}\left(  z+x_{\bar{q}}\right)
}\right)  }\nonumber\\
&  +\left.  \frac{z\left(  (\vec{X} \cdot \vec{p}_{q1})\Delta_{\bar{q}}^{i}{}_{\bot
}-(\vec{p}_{q1} \cdot \vec{\Delta}_{\bar{q}})X_{\bot}^{i}\right)  \left(
dz+4x_{q}-4\right)  -(\vec{X} \cdot \vec{\Delta}_{\bar{q}})p_{q1}^{i}{}_{\bot}\left(
2x_{q}-1\right)  \left(  4\left(  x_{q}-1\right)  x_{\bar{q}}-dz^{2}\right)
}{z\left(  z+x_{\bar{q}}\right)  {}^{3}\vec{\Delta}{}_{\bar{q}}^{2}\left(
Q^{2}+\frac{\vec{p}_{g}^{\,\,2}}{z}+\frac{\vec{p}{}_{q1}^{\,\,2}}{x_{q}}%
+\frac{\vec{p}{}_{\bar{q}2}^{\,\,2}}{x_{\bar{q}}}\right)  \left(  Q^{2}%
+\frac{\vec{p}{}_{q1}^{\,\,2}}{x_{q}\left(  z+x_{\bar{q}}\right)  }\right)
\left(  Q^{2}+\frac{\vec{p}{}_{q1^{\prime}}^{\,\,2}}{x_{q}\left(  z+x_{\bar
{q}}\right)  }\right)  }\right) \nonumber\\
&  +(q\leftrightarrow\bar{q}) \; , \numberthis[phi_tilde_phi_4_dipole_dipole_LT]
\end{align*}
where 

\begin{equation}
\vec{\Delta}_{q} = \frac{x_q \vec{p}_g - x_g \vec{p}_q}{x_q + x_g} \; , \hspace{1 cm}
\vec{\Delta}_{\bar{q}} = \frac{x_{\bar{q}} \vec{p}_g - x_g \vec{p}_{\bar{q}}}{x_q + x_g}
\end{equation}
\begin{align}
X_{\bot}^{i}   =x_{\bar{q}}p_{g\bot}^{i}-zp_{\bar{q}2\bot}^{i} = & P_{\bot}%
^{i}|_{p_{3}=0},\quad J_{\bot}^{i}=x_{q}p_{g\bot}^{i}-zp_{q1^{\prime}\bot}%
^{i}=H_{\bot}^{i}|_{p_{3}^{\prime}=0},\nonumber\\
K_{\bot}^{i}  &  =x_{\bar{q}}p_{g\bot}^{i}-zp_{\bar{q}2^{\prime}\bot}%
^{i}=G_{\bot}^{i}|_{p_{3}^{\prime}=0}.
\end{align}
The TL transition is obtained from above by complex conjugation and inverting the naming of the different momenta in \eqref{eq:phi_tilde_phi_4_dipole_dipole_LT} and \eqref{eq:phi_4_phi_4_LT}. \\
The double dipole $\times$ dipole have, respectively, the form 
\begin{equation}
\Phi_4^{i}(\vec{p}_{1}, \vec{p}_{2}, \vec{p}_{3}) \Phi_3^{+*}(\vec{p}_{1'}, \vec{p}_{2'}) = \Phi_4^i(\vec{p}_1, \vec{p}_2, \vec{p}_3) \Phi_4^{+*}(\vec{p}_{1'}, \vec{p}_{2'}, 0) + \Phi_4^i(\vec{p}_1, \vec{p}_2, \vec{p}_3) \tilde{\Phi}_3^{+*}(\vec{p}_{1'}, \vec{p}_{2'})  \; ,
\end{equation}
\begin{equation}
   \Phi_4^{+} (\vec{p}_{1}, \vec{p}_{2}, \vec{p}_{3}) \Phi_3 ^{i*}(\vec{p}_{1'}, \vec{p}_{2'}) = \Phi_4^{+} (\vec{p}_{1}, \vec{p}_{2}, \vec{p}_{3}) \Phi_4^{i*}(\vec{p}_{1'}, \vec{p}_{2'}, \vec{0}) + \Phi_4^{+} (\vec{p}_{1}, \vec{p}_{2}, \vec{p}_{3}) \tilde{\Phi}_3^{i*}(\vec{p}_{1'}, \vec{p}_{2'}) \; ,
\end{equation}
where 
\begin{align}
& \Phi_4^i(\vec{p}_1, \vec{p}_2, \vec{p}_3) \tilde{\Phi}_3^{+*}(\vec{p}_{1'}, \vec{p}_{2'})  =\frac{4p_{\gamma}^{+}{}^{3}}{\left(  x_{q}+z\right)  \vec{\Delta}_{q}^{2}\left(  \frac{\vec{p}{}_{\bar{q}2^{\prime}}^{\,\,2}}%
{x_{\bar{q}}\left(  x_{q}+z\right)  }+Q^{2}\right)  \left(  \frac{\vec{p}%
{}_{q1}^{\,\,2}}{x_{q}}+\frac{\vec{p}{}_{\bar{q}2}^{\,\,2}}{x_{\bar{q}}}%
+\frac{\vec{p}_{g3}^{\; 2}}{z}+Q^{2}\right)  } \nonumber\\
&  \times\left\{  \frac{x_{q}x_{\bar{q}}\Delta_{q}^{i}\left(
dz(z+1)-2\left(  1-2x_{q}\right)  \left(  x_{q}+z\right)  \right)  }{\left(
x_{q}+z\right)  {}\left(  x_{\bar{q}}+z\right) } + \frac{\left(  dz+4x_{q}-2\right)  \left(  \Delta_{q}^{i}
\vec{P} \cdot \vec{p}_{q1} -P^{i}   \vec{p}_{q1} \cdot
\vec{\Delta}_{q}  \right)  }{\left(  x_{\bar{q}}+z\right)  {}%
^{2}\left(  \frac{\vec{p}_{q1}^{ \; 2}}{x_{q}\left(  x_{\bar{q}}+z\right)
}+Q^{2}\right)  } \right. \nonumber\\
&  +\frac{\left(  2x_{q}-1\right)  p_{q1}^{i} \vec{P} \cdot
\vec{\Delta}_{q} \left(  z(dz-2)-4x_{q}x_{\bar{q}}\right)
}{z\left(  x_{\bar{q}}+z\right)  {}^{2}\left(  \frac{\vec{p}_{q1}^{\; 2}%
}{x_{q}\left(  x_{\bar{q}}+z\right)  }+Q^{2}\right)} - \frac{\left(  (d-4)z-4x_{q}\right)  \left( W^{i}  \vec{p}_{\bar{q}2} \cdot \vec{\Delta}_{q}  -\Delta_{q}^{i} 
\vec{W} \cdot \vec{p}_{\bar{q}2} \right) }{\left(  x_{q}+z\right)
{}^{2}\left(  \frac{\vec{p}_{\bar{q}2}^{ \; 2}}{x_{\bar{q}}\left(
x_{q}+z\right)  }+Q^{2}\right)  } \nonumber\\
&  +\left.  \frac{\left(  2x_{\bar{q}}-1\right)  \left(  dz^{2}+4x_{q}\left(
x_{q}+z\right)  \right)  p_{\bar{q}2}^{i}  \vec{W} \cdot \vec{\Delta}_{q} }{z\left(  x_{q}+z\right)  {}^{2}\left(  \frac
{\vec{p}_{\bar{q}2}^{\; 2}}{x_{\bar{q}}\left(  x_{q}+z\right)  }%
+Q^{2}\right)  }\right\} +(q \leftrightarrow \bar{q}) \; ,
\end{align}%
and
\begin{align}
\Phi_4^{+} (\vec{p}_{1}, \vec{p}_{2}, \vec{p}_{3}) & \tilde{\Phi}_3^{i*}(\vec{p}_{1'}, \vec{p}_{2'}) =\frac{4p_{\gamma}^{+}{}^{3}}{z\vec{\Delta}{}_{q}^{2}\left(
x_{q}+z\right)  {}^{2}\left(  Q^{2}+\frac{\vec{p}_{g3}^{\,\,2}}{z}+\frac
{\vec{p}{}_{q1}^{\,\,2}}{x_{q}}+\frac{\vec{p}{}_{\bar{q}2}^{\,\,2}}{x_{\bar
{q}}}\right)  \left(  Q^{2}+\frac{\vec{p}{}_{\bar{q}2^{\prime}}^{\,\,2}%
}{\left(  z+x_{q}\right)  x_{\bar{q}}}\right)  }\nonumber\\
&  \times\left[  \frac{x_{q}z\left(  (d-4)z-4x_{q}+2\right)  \left(
P^{i}\left(  \vec{p}_{\bar{q}2^{\prime}} \cdot \vec{\Delta}_{q}\right)
-\Delta_{q}^{i}\left( \vec{P} \cdot \vec{p}_{\bar{q}2^{\prime}}\right)
\right)  }{x_{\bar{q}}\left(  x_{\bar{q}}+z\right)  \left(  \frac
{\vec{p}_{q1}^{\; 2}}{x_{q}\left(  x_{\bar{q}}+z\right)  }+Q^{2}\right)
}\right.  \nonumber\\
&  -\frac{x_{q}\left(  x_{q}-x_{\bar{q}}+z\right)  p_{\bar{q}2^{\prime}}^{i}\left(  \vec{P} \cdot \vec{\Delta}_{q} \right)  \left(  z(dz-2)-4x_{q}%
x_{\bar{q}}\right)}{x_{\bar{q}}\left(  x_{\bar{q}}+z\right)  \left(
\frac{\vec{p}_{q1}^{\; 2}}{x_{q}\left(  x_{\bar{q}}+z\right)  }%
+Q^{2}\right)  }\nonumber\\
&  -\frac{\left(  x_{q}-x_{\bar{q}}+z\right)  \left(  dz^{2}+4x_{q}\left(
x_{q}+z\right)  \right)  p_{\bar{q}2^{\prime}}^{i}\left(  \vec{W} \cdot \vec{\Delta}_{q} \right)  }{\left(  x_{q}+z\right)  {}\left(
\frac{\vec{p}_{\bar{q}2}^{\; 2}}{x_{\bar{q}}\left(  x_{q}+z\right)
}+Q^{2}\right)  }\nonumber\\
&  -\left.  \frac{z\left(  (d-4)z-4x_{q}\right)  \left(  \Delta_{q}^{i} \left( \vec{W} \cdot \vec{p}_{\bar{q}2^{\prime}} \right) - W^{i} \left(
\vec{p}_{\bar{q}2^{\prime}} \cdot \vec{\Delta}_{q} \right)  \right)
}{\left(  x_{q}+z\right)  {}\left(  \frac{\vec{p}_{\bar{q}2}^{\,\, 2}%
}{x_{\bar{q}}\left(  x_{q}+z\right)  }+Q^{2}\right)  }\right] +(q \leftrightarrow \bar{q}) .
\end{align}
Here, we introduced 
\begin{equation}
W_{\bot}^{i}=x_{q}p_{g3\bot}^{i}-zp_{q1\bot}^{i}.
\end{equation}

\subsection{TT transition}
The double dipole $\times$ double dipole contribution is 

\begin{align}
&  \Phi_{4}^{i}(p_{1\bot},p_{2\bot},p_{3\bot})\Phi_{4}^{k}(p_{1\bot}^{\prime
},p_{2\bot}^{\prime},p_{3\bot}^{\prime})^{\ast}=\left(  \frac{p_{\gamma}^{+}%
{}^{2}}{\left(  Q^{2}+\frac{\vec{p}_{g3}^{\;2}}{z}+\frac{\vec{p}_{q1}^{\;2}%
}{x_{q}}+\frac{\vec{p}_{\bar{q}2}^{\,\,2}}{x_{\bar{q}}}\right)  \left(
Q^{2}+\frac{\vec{p}_{g3^{\prime}}^{\;2}}{z}+\frac{\vec{p}_{q1^{\prime}}^{\;2}%
}{x_{q}}+\frac{\vec{p}_{\bar{q}2^{\prime}}^{\,\,2}}{x_{\bar{q}}}\right)
}\right. \nonumber\\
&  \times\left[  -\frac{g_{\bot}^{ik}x_{q}x_{\bar{q}}\left(zd+d-2+2x_{\bar
{q}}\right)  }{\left(  z+x_{q}\right)^{2}\left(z+x_{\bar{q}}\right)
}-\frac{2P_{\bot}^{k} p_{q1\bot}^{i}\left( 1-2x_{q}\right)  }{z\left(
z+x_{\bar{q}}\right)^{2}\left( Q^{2}+\frac{\vec{p}_{q1}^{\;2}}%
{x_{q}\left(  z+x_{\bar{q}}\right)  }\right)  }\left(  \frac{(d-2)z-2x_{\bar
{q}}}{z+x_{\bar{q}}}+\frac{dz+2x_{\bar{q}}}{z+x_{q}}\right)  \right.
\nonumber\\
&  -\frac{2\left(  g_{\bot}^{ik}(\vec{P} \cdot \vec{p}_{q1})+P_{\bot}^{i}p_{q1\bot}%
{}^{k}\right)  }{z\left(  z+x_{\bar{q}}\right)^{2}\left(  Q^{2}+\frac
{\vec{p}_{q1}^{\;2}}{x_{q}\left(  z+x_{\bar{q}}\right)  }\right)  }\left(
\frac{(d-4)z-2x_{\bar{q}}}{z+x_{q}}+\frac{(d-2)z-2x_{\bar{q}}}{z+x_{\bar{q}}%
}\right) \nonumber
\end{align}%
\begin{align}
&  -\frac{1}{z^{2} x_{q}\left(  z+x_{q}\right)^{2} x_{\bar{q}}\left(
z+x_{\bar{q}}\right)^{2}\left(  Q^{2}+\frac{\vec{p}_{\bar{q}2^{\prime}%
}^{\,\,2}}{\left(  z+x_{q}\right)  x_{\bar{q}}}\right)  \left(  Q^{2}%
+\frac{\vec{p}_{q1}^{\;2}}{x_{q}\left(  z+x_{\bar{q}}\right)  }\right)
}\left\{  (\vec{H}  \cdot \vec{P})\left[  p_{q1\bot}^{i}{}p_{\bar{q}2^{\prime}\bot
}^{k}{}\left(  1-2x_{q}\right)  \right.  \right. \nonumber\\
&  \times\left.  \left(  1-2x_{\bar{q}}\right)  \left(  z(2-dz)+4x_{q}%
x_{\bar{q}}\right)  +(g_{\bot}^{ik}(\vec{p}_{q1}  \cdot \vec{p}_{\bar{q}2^{\prime}%
})+p_{q1\bot}^{k} p_{\bar{q}2^{\prime}\bot}^{i})\left(  z(2-(d-4)z)+4x_{q}%
x_{\bar{q}}\right)  \right] \nonumber\\
&  +((d-4)z-2)\left[  z(\vec{H}  \cdot  \vec{p}_{\bar{q}2^{\prime}})(g_{\bot}^{ik}%
(\vec{P}  \cdot  \vec{p}_{q1})+P_{\bot}^{i}p_{q1\bot}^{k})+z H_{\bot}^{k}\left(
(\vec{P}  \cdot \vec{p}_{q1})p_{\bar{q}2^{\prime}\bot}^{i}-(\vec{p}_{q1}  \cdot \vec{p}_{\bar{q}2^{\prime}})P_{\bot}^{i}\right)  \right] \nonumber\\
&  +((d-4)z+2)\left[  zH^{i}\left(  (\vec{P}  \cdot  \vec{p}_{\bar{q}2^{\prime}%
})p_{q1\bot}^{k}-(\vec{p}_{q1}  \cdot \vec{p}_{\bar{q}2^{\prime}}) P_{\bot}%
^{k}\right)  +z(\vec{H}  \cdot  \vec{p}_{q1})(g_{\bot}^{ik}(\vec{P}  \cdot \vec{p}_{\bar{q}2^{\prime}})+P_{\bot}^{k}p_{\bar{q}2^{\prime}\bot}^{i})\right]
\nonumber\\
&  +\left.  2z\left(  (\vec{H}  \cdot  \vec{p}_{\bar{q}2^{\prime}})P_{\bot}^{k}%
-(\vec{P}  \cdot  \vec{p}_{\bar{q}2^{\prime}}) H_{\bot}^{k}\right)  p_{q1\bot}{}%
^{i}\left(  1-2x_{q}\right)  \left(  dz+4x_{\bar{q}}-2\right)  \right\}
\nonumber
\end{align}%
\begin{align}
&  -\frac{1}{z^{2}x_{q}x_{\bar{q}}\left(  z+x_{\bar{q}}\right)  {}^{4}\left(
Q^{2}+\frac{\vec{p}_{q1}^{\;2}}{x_{q}\left(  z+x_{\bar{q}}\right)  }\right)
\left(  Q^{2}+\frac{\vec{p}_{q1^{\prime}}^{\;2}}{x_{q}\left(  z+x_{\bar{q}%
}\right)  }\right)  }\left\{  z\left(  (d-4)z-4x_{\bar{q}}\right)  \frac{{}%
}{{}}\right. \nonumber\\
&  \times\left[  g_{\bot}^{ik}\left(  (\vec{G}  \cdot \vec{p}_{q1^{\prime}}) (\vec{P}  \cdot  \vec{p}_{q1})-(\vec{G}  \cdot  \vec{p}_{q1})(\vec{P}  \cdot  \vec{p}_{q1^{\prime}})\right) +(\vec{p}_{q1}  \cdot  \vec{p}_{q1^{\prime}})\left(  G_{\bot}^{i}P_{\bot}^{k}-G_{\bot}^{k}P_{\bot}^{i}\right)  \right. \nonumber\\
&  +\left.  2(\vec{G}  \cdot  \vec{p}_{q1^{\prime}})\left(  P_{\bot}^{i}p_{q1\bot}^{k}+P_{\bot}^{k} p_{q1\bot}^{i}\left(  1-2x_{q}\right)  \right)
-2(\vec{G}  \cdot  \vec{p}_{q1}) \left(  P_{\bot}^{k} p_{q1^{\prime}\bot}^{i}+P_{\bot}^{i} p_{q1^{\prime}\bot}^{k}\left(  1-2x_{q}\right)  \right)  \right] \nonumber\\
&  +\left.  \left.  (\vec{G}  \cdot  \vec{P})\left[  p_{q1\bot}^{k} p_{q1^{\prime}\bot}^{i}-p_{q1\bot}^{i} p_{q1^{\prime}\bot}^{k}\left(  1-2x_{q}\right)^{2}+g_{\bot}^{ik}\left(  \vec{p}_{q1}  \cdot  \vec{p}_{q1^{\prime}}\right)  \right] \left(  dz^{2}+4x_{\bar{q}}\left(  z+x_{\bar{q}}\right)  \right)  \right\}
\right] \nonumber\\
&  +\left.  \frac{{}}{{}}(1\leftrightarrow1^{\prime},2\leftrightarrow
2^{\prime},3\leftrightarrow3^{\prime},i\leftrightarrow k)\right)
+(q\leftrightarrow\bar{q}).
\end{align}
The interference term in the dipole $\times$ dipole contribution reads
\begin{align*}
& \left( \tilde{\Phi}_3^i(\vec{p}_1,\vec{p}_2)\Phi_4^{k*}(\vec{p}_{1'}, \vec{p}_{2'}, \vec{0}) + \Phi_4^i(\vec{p}_1, \vec{p}_{2}, \vec{0}) \tilde{\Phi}_3^{k*}(\vec{p}_{1'}, \vec{p}_{2'}) \right) \\
&  =\left(  \frac{2p_{\gamma}^{+}{}^{2}}{\vec{\Delta}{}_{q}^{2}\left(
Q^{2}+\frac{\vec{p}_{g}^{\,\,2}}{z}+\frac{\vec{p}{}_{q1}^{\,\,2}}{x_{q}}%
+\frac{\vec{p}{}_{\bar{q}2}^{\,\,2}}{x_{\bar{q}}}\right)  \left(  Q^{2}%
+\frac{\vec{p}{}_{\bar{q}2^{\prime}}^{\,\,2}}{\left(  z+x_{q}\right)
x_{\bar{q}}}\right)  }\right. \\
& \times \left[  \frac{\left(  (d-2)z-2x_{q}\right)  x_{q}}{\left(
z+x_{q}\right)  {}^{3}}\left(  g_{\bot}^{ik}(\vec{p}_{\bar{q}2^{\prime}} \cdot \vec{\Delta}_{q})+p_{\bar{q}2^{\prime}}{}_{\bot}^{i}\Delta_{q\bot}^{k} +p_{\bar{q}2^{\prime}\bot}^{k}\Delta_{q\bot}^{i}\left(  1-2x_{\bar{q}}\right)  \right)  \right. \\
&  +\frac{x_{q}\left(  \left(  (d-4)z-2x_{q}\right)  \left(  g_{\bot}^{ik}(\vec{p}_{\bar{q}2^{\prime}} \cdot \vec{\Delta}_{q})+p_{\bar{q}2^{\prime}\bot}^{i}{}\Delta_{q\bot}^{k}{}\right)  +p_{\bar{q}2^{\prime}\bot}^{k} \Delta_{q\bot}^{i} \left(  dz+2x_{q}\right)  \left(  1-2x_{\bar{q}}\right) \right)  }{\left(  z+x_{q}\right)  {}^{2}\left(  z+x_{\bar{q}}\right)  } \\
& -\frac{1}{z\left(  z+x_{q}\right)  {}^{2}x_{\bar{q}}\left(  z+x_{\bar{q}%
}\right)  {}^{2}\left(  Q^{2}+\frac{\vec{p}_{q1}^{\;2}}{x_{q}\left(
z+x_{\bar{q}}\right)  }\right)  }\left\{  z((d-4)z+2)\frac{{}}{{}}\right. \\
& \hspace{-0.1 cm} \times \hspace{-0.1 cm} \left[  p_{q1}{}_{\bot}^{i} \left(  (\vec{p}_{\bar{q}2^{\prime}}\cdot \vec{\Delta}_{q})X_{\bot}^{k}-(\vec{X} \cdot \vec{p}_{\bar{q}2^{\prime}}) \Delta_{q \bot}^{k}\right)  \left( 2x_{q}-1\right)  -(\vec{X} \cdot \vec{p}_{\bar{q}2^{\prime}})\left(  g_{\bot}^{ik}(\vec{p}_{q1} \cdot \vec{\Delta}_{q})+p_{q1\bot}^{k} \Delta_{q\bot}^{i}\right)  \right. \\
&  \left. - \hspace{-0.05 cm}  X_{\bot}^{k} \hspace{-0.05 cm} \left(  (\vec{p}_{q1} \hspace{-0.05 cm} \cdot \hspace{-0.05 cm} \vec{\Delta}_{q}) p_{\bar
{q}2^{\prime} \bot}^{i} \hspace{-0.1 cm} -(\vec{p}_{q1} \hspace{-0.05 cm} \cdot \vec{p}_{\bar{q}2^{\prime}})\Delta_{q}{}_{\bot}^{i}\right)  \right] \hspace{-0.05 cm} + \hspace{-0.05 cm} 4x_{q}z \left(  1 \hspace{-0.05 cm} - \hspace{-0.05 cm} 2x_{q}\right)
p_{q1\bot}^{i} \hspace{-0.1 cm} \left(  (\vec{p}_{\bar{q}2^{\prime}} \hspace{-0.05 cm} \cdot \hspace{-0.05 cm} \vec{\Delta}%
_{q})X_{\bot}^{k} \hspace{-0.1 cm} - \hspace{-0.1 cm}(\vec{X} \hspace{-0.05 cm} \cdot \vec{p}_{\bar{q}2^{\prime}}) \Delta_{q\bot}%
^{k}\right)  \\
&  +z\left(  1-2x_{\bar{q}}\right)  \left(  dz+4x_{q}-2\right)  p_{\bar
{q}2^{\prime}\bot}^{k}\left(  (\vec{p}_{q1} \cdot \vec{\Delta}_{q}) X_{\bot}%
^{i}-(\vec{X} \cdot \vec{p}_{q1}) \Delta_{q\bot}^{i}\right)  -z((d-4)z-2)\\
&  \times \left[  \left(  g_{\bot}^{ik} (\vec{X} \cdot \vec{p}_{q1})+X_{\bot}^{i} 
p_{q1\bot}^{k}\right)  (\vec{p}_{\bar{q}2^{\prime}} \cdot \vec{\Delta}%
_{q})+ \left(  (\vec{X}\vec{p}_{q1})p_{\bar{q}2^{\prime}}{}_{\bot}^{i}-(\vec
{p}_{q1}\vec{p}_{\bar{q}2^{\prime}})X_{\bot}^{i}\right)  \Delta_{q\bot
}^{k}\right] \\
&  +(\vec{X} \cdot \vec{\Delta}_{q})p_{q1\bot}^{i}p_{\bar{q}2^{\prime} \bot
}^{k}\left( 1-2x_{q}\right)  \left( 1-2x_{\bar{q}}\right)  \left(
z(d z-2)-4x_{q}x_{\bar{q}}\right) \\
&  - \left.  (\vec{X} \cdot \vec{\Delta}_{q})\left(  g_{\bot}^{ik}(\vec{p}_{q1} \cdot \vec{p}_{\bar{q}2^{\prime}})+p_{q1\bot}^{k} p_{\bar{q}2^{\prime}\bot}^{i}\right)  \left(  z(2-(d-4)z)+4x_{q}x_{\bar{q}}\right)  \right\} \\
& -    \frac{1}{z\left(  z+x_{q}\right)  {}^{4}\left(  Q^{2}+\frac{\vec{p}%
{}_{\bar{q}2}^{\,\,2}}{\left(  z+x_{q}\right)  x_{\bar{q}}}\right)  x_{\bar
{q}}}\left\{  z\left(  dz+4x_{\bar{q}}-4\right)  \left[  \left(  1-2x_{\bar
{q}}\right)  \frac{{}}{{}}\right.  \right. \nonumber\\
&  \times  \left(  p_{\bar{q}2^{\prime}\bot}^{k} \left(  (\vec{p}_{\bar
{q}2} \cdot \vec{\Delta}_{q}) V_{\bot}^{i}-(\vec{V} \cdot \vec{p}_{\bar{q}2})\Delta_{q\bot}^{i} \right)  +p_{\bar{q}2\bot}^{i}\left(  (\vec{V}\vec{p}_{\bar{q}2^{\prime}})\Delta_{q\bot}^{k}-(\vec{p}_{\bar{q}2^{\prime}} \cdot \vec{\Delta}_{q}) V_{\bot}^{k}\right)  \right) \nonumber\\
& +    V_{\bot}^{k} \left(  (\vec{p}_{\bar{q}2} \cdot \vec{\Delta}_{q}) p_{\bar
{q}2^{\prime}\bot}^{i}-(\vec{p}_{\bar{q}2} \cdot \vec{p}_{\bar{q}2^{\prime}}) \Delta_{q\bot}^{i}\right)  + \left(  (\vec{p}_{\bar{q}2} \cdot \vec{p}_{\bar
{q}2^{\prime}}) V_{\bot}^{i}-(\vec{V} \cdot \vec{p}_{\bar{q}2}) p_{\bar{q}2^{\prime}\bot}^{i}\right)  \Delta_{q \bot}^{k}\nonumber\\
& +    \left.  g_{\bot}^{ik} \left(  (\vec{V} \cdot \vec{p}_{\bar{q}2^{\prime}}%
)(\vec{p}_{\bar{q}2} \cdot \vec{\Delta}_{q})-(\vec{V} \cdot \vec{p}_{\bar{q}2} )(\vec
{p}_{\bar{q}2^{\prime}} \cdot  \vec{\Delta}_{q}) \right)  + p_{\bar{q}2\bot}^{k}\left(  (\vec{V} \cdot \vec{p}_{\bar{q}2^{\prime}}) \Delta_{q\bot}^{i} 
- (\vec{p}_{\bar{q}2^{\prime}} \cdot \vec{\Delta}_{q}) V_{\bot}^{i}\right)  \right]
\nonumber\\
& +    \left.  \left.  (\vec{V} \cdot \vec{\Delta}_{q})\left(  p_{\bar{q}2\bot
}^{i}p_{\bar{q}2^{\prime}\bot}^{k}\left( 1-2x_{\bar{q}}\right)^{2}-g_{\bot}^{ik}(\vec{p}_{\bar{q}2} \cdot \vec{p}_{\bar{q}2^{\prime}})-p_{\bar{q}2\bot}^{k} p_{\bar{q}2^{\prime} \bot}^{i}\right)  \left(d z^{2}-4x_{q}\left( x_{\bar{q}}-1\right)  \right)  \right\}  \frac{{}}{{}%
}\right] \nonumber \\
&  +   \left.  \frac{{}}{{}}(1\leftrightarrow1^{\prime},2\leftrightarrow
2^{\prime},i\leftrightarrow k)\right)  +(q\leftrightarrow\bar{q}). \numberthis
\end{align*}
Here,
\begin{equation}
V_{\bot}^{i}=x_{q}p_{g\bot}^{i}-zp_{q1\bot}^{i}.
\end{equation}
The double dipole $\times$ dipole contribution has the form  
\begin{equation}
\Phi_{4}^{i}(\vec{p}_1, \vec{p}_2, \vec{p}_3)\,\Phi_{3}^{k*}(\vec{p}_{1'}, \vec{p}_{2'})=\Phi_{4}^{i}(\vec{p}_1, \vec{p}_2, \vec{p}_3)\Phi_{4}^{k*}(\vec{p}_{1'}, \vec{p}_{2'}, \vec{0})+\Phi_4^i(\vec{p}_1, \vec{p}_2, \vec{p}_3) \tilde{\Phi}_3^{k*}(\vec{p}_{1'}, \vec{p}_{2'}),
\end{equation}
where
\begin{align*}
& \Phi_4^i(\vec{p}_1, \vec{p}_2, \vec{p}_3) \tilde{\Phi}_3^{k*}(\vec{p}_{1'}, \vec{p}_{2'})  \\
&  =\frac{2p_{\gamma}^{+}{}^{2}}{\vec{\Delta}{}_{q}^{2}\left(
Q^{2}+\frac{\vec{p}_{g3}^{\,\,2}}{z}+\frac{\vec{p}{}_{q1}^{\,\,2}}{x_{q}%
}+\frac{\vec{p}{}_{\bar{q}2}^{\,\,2}}{x_{\bar{q}}}\right)  \left(  Q^{2}%
+\frac{\vec{p}{}_{\bar{q}2^{\prime}}^{\,\,2}}{\left(  z+x_{q}\right)
x_{\bar{q}}}\right)  }\nonumber\\
&  \times\left[  \frac{\left(  (d-2)z-2x_{q}\right)  x_{q}}{\left(
z+x_{q}\right)  {}^{3}}\left(  g_{\bot}^{ik}(\vec{p}_{\bar{q}2^{\prime}}%
\vec{\Delta}_{q})+p_{\bar{q}2^{\prime}}{}_{\bot}^{i}\Delta_{q\bot}{}%
^{k}+p_{\bar{q}2^{\prime}\bot}{}^{k}\Delta_{q\bot}{}^{i}\left(  1-2x_{\bar{q}%
}\right)  \right)  \right. \nonumber\\
&  +\frac{x_{q}\left(  \left(  (d-4)z-2x_{q}\right)  \left(  g_{\bot}%
^{ik}(\vec{p}_{\bar{q}2^{\prime}}\vec{\Delta}_{q})+p_{\bar{q}2^{\prime}\bot
}^{i}{}\Delta_{q\bot}^{k}{}\right)  +p_{\bar{q}2^{\prime}\bot}^{k}{}%
\Delta_{q\bot}^{i}{}\left(  dz+2x_{q}\right)  \left(  1-2x_{\bar{q}}\right)
\right)  }{\left(  z+x_{q}\right)  {}^{2}\left(  z+x_{\bar{q}}\right)
}\nonumber\\
&  -\frac{1}{z\left(  z+x_{q}\right)  {}^{2}x_{\bar{q}}\left(  z+x_{\bar{q}%
}\right)  {}^{2}\left(  Q^{2}+\frac{\vec{p}_{q1}^{\;2}}{x_{q}\left(
z+x_{\bar{q}}\right)  }\right)  }\left\{  z((d-4)z+2)\frac{{}}{{}}\right.
\nonumber\\
&  \times\left[  p_{q1}{}_{\bot}^{i}\left(  (\vec{p}_{\bar{q}2^{\prime}}%
\vec{\Delta}_{q})P_{\bot}^{k}-(\vec{P}\vec{p}_{\bar{q}2^{\prime}})\Delta_{q}%
{}_{\bot}^{k}\right)  \left(  2x_{q}-1\right)  -(\vec{P}\vec{p}_{\bar
{q}2^{\prime}})\left(  g_{\bot}^{ik}(\vec{p}_{q1}\vec{\Delta}_{q})+p_{q1}%
{}_{\bot}^{k}\Delta_{q}{}_{\bot}^{i}\right)  \right. \nonumber\\
&  -\!\left.  P_{\bot}^{k}\!\left(  \!(\vec{p}_{q1}\vec{\Delta}_{q})p_{\bar
{q}2^{\prime}}{}_{\bot}^{i}\!-\!(\vec{p}_{q1}\vec{p}_{\bar{q}2^{\prime}%
})\Delta_{q}{}_{\bot}^{i}\!\right)  \!\right]  +4x_{q}z\left(  1-2x_{q}%
\right)  p_{q1}{}_{\bot}^{i}\!\left(  \!(\vec{p}_{\bar{q}2^{\prime}}%
\vec{\Delta}_{q})P_{\bot}^{k}-(\vec{P}\vec{p}_{\bar{q}2^{\prime}})\Delta_{q}%
{}_{\bot}^{k}\!\right) \nonumber\\
&  +z\left(  1-2x_{\bar{q}}\right)  \left(  dz+4x_{q}-2\right)  p_{\bar
{q}2^{\prime}}{}_{\bot}^{k}\left(  (\vec{p}_{q1}\vec{\Delta}_{q})P_{\bot}%
^{i}-(\vec{P}\vec{p}_{q1})\Delta_{q}{}_{\bot}^{i}\right)
-z((d-4)z-2)\nonumber\\
&  \times\left[  \left(  g_{\bot}^{ik}(\vec{P}\vec{p}_{q1})+P_{\bot}^{i}%
p_{q1}{}_{\bot}^{k}\right)  (\vec{p}_{\bar{q}2^{\prime}}\vec{\Delta}%
_{q})+\left(  (\vec{P}\vec{p}_{q1})p_{\bar{q}2^{\prime}}{}_{\bot}^{i}-(\vec
{p}_{q1}\vec{p}_{\bar{q}2^{\prime}})P_{\bot}^{i}\right)  \Delta_{q}{}_{\bot
}^{k}\right] \nonumber\\
&  +(\vec{P}\vec{\Delta}_{q})p_{q1}{}_{\bot}^{i}p_{\bar{q}2^{\prime}}{}_{\bot
}^{k}\left(  1-2x_{q}\right)  \left(  1-2x_{\bar{q}}\right)  \left(
z(dz-2)-4x_{q}x_{\bar{q}}\right) \nonumber\\
&  -\left.  (\vec{P}\vec{\Delta}_{q})\left(  g_{\bot}^{ik}(\vec{p}_{q1}\vec
{p}_{\bar{q}2^{\prime}})+p_{q1}{}_{\bot}^{k}p_{\bar{q}2^{\prime}}{}_{\bot}%
^{i}\right)  \left(  z(2-(d-4)z)+4x_{q}x_{\bar{q}}\right)  \right\} \; . \numberthis
\end{align*}%
As above, the dipole $\times$ double dipole contribution is obtained by complex conjugation and changing the momenta.

\providecommand{\href}[2]{#2}\begingroup\raggedright\endgroup

\end{document}